\documentclass[12pt]{article}
\usepackage{a41}

\usepackage{palatino}
\usepackage{xcolor}
\usepackage{amsmath}
\usepackage{booktabs}
\usepackage{microtype}

\usepackage{cite}
\usepackage{amsmath}
\usepackage{amssymb}
\usepackage{color}
\usepackage[rflt]{floatflt}
\usepackage{float}

\usepackage[defaultcolor=red]{changes}

\newcommand{\centeredgraphics}[2][]{\vcenter{\hbox{\includegraphics[#1]{#2}}}}

\newcommand{\PI}{p_\infty}

\usepackage[rflt]{floatflt}
\usepackage{float}
\usepackage{slashed}

\def\b0{\beta_0}

\usepackage{array}

\DeclareMathOperator{\tr}{tr}

\usepackage[english]{babel}
\usepackage[latin1]{inputenc}
\usepackage[T1]{fontenc}
\usepackage{ae}

\usepackage{url}

\usepackage{amsmath, amsthm, amssymb}
\newtheorem{thm}{Theorem}[section]

\newtheorem{definition}[thm]{Definition}

\newcommand{\ep}{\varepsilon}

\usepackage{rotating}

\usepackage{graphicx}

\newcounter{mmacnt}
\def\restartmma{\setcounter{mmacnt}{0}}
\restartmma \catcode`|=\active
\def|#1|{\mathrm{#1}}
\catcode`|=12
\newenvironment{mma}{
 \par\smallskip
 \catcode`|=\active
 \parskip=0pt\parindent=0pt 
 \small
 \def\In##1\\{%
\def\linebreak{\hfill\break\null\qquad}%
\refstepcounter{mmacnt}
\hangindent=2.5em\hangafter=0
\leavevmode
\llap{\tiny\sffamily n[\arabic{mmacnt}]:=\kern.5em}%
\mathversion{bold}\footnotesize$\displaystyle##1$\normalsize
\mathversion{normal}\par
 }%
 \def\Print##1\\{%
\def\linebreak{\hfill\break}%
\hangindent=2.5em\hangafter=0
\leavevmode ##1\par}%
 \def\Out##1\\{%
\def\linebreak{$\hfill\break\null\hfill$}%
\kern\abovedisplayskip\par
\hangindent=2.5em\hangafter=0
\leavevmode
\llap{\tiny\sffamily Out[\arabic{mmacnt}]=\kern.5em}
\footnotesize$\displaystyle##1$\normalsize\hfill\null\par
\kern\belowdisplayskip
 }%
 \def\Warning##1##2\\{%
\def\linebreak{\hfill\break}%
\hangindent=2.5em\hangafter=0
\leavevmode
{\scriptsize##1 : ##2}\par}%
}{%
 \par\smallskip
}

\usepackage{color}

\newenvironment{fshaded}{%
\MakeFramed {\FrameRestore}
}%
{\endMakeFramed}

\newcommand*\pFqskip{8mu}
\catcode`,\active
\newcommand*\pFq{\begingroup
        \catcode`\,\active
        \def ,{\mskip\pFqskip\relax}%
        \dopFq
}
\catcode`\,12
\def\dopFq#1#2#3#4#5{%
        {}_{#1}F_{#2}\biggl[\genfrac..{0pt}{}{#3}{#4};#5\biggr]%
        \endgroup
}

\def\b0{\beta_0}

\def\Gp0{{\Gamma^{'}_0}}
\def\Gp1{{\Gamma^{'}_1}}
\def\Gp2{{\Gamma^{'}_2}}

\allowdisplaybreaks[4]
\allowdisplaybreaks[4]

\begin{document}
\setlength{\baselineskip}{0.515cm}

\sloppy
\thispagestyle{empty}
\begin{flushleft}
DESY 21--151
\hfill  {\tt arXiv:2110.13822 [gr-qc]}
\\
DO--TH 21/27\\
SAGEX--21--30\\
October 2021
\end{flushleft}

\mbox{}
\vspace*{\fill}
\begin{center}

{\Large\bf
The fifth-order post-Newtonian Hamiltonian dynamics of}

\vspace*{4mm}
{\Large\bf
two-body systems from an effective field theory approach}

\vspace{3cm} \large
{\large J.~Bl\"umlein$^a$, A.~Maier$^a$, P.~Marquard$^a$, and G.~Sch\"afer$^b$}

\vspace{1.cm}
\normalsize
{\it $^a$Deutsches Elektronen--Synchrotron DESY,}\\
{\it   Platanenallee 6, 15738 Zeuthen, Germany}

\vspace*{2mm}
{\it $^b$Theoretisch-Physikalisches Institut, Friedrich-Schiller-Universit\"at, \\
Max-Wien-Platz 1, D--07743 Jena, Germany}\\


\end{center}
\normalsize
\vspace{\fill}
\begin{abstract}
\noindent
Within an effective field theory method to general relativity, we calculate the fifth-order post--Newtonian (5PN)
Hamiltonian dynamics also for the tail terms, extending earlier work  on the potential contributions, working in
harmonic coordinates. Here we calculate independently all (local) 5PN {far-zone} contributions using the
in--in formalism, on which we give a detailed account. The five expansion terms of the Hamiltonian in the effective
one body (EOB) approach, $q_{82}, q_{63}, q_{44}, \bar{d_5}$ and $an$, can all be determined from the local
contributions to the periastron advance $K^{\rm loc,h}(\hat{E},j)$, without further assumptions on the structure of
the symmetric mass ratio, $\nu$, of  the expansion coefficients of the scattering angle $\chi_k$. The $O(\nu^2)$
contributions to the 5PN EOB parameters have been unknown in part before. We perform comparisons of our analytic
results with the literature and also present numerical results on some observables.
\end{abstract}

\vspace*{\fill}
\noindent
\newpage

\section{Introduction}
\label{sec:1}

\vspace{1mm}
\noindent
The discovery of gravitational wave signals from merging black holes and neutron stars \cite{LIGO}  has
been a recent milestone for general relativity and astrophysics. The different present and planned
gravitational wave detectors are reaching higher and higher sensitivity \cite{PROJECT}, which requires
more detailed predictions on the theoretical side than currently available. The motion of gravitating
massive binary systems was studied perturbatively expanding in higher post--Newtonian (PN) orders right
after the theory of general relativity has been found \cite{1PN,Kol:2007bc}. Later on the corrections at
2PN~\cite{2PN}, 3PN \cite{3PN} and 4PN \cite{Damour:2014jta, 4PN,Foffa:2019rdf,Foffa:2019yfl,
Blumlein:2020pog}
have been calculated using a variety of techniques. First results at 5PN have been
obtained in Refs.~\cite{Foffa:2019hrb,Blumlein:2019zku,Bini:2019nra,Bini:2020wpo,Blumlein:2020pyo,Foffa:2020nqe} and
at 6PN in Refs.~\cite{Blumlein:2020znm,Bini:2020nsb,Bini:2020hmy}.\footnote{There is also a lot of activity in
calculating post--Minkowskian corrections, cf.~\cite{Blumlein:2020znm}, Ref.~[12], and
\cite{PM,Bern:2019nnu,Kalin:2019inp,Damour:2019lcq}.}
In the Schwarzschild limit the contributions of $O(\nu^0)$ are obtained to all post--Newtonian orders and
the terms of $O(\nu)$ are know to $O(G_N^{21.5})$
\cite{Bini:2014nfa,WARD1,Blanchet:2019rjs} from the self--force formalism
\cite{SF}\footnote{Also in self--force calculations regularizations are applied. In the present paper 
we are not
discussing this aspect, but assume that the final results derived are regularization--independent.},
where $G_N$ denotes Newton's constant. The following kinematic variables are used
\begin{eqnarray}
\nu = \frac{m_1 m_2}{M^2} \in \left[0,\tfrac{1}{4}\right],~~~M = m_1 + m_2,~~~m_{1,2} = \frac{1}{2} M\left[
1 \pm \sqrt{1 - 4 \nu}\right],~~\mu = M \nu,
\end{eqnarray}
where $m_1$ and $m_2$ are the two gravitating masses.

In this paper we complete the calculation of the 5PN corrections of the conservative dynamics for binary systems
using an effective field theory (EFT) method~{\cite{Goldberger:2004jt,Kol:2007bc}} and present main physical results. The calculation needs to be
performed in $D = 4 - 2 \ep$ dimensions because pole terms of $O(1/\ep)$ are occurring in intermediate steps.
One first considers gravity in $d = D - 1$ integer spatial dimensions for all contributing pieces in the
Lagrangian and then performs an analytic continuation. The multipole moments are dealt with in configuration
space, while the graviton dynamics is calculated in momentum--space, after the post--Newtonian expansion has been performed via a
Fourier--transform. Concerning the Feynman-diagrammatic representation of the {far-zone} contributions
we follow  Ref.~\cite{Foffa:2019eeb} and
extend it.
The conservative Hamiltonian of the binary system, $H$, consists of the potential,
$H_{\rm pot}$, and the {far-zone} terms {involving back scattered radiation},
$H_{\text{{rad}}}$, or can be likewise decomposed into the local, $H_{\rm loc}$,
and non--local,  $H_{\rm nl}$, contributions
\begin{eqnarray}
H = H_{\rm pot} + H_{\text{{rad}}} = H_{\rm loc} + H_{\rm nl}.
\label{eq:H}
\end{eqnarray}
We calculate the 5PN Hamiltonian {\it ab initio} starting with the path integral
for classical gravity, for which a post--Newtonian expansion is performed by applying an EFT description
\cite{Goldberger:2004jt,Kol:2007bc}. Methods originally developed for Quantum Field Theories are applied.
About 190.000 Feynman diagrams contribute. They are generated by {\tt QGRAF} \cite{Nogueira:1991ex}. Main
parts of the calculation are performed using {\tt FORM} \cite{FORM}. The reduction to a very small set of
master integrals is performed using {\tt Crusher} \cite{CRUSHER} using the integration-by-parts relations
\cite{IBP}.
The 5PN potential terms have been calculated by us in  Ref.~\cite{Blumlein:2020pyo}.\footnote{We compared
also to the factorizing contributions to the potential terms, which have been obtained in \cite{Foffa:2020nqe},
version 2, very recently, to which the corresponding subset of our results is agreeing.} There we also
described how the singularities in the potential and {far-zone} terms
\cite{Damour:2014jta,Bini:2020wpo,Bini:2020hmy,Thorne:1980ru,TAIL,Ross:2012fc,Marchand:2020fpt,Foffa:2019eeb}
are canceling, together with an additional canonical transformation. The whole 5PN calculation is performed
starting in the Lagrange formalism and finally deriving the Hamiltonian.\footnote{As outlined in detail
in Refs.~\cite{Blumlein:2020pog,Blumlein:2020pyo}, the treatment of higher time derivatives has to be
performed without applying the equation of motion. Note, that there are different approaches in the
literature.} Also the non--local 5PN contributions were
presented, see also \cite{Bini:2020wpo}.
What remained to be calculated beyond the contributions given in \cite{Blumlein:2020pyo} is a series of local
{far-zone}
contributions. These terms contribute at $O(\nu)$ and $O(\nu^2)$. They are due to the 1PN correction
to the electric quadrupole moment, written symbolically, $E Q_{ij} Q_{ji}$, with $E$ the energy,
\cite{Marchand:2020fpt}, the octupole moment,
$ E O_{ijk} O_{ijk}$, the magnetic quadrupole moment, $E J_{ij} J_{ji}$, the angular momentum failed tail, $L_k
\varepsilon_{ijk}
Q_{il} Q_{jl}$, with $\vec{L}$ the angular momentum, and the memory terms, $Q_{ij} Q_{jk} Q_{ki}$,  which
have also
been considered in Ref.~\cite{Foffa:2019eeb}.\footnote{For a definition of the multipole moments see
e.g.~\cite{Lins:2020omt}.}  These multipole contributions are calculated in $D$ dimensions starting
with
harmonic coordinates. As has been outlined in \cite{Foffa:2019eeb,Blumlein:2020pyo}
only the {contributions involving two} electric quadrupole, octupole, {or}
magnetic quadrupole moments have poles of $O(1/\ep)$ and receive logarithmic contributions. All other
multipole
moments can be calculated in $D=4$ dimensions and form rational contributions to the 5PN Hamiltonian. We have
performed an independent calculation of these and related contributions and performed a detailed comparison
with \cite{Foffa:2019eeb}.\footnote{
The result in \cite{Foffa:2019eeb} for the angular momentum failed tail still needs a sign change, as
communicated to us by the authors.}

The Hamiltonian $H$ in Eq.~(\ref{eq:H}) is  gauge dependent and singular. To compare different
approaches one has to either relate the Hamiltonians by canonical transformations or to compare the predictions for
the resulting observables. This applies to the case of harmonic coordinates in the present approach and
effective one body coordinates in \cite{Bini:2020wpo}. In Ref.~\cite{Bini:2020wpo} different methods
and constraints  implied by the $\nu$-structure of the expansion coefficients of the scattering angle $\chi_k$,
(\ref{eq:SCA}), have been used to construct the
Hamiltonian and all but two $O(\nu^2)$ parameters, $\bar{d}_5$ and $a_6$, were determined
in this way for $H_{\rm EOB}$ up to 5PN.

In Ref.~\cite{Blumlein:2020pyo} we obtained all terms except the rational terms of $O(\nu^2)$, the
calculation of which we had not yet
completed, as far as local {far-zone} contributions
were concerned. For $\bar{d}_5$ and $a_6$ we have,
more than in Ref.~\cite{Bini:2020wpo}, obtained the $O(\nu^2 \pi^2)$ terms, which stem from the potential
contributions. We remind that the magnetic quadrupole contribution necessitates a
finite renormalization\footnote{In renormalizable Quantum Field Theories the related problem is the
so--called $\gamma_5$ problem, cf.~\cite{Larin:1993tq}, and the corresponding operation is called a finite
renormalization which is well--known from numerous calculations.}
due to the analytic continuation of $\varepsilon_{ijk}$ in dimensional regularization,
which implies the contribution $\delta_J H$, Eq.~(61) in \cite{Blumlein:2020pyo}.\footnote{A slightly different
$D$--dimensional representation than used in \cite{Blumlein:2020pyo} has been presented later in
Ref.~\cite{Henry:2021cek}, cf. also Ref.~\cite{Bini:2021gat}, leading to the same contribution $\delta_J H$, however.
There also multipole moments, vanishing in $d=3$ dimensions have been discussed. They do not contribute
in the present case \cite{BLANCHET21}.
} We observe the same
finite renormalization in the $O(\nu)$ term for the binding energy and the local contribution to periastron
advance $K^{\rm loc,h}(\hat{E},j)$, cf.~\cite{Blumlein:2020pyo}. Here `h' stands for harmonic coordinates.

In the present paper we calculate the 5PN $O(\nu^2)$ terms and perform a series of comparisons to the literature.
In Section~\ref{sec:2} we calculate the 5PN local {far-zone} terms. To compare with Ref.~\cite{Bini:2020wpo} we express
our results for the local contributions obtained in the harmonic gauge in terms of effective one body (EOB)
potentials in Section~\ref{sec:3}. To fix the five new EOB parameters at 5PN we are solely using
the local contributions to
periastron advance
$K^{\rm loc,h}(\hat{E},j)$, which is an observable \cite{Heisenberg:1925zz} since the non--local terms are known
in explicit form via an eccentricity expansion, cf.~\cite{Bini:2020wpo,Blumlein:2020pyo}, and does not require any
regularization.\footnote{Regularizations may imply scheme dependencies.} Furthermore, we will not use assumptions
on the $\nu$ dependence of the observables used but perform a direct calculation within the framework of an
effective field theory approach. The binding energy and periastron advance in the circular case are used for
consistency checks for two parameters. Finally, we also discuss the determination of the 5PN parameters using
the expansion coefficients of the scattering angle and compare to the literature. We summarize phenomenological
results in Section~\ref{sec:4} and Section~\ref{sec:5} contains the conclusions. Some technical
aspects are given in the Appendices \ref{sec:A}--\ref{sec:E} on the Feynman rules, invariant functions, the
in--in formalism, calculation of the {far-zone} terms and relations for the scattering angle.
\section{The local {far-zone} terms of the pole--free Hamiltonian}
\label{sec:2}

\vspace*{1mm}
\noindent
The {far-zone} terms can be derived both classically and by using EFT methods applying methods from Quantum Field Theory.
Both methods have to lead to the same result. There are two types of {far-zone} contributions, {\it i)} singular
ones, which have to be calculated in $D$ dimensions and {\it ii)} non--singular ones, which can be calculated in $D = 4$
space--time dimensions. The corresponding contributions are expressed in terms of a multipole expansion, known in gravity
since long, starting with the quadrupole moment \cite{Einstein:1918btx}, the current quadrupole moment and the mass octupole
moment \cite{PAPAPETROU}\footnote{See \cite{Thorne:1980ru} for
details.}. 

We will use this {\it paradigm} in the following as working assumption. To 4PN and 5PN we observe a cancellation 
of the pole terms with those from the potential contribution (up to a canonical transformation from harmonic coordinates 
\cite{Blumlein:2020pog,Blumlein:2020pyo}). Moreover, the logarithmic contributions come out correctly and the $O(\nu)$
terms all come out correctly and in accordance with the self--force predictions, as well as, all terms of $O(\nu^k),~~k \geq 
3$. What remains is the detailed understanding of the 5PN non--singular contributions of $O(\nu^2)$.
Despite of this success, it still may turn out in the future, that this paradigm has to be extended from 5PN onward.
Let us also stress that we will fix the kinematics, e.g. in form of the EOB parameters, using information on elliptic orbits 
only, but not also using the kinematically different scattering process, as done 
in \cite{Bini:2020wpo,Bini:2020nsb,Bini:2020hmy}. There a different 
orbit average is performed, making it difficult to perform the analytic continuation to the inspiraling process.

The non--local {far-zone} terms, if viewed from a $D$ dimensional calculation, arise as logarithmic corrections in the $\ep$
expansion. To 5PN their structure was derived by classical methods in \cite{Thorne:1980ru,Blanchet:1989cu}, see
also
\cite{Bini:2020wpo}, and using EFT methods in \cite{Foffa:2019eeb,Blanchet:2019rjs,Blumlein:2020pyo}. Both methods lead
to the same results for the electric quadrupole moment, $E Q_{ij} Q_{ji}$,\cite{Marchand:2020fpt},
the octupole moment, $O_{ijk} O_{ijk}$, and the magnetic quadrupole moment, $E J_{ij} J_{ji}$, which develop
logarithmic and pole contributions in the
EFT--approach.
Their normalization coefficients are the same as for their imaginary part, contributing to $dE/dt$,
\cite{Thorne:1980ru},
Eq.~(4.16').\footnote{For other classical calculations see \cite{Blanchet:1989cu}. To our knowledge, there is
no other derivation yet for the failed angular momentum and memory term, but that by using EFT methods. We thank
L.~Blanchet for a corresponding remark.}

The associated $D$--dimensional  terms up to the constant parts were calculated in \cite{Foffa:2019eeb,
Blumlein:2020pyo}, see also \cite{Marchand:2020fpt,Blanchet:2020ngx}, and do also agree. This concerns the 1PN correction
to the electric quadrupole moment, $E Q_{ij} Q_{ji}$,\cite{Marchand:2020fpt}, the octupole moment, $E O_{ijk}
O_{ijk}$, and the
magnetic quadrupole moment, $E J_{ij} J_{ji}$. The angular momentum failed tail,
$L_k \varepsilon_{ijk} Q_{il} Q_{jl}$,
and the memory term, $Q_{ij} Q_{jk} Q_{ki}$, are non singular and contribute only local $O(\nu^2)$ terms.
The latter two contributions were calculated in
\cite{Foffa:2019eeb,Foffa:2021pkg}, however, not using the
Schwinger--Keldysh
formalism (also called in--in or closed time path formalism), which
has been developed
in Refs.~\cite{Schwinger:1960qe,Bakshi:1962dv,Keldysh:1964ud,Buchbinder:1981hu,Chou:1984es,DEWITT,Jordan:1986ug,
Hu:2003qn,DEWITT1,KLEINERT,Korenman:66,Galley:2009px,Galley:2015kus,Foffa:2021pkg}.
The angular momentum failed tail had been calculated by us at the time of  \cite{Blumlein:2020pyo} obtaining
the same  result as in \cite{Foffa:2019eeb}.

The action, by which the vertices of the  multipole moments in the diagrams Ref.~\cite{Foffa:2019eeb} are defined has been
derived using group--theoretical methods in $D=4$ dimensions in Ref.~\cite{Ross:2012fc}, Eq.~(100), e.g., and reads
\begin{eqnarray}
S_{\rm mp} &=& - 
\int_{-\infty}^{+\infty} dt \left[{\frac{1}{2}}E h_{00} + {\frac{1}{2}} L^i \ep_{ijk} \partial_j h_{0k}
+ \sum_{l=2}^\infty \frac{1}{l!} I^L \partial_{L-2} E_{k_{l-1}, k_l}\right.
\nonumber\\ && \hspace*{2cm}
\left.- \sum_{l=2}^\infty \frac{2l}{(l+1)!} J^L \partial_{L-2} B_{k_{l-1}, k_l} + 
{\dots}\right],
\label{eq:Smp}
\end{eqnarray}
with
\begin{eqnarray}
h^{\mu\nu} =  g^{\mu\nu} - \eta^{\mu\nu},
\end{eqnarray}
$g^{\mu\nu}$ the metric, $\eta^{\mu\nu}$ the Minkowski metric, and $L$ a
multi--index. We set the velocity of light, $c$, in many places to $c = 1$, except of those at which we would like
to use this parameter for the explicit counting of the post--Newtonian orders.
The tensors $E_{ij}$ and $B_{ij}$ are given by
\begin{eqnarray}
E_{ij} &=& R_{0i0j} \simeq \frac{1}{2}\left[h_{00,ij} + \ddot{h}_{ij}
- \dot{h}_{0i,j} - \dot{h}_{0j,i}\right] + O(h_{\mu\nu}^2)
\\
B_{ij} &=& \frac{1}{2} \ep_{ikl} R_{0jkl} \simeq \frac{1}{4} \ep_{ikl}
\left[\dot{h}_{jk,l} - \dot{h}_{jl,k} + h_{0l,jk} - h_{0k,jl}\right] + O(h_{\mu\nu}^2),
\end{eqnarray}
$R^\mu_{\nu\rho\sigma}$ denotes the Riemann tensor and $\dot{x} = dx/dt$. Later we will also consider
vertices of multipole  moments and two gravitons, see also Appendix~\ref{sec:A}.

The multipole--moment terms $K_L^{(k)} K_L^{(k)}$ {where}
\begin{eqnarray}
{X^{(k)} = \frac{d^k}{dt^k} X(t)},
\end{eqnarray}
$K = I, J, O$, and $L$ the corresponding multi--index, have the following structure
\begin{eqnarray}
[K_L^{(k)} K_L^{(k)}](\ep) = [K_L^{(k)} K_L^{(k)}] + \ep \overline{[K_L^{(k)} K_L^{(k)}]} + O(\ep^2).
\end{eqnarray}

We calculate the corresponding contributions to the action{, obtaining}
\begin{equation}
  \label{eq:Srad}
  \begin{split}
  {S_{\text{rad}}^{\text{loc}} = \int dt\ G_N^2\Biggl[}&
    {\frac{1}{10}\Biggl(\frac{1}{\varepsilon} + \frac{41}{15}\Biggr) E Q^{(4)}_{ij}Q^{(4)}_{ij}
    + \frac{1}{378}\Biggl(\frac{1}{\varepsilon} + \frac{164}{35}\Biggr) M O^{(4)}_{ikl}O^{(4)}_{ikl}
    + \frac{8}{45}\Biggl(\frac{1}{\varepsilon} + \frac{67}{30}\Biggr)} \\
    &{\times M J^{(3)}_{ij}J^{(3)}_{ij}
- \frac{8}{15}\ep_{ijk} L_k Q_{il}^{(4)} Q_{jl}^{(3)}
    + \Biggl(\frac{1}{24} - \frac{1}{20}\Biggr) Q_{ij} Q_{jk}^{(4)} Q_{ki}^{(4)}
    - \frac{1}{35} Q_{ij}^{(2)} Q_{jk}^{(3)} Q_{ki}^{(3)}\Biggr].}
\end{split}
\end{equation}
{In the contribution proportional to $Q_{ij} Q_{jk}^{(4)} 
Q_{ki}^{(4)}$ we distinguish between the contributions from two diagram topologies, see below}.
A Legendre transformation
leads to the following Hamiltonian for the finite {far-zone}
contributions
\begin{eqnarray}
\label{eq:HA1}
H_{\text{rad, finite}}^{\rm loc, 5PN} &=& \eta^{10} \Biggl\{
  H_{\rm el. quad}^{\rm 5PN} +
G_N^2 \Biggl[
- M \Biggl(\frac{82}{6615} O_{ikl}^{(4)} O_{ikl}^{(4)}
+ \frac{1}{378} \overline{O_{ikl}^{(4)} O_{ikl}^{(4)}}
+ \frac{268}{675} J_{ij}^{(3)} J_{ij}^{(3)}
\nonumber\\ &&
+ \frac{8}{45} \overline{J_{ij}^{(3)} J_{ij}^{(3)}}\Biggr)
+  \frac{8}{15} \ep_{ijk} L_k Q_{il}^{(4)} Q_{jl}^{(3)}
\Biggr]
+ H_{\rm mem. 1}
+ H_{\rm mem. 2}
+ H_{\rm mem. 3}
\Biggr\}.
\end{eqnarray}
We performed the canonical transformation to eliminate the $1/\ep$ pole terms in Ref.~\cite{Blumlein:2020pyo}
before, where the corresponding terms have been given already. Here the multipole--moments have to be used in
$D$ dimensions, except for the last four terms in (\ref{eq:HA1}).
One obtains
\begin{eqnarray}
\label{eq:HA2}
H_{\text{{rad}, finite}}^{\rm loc, 5PN} &=& \eta^{10} \Biggl\{
  H_{\rm el. quad}^{\rm 5PN}
+ H_{\rm oct.}
+ H_{\rm mag. quad.}
+ H_{\rm ang.}
+ H_{\rm mem. 1}
+ H_{\rm mem. 2}
+ H_{\rm mem. 3}
\Biggr\}.
\end{eqnarray}
The calculation is performed in the in--in formalism.
For the discussion in Section~\ref{sec:4} it is essential to quantify the local 5PN {far-zone} terms to the pole--free
Hamiltonian.

\begin{figure}[H]
  \centering
  \hskip-0.8cm
  \includegraphics[width=.45\linewidth]{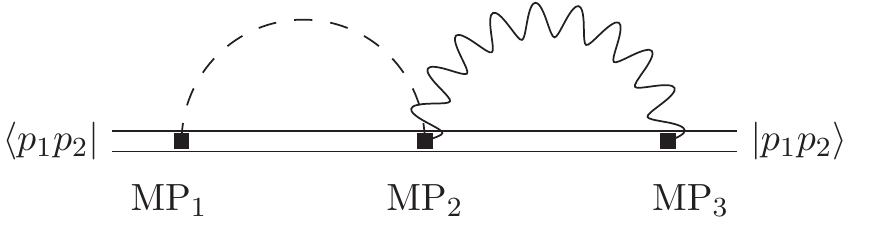}~~~~~~~~
  \includegraphics[width=.45\linewidth]{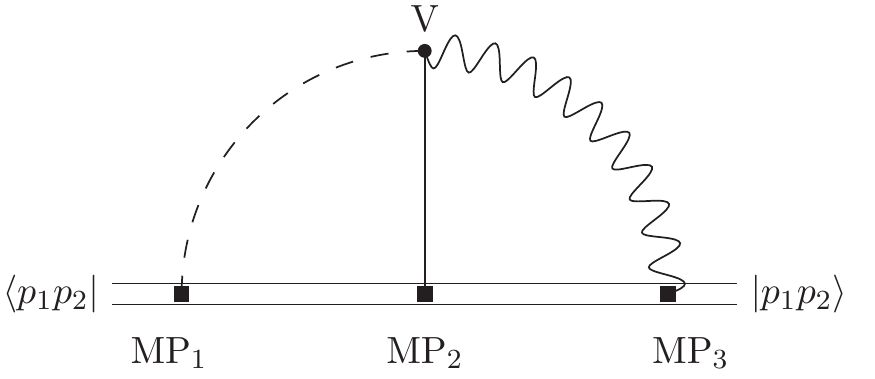}
  \caption[]{\sf Schematic diagrams for the 5PN {far-zone} contributions in the in--in formalism. The symbols MP$_i$
  denote the three multipole moments, $V$ the triple vertex between the different contributing fields, the propagators
  in the decomposition of Ref.~\cite{Kol:2007bc} of which have been depicted as dashed, full and curly lines,
  although some of them can denote the same field. Left: 2-graviton diagrams; Right: 3-graviton diagrams.}
  \label{fig:1}
\end{figure}

The structure of the diagrams contributing to the {far-zone} terms of the Hamiltonian in EFT is illustrated in
Figure~\ref{fig:1}. The following velocity counting holds,
\begin{itemize}
\item each graviton propagator scales with $v^{-2}$
\item each momentum integral yields a factor of $v^3$
\item each post--Newtonian correction more adds a factor of $v^2$
\item the graviton triple vertex is $\sim v^3$
\item graviton coupling to $E \sim v$, $L_k \sim v^2$, $Q_{ij}, J_{ij} \sim v^3$, and $O_{ijk} \sim v^4$
\item multipole moments $L_k, J_{ij} \sim v$
\item the double graviton vertex to $Q_{ij} \sim v^4$.
\end{itemize}

The action (\ref{eq:Smp}) allows for a wide variety of triple multipole diagrams. The lowest contributing terms
are those containing two electric {quadrupole} moments, because $E$ and $L_k$ are conserved in the stationary case we are
considering. The diagram {\sf QEQ} is of $O(v^{10})$ or of lowest order 4PN. The diagrams {\sf QLQ, JEJ, OEO} and
{\sf QQQ} are of $O(v^{12})$ and of lowest order 5PN.
Other valid combinations
are of higher than 5PN order or vanish, except the {\sf QQQ} combination with two gravitons and one
two--graviton $Q_{ij}$ interaction.

In the following we list the local finite 5PN contributions from the electric quadrupole moment, $H_{\rm el.
quad}^{\rm 5PN}$,
the octupole term, $H_{\rm oct.}$, the magnetic quadrupole moment, $H_{\rm mag. quad.}$,
the angular momentum term $H_{\rm ang.}$, and the three contributions
to the memory term, $H_{\rm mem. 1,2,3}$,
{expressed in terms of the rescaled orbital distance $r = r_{\text{phys}} c^2/(G_N M)$.}
\begin{eqnarray}
H_{\rm el. quad}^{\rm 5PN} &=&
\nu  \Biggl[
         -(1-3\nu) \frac{1}{r^3} \Biggl(
                \frac{392}{75}  p^6
                -\frac{3508}{75}  p^4 (p.n)^2
                +\frac{124}{3}  p^2 (p.n)^4
        \Biggr)
\nonumber\\ &&
        +\frac{1}{r^4} \Biggl(
                -\frac{2 (4573+32831 \nu ) p^4}{3675}
                -\frac{2 (-598985+122233 \nu ) p^2 (p.n)^2}{3675}
\nonumber\\ &&
     +\frac{4 (-286508+54209 \nu ) (p.n)^4}{3675}
        \Biggr)
        +\frac{1}{r^5}
\Biggl(
\Biggl(
                \frac{696}{1225}+\frac{77428 \nu }{3675}\Biggr) p^2
\nonumber\\ &&
        +\Biggl(
                \frac{30904}{3675}+\frac{19168 \nu }{735}\Biggr)
(p.n)^2 \Biggr)
        + \frac{8 (-244+137 \nu )}{525 r^6}
\Biggr],
\\
H_{\rm oct.} &=&
(1-4 \nu ) \nu  \Biggl[
         \frac{1}{r^4} \Biggl(\frac{17657 p^4}{2205}
        -\frac{130058 p^2 (p.n)^2}{3675}
        +\frac{60409 (p.n)^4}{2205}
        \Biggr)
\nonumber\\ &&
        +\frac{1}{r^5} \Biggl(-\frac{2848 p^2}{1575}
        +\frac{20848 (p.n)^2}{11025}
        \Biggr)
        + \frac{4}{7 r^6}
\Biggr],
\\
H_{\rm mag. quad.} &=&
(1-4 \nu ) \nu  \Biggl[
        \frac{1}{r^4} \Biggl(\frac{58 p^4}{45}
        -\frac{604 p^2 (p.n)^2}{75}
        +\frac{1522 (p.n)^4}{225}
        \Biggr) -
\frac{1}{r^5} \frac{488}{225}
        (p^2 - (p.n)^2)
\Biggr],
\nonumber\\
\\
H_{\rm ang.} &=&
\nu ^2 \Biggl[
        \frac{1}{r^4}\Biggl(-\frac{256 p^4}{5}
        +\frac{1056 p^2 (p.n)^2}{5}
        -160 (p.n)^4 \Biggr)
        + \frac{1}{r^5} \frac{256}{15} (p^2 - (p.n)^2)
\Biggr].
\end{eqnarray}
Note that the terms $\propto \ln(r/r_0)$ occurring have been considered together with the non--local 
terms here.

For all these terms we agree with the principle structure given in
Ref.~\cite{Foffa:2019eeb}, where the multipole moments were not
inserted, with the exception that $H_{\rm mag. quad.}$ still needs
the finite renormalization as described in
Ref.~\cite{Blumlein:2019zku}, which here has already been considered.
In the same way we agree with the 4PN tail term ${\sf QEQ}$.

In the EFT calculation
the choice of the time dependence of the propagators in these 5PN cases turns out to be irrelevant in
the explicit calculation, since the causal and the in--in formalism lead to the same result.
The reason for this is that the internal multipole moment, either $E$ or $L^k$, is a conserved quantity, implying a
$\delta$--distribution for the energy components at their vertex. The
associated propagator is therefore an (effective) space--like potential and the corresponding diagram only
depends on a single
energy variable, $k_0$. In the case of the potential terms, cf.~\cite{Foffa:2019rdf,Blumlein:2020pog},
the reason for the same agreement is different. Here the propagators in Fourier--space are expanded as
\begin{eqnarray}
\frac{1}{\vec{k}^2 - k_0^2 \pm i \ep} = \frac{1}{\vec{k}^2} \sum_{l=0}^\infty
\left(\frac{k_0^2}{\vec{k}^2}\right)^l,
\end{eqnarray}
in which the $i \ep$--prescription does not play a role. When the in--in formalism is used for
all these terms,
it leads to the same results as using the usual (causal) path integral \cite{FH,STERMAN}.

Indeed, a fundamental argument has been raised for the general use of the in--in formalism by B.~DeWitt \cite{DEWITT1}.
In using $S$--matrix theory the LSZ--formalism \cite{Lehmann:1954rq} requires a clear definition of the Hilbert spaces
both
for the initial state at $t = - \infty$ and for the final state at $t = + \infty$, made of (interaction)
free states in
both cases. This is fulfilled in elementary particle scattering processes, however, not in inspiraling processes
like the merging of two large masses. While the initial state can be very well defined as
\begin{eqnarray}
\label{eq:inin}
|p_1 p_2\rangle = |p_1\rangle |p_2\rangle,
\end{eqnarray}
with $p_1$ and $p_2$ the 4--momenta of the non--interacting masses at $t = -\infty$, the synonymous information on
the final state of the merging process at $t = + \infty$ is not really known. The in--in formalism, however, requires
only to know the initial state.\footnote{The method has some similarity to descriptions used in deep--inelastic scattering,
like the forward Compton amplitude \cite{FC}, referring to the optical theorem \cite{Veltman:1963th}, using cutting
methods in elementary particle physics.}

The contributions to the {far-zone} term require retarded boundary conditions, as known from the Feynman--Wheeler
formalism \cite{FW,Dirac:1938nz} in Quantum Electrodynamics. The Schwinger formalism using in--in states
provides this description.
As well--known, both the usual path--integral formalism \cite{FH,STERMAN}, leading to causal Green's
functions with
$T^*$--ordering, and, analogously, the so-called in--in formalism, cf.~\cite{Schwinger:1960qe,Galley:2009px},
are exactly defined, also concerning the type of the contributing propagators. The in--in formalism has
also applications in statistical physics, cf.~\cite{Keldysh:1964ud,Chou:1984es}. A critical question concerns
the unitarity of the respective formalism. As has been shown in \cite{Jordan:1986ug} this is obeyed for the
in--in formalism at least at two--loop order, the level necessary at 5PN. In the EFT approach the
{far-zone} diagrams
up 5PN read structurally as shown in Figure~\ref{fig:1} using the in--in formalism: these are diagrams
containing three multipole moments with single or double graviton interaction between two in--states.
Also here the gravitons have at most self--couplings
and all end at the worldline being connected to one of the multipole moment insertions.\footnote{Please note  that these 
diagrams appear not at the same footing as diagrams with just ultrasoft lines in the approach of Ref.~\cite{Bern:2021yeh}
since they contain multipole insertions here, unlike the case in \cite{Bern:2021yeh}.}

The only contributing diagrams at
5PN are two--loop diagrams. In the action, the 2--loop graviton exchange is  integrated out. The necessary Feynman
rules are listed in Appendix~\ref{sec:A}. One then may read off the contributions to the conservative
{far-zone} Lagrangian
(Hamiltonian) from the action directly.
The in--in formalism is described in detail in Appendix~\ref{sec:C}.

In the following we calculate {the contributions to the memory term in
the in--in formalism. This differs from the previous work
\cite{Foffa:2019eeb}, which employed a variant of the in--out formalism with
Feynman Green's functions replaced by advanced or retarded ones.
Further differences may arise in the Feynman rules, which are not stated explicitly
in \cite{Foffa:2019eeb}.}
The $D$ dimensional tensor integrals are decomposed using the
Passarino--Veltman representation \cite{Passarino:1978jh} and the momentum integrals are performed using
hypergeometric techniques \cite{HYP,Blumlein:2018cms} after a Feynman parameterization.
In the final result the $\ep$
dependence
cancels, as the case of $H_{\rm ang.}$, leaving the integrals over the energy components.

The results for the diagrams in Figure~\ref{fig:1} are
\begin{eqnarray}
\label{eq:Mmem1}
H_{\rm mem. 1} &=& \nu^2 \Biggl[
\frac{1}{r^4} \Biggl(
        {-}\frac{10 p^4}{3}
        {+}\frac{967 p^2 (p.n)^2}{54}
        {-}\frac{83 (p.n)^4}{6}
\Biggr)
+ \frac{1}{r^5} \Biggl(
        {+}\frac{77 p^2}{27}
        {-}\frac{97 (p.n)^2}{27}
\Biggr)
{-} \frac{11}{27 r^6}
\Biggr]
\nonumber\\
\end{eqnarray}
for the term in the l.h.s. and we obtain for the two contributions of the second diagram
\begin{eqnarray}
\label{eq:Mmem2}
H_{\rm mem. 2} &=&
\nu ^2 \Biggl[
        \frac{1}{r^4} \Biggl(4 p^4
        {-}\frac{967 p^2 (p.n)^2}{45}
        {+}\frac{83 (p.n)^4}{5} \Biggr)
        +\frac{1}{r^5}\Biggl({-}\frac{154 p^2}{45}
        {+}\frac{194 (p.n)^2}{45}\Biggr)
        {+}\frac{22}{45 r^6}
\Biggr],
\nonumber\\
\\
\label{eq:Mmem3}
H_{\rm mem. 3} &=&
\nu ^2 \Biggl[
        \frac{1}{r^4}\Biggl(
\frac{16 p^4}{315}
{+}\frac{152 p^2 (p.n)^2}{105}
{-}\frac{8 (p.n)^4}{7} \Biggr)
+\frac{1}{r^5} \Biggl(
         {-}\frac{16 p^2}{105}
        {+}\frac{256 (p.n)^2}{315} \Biggr)
{+}\frac{32}{315 r^6}
\Biggr].
\nonumber\\
\end{eqnarray}
Details of the calculation of the $H_{\rm mem.}$ terms are given in Appendix~\ref{sec:D}.
The results in (\ref{eq:Mmem2}--\ref{eq:Mmem3}) differ from those given in \cite{Foffa:2019eeb}
by a factor
and the
calculation of the contribution (\ref{eq:Mmem1}) is new. We have repeated the calculation of
(\ref{eq:Mmem2}--\ref{eq:Mmem3}) in the same manner as described in \cite{Foffa:2019eeb}\footnote{See the
remark below Eq.~(25) there.} and agree. However,
our result is based on the derivation of the corresponding graph using the path integral in the in--in
formalism and differs from that in \cite{Foffa:2019eeb}. There a
specific choice for linking the advanced and
retarded propagators has been made, which is not confirmed.

We also would like to mention that we have slightly modified the Hamiltonian given in \cite{Blumlein:2019zku},
by treating
higher time derivatives in the action, which are now eliminated by partial integration. This change, however, just
corresponds to a canonical transformation, as we will show below in calculating observables which come out the same.
The complete local pole--free
Hamiltonian, with tags on different contributions, is given in computer readable form in the ancillary file
{\tt HAMILTONIAN.m} to this paper.
\section{Determining the EOB potentials from the Hamiltonian in harmonic coordinates}
\label{sec:3}

\vspace*{1mm}
\noindent
In Ref.~\cite{Blumlein:2020pyo} we have calculated the 5PN potential contributions to the Hamiltonian using
dimensional regularization in $D = 4 - 2\ep$ space--time dimensions, including the pole terms and the
non--local {far-zone} contribution in complete form using harmonic coordinates.
The sum of these terms does still contain a pole contribution $1/\ep$.
A canonical transformation leads to a pole--free Hamiltonian. In \cite{Blumlein:2020pyo} we have
left out finite rational 5PN terms of $O(\nu^2)$, i.e. local {far-zone} contributions which are purely rational. We have
presented
already all the $\pi^2$ terms of $O(\nu^2)$ since they stem from the potential terms.

A central point of our investigation is to compare to previous results given in Ref.~\cite{Bini:2020wpo}, which have
been presented based on a 5PN EOB Hamiltonian. For the local parts of both Hamiltonians one may either perform a canonical
transformation, as done in Ref.~\cite{Blumlein:2020pyo}, or determine the EOB potentials by an observable.
We will choose the latter way and use the local contribution to periastron advance $K^{\rm loc,h}(\hat{E},j)$.
Corresponding consistency checks can be performed by additional observables, such as the binding energy and  periastron
advance for circular motion and also the expansion coefficients of the scattering angle.
\subsection{The EOB parameters}

\vspace*{1mm}
\noindent
The EOB Hamiltonian is given by \cite{Bini:2020wpo}
\begin{eqnarray}
  \label{H_EOB}
H_{\rm EOB}^{\rm loc,eff} = \sqrt{A (1 + A D \eta^2 (p.n)^2 + \eta^2 (p^2 - (p.n)^2) + Q)},
\end{eqnarray}
which we consider up to the contributions to 5PN. Here $\eta = 1/c$ denotes the post--Newtonian expansion
parameter, with $c$ the velocity of light.
The potentials $A, D$ and $Q$ are parameterized by\footnote{The 5PN level in EOB coordinates are of $O(\eta^{12})$.}
\begin{eqnarray}
A &=& 1 + \sum_{k=1}^6 a_k(\nu) \eta^{2k} u^k,~~a_2 = 0,
\\
D &=& 1 + \sum_{k=2}^5 d_k(\nu) \eta^{2k} u^k,
\\
Q &=&
\eta^4 (p.n)^4 [q_{42}(\nu) \eta^4 u^2 + q_{43}(\nu) \eta^6 u^3 + q_{44}(\nu) \eta^8 u^4] +
\eta^6 (p.n)^6 [q_{62}(\nu) \eta^4 u^2 + q_{63}(\nu) \eta^6 u^3]
\nonumber\\ &&
+
\eta^{12} (p.n)^8 u^2 q_{82}(\nu),
\end{eqnarray}
where $u = 1/r$.

The known 4PN Hamiltonians \cite{Damour:2014jta,4PN,Foffa:2019rdf,Foffa:2019yfl,Blumlein:2020pog} in harmonic
and ADM coordinates are connected by canonical transformations to $H_{\rm EOB}$, cf.~\cite{Blumlein:2020pog},which
imply the expansion parameters
\begin{align}
{\rm N},   u~:  &~a_1    &=&~-2,
\\
{\rm 2PN}, u^2: &~d_2    &=&~6 \nu,
\\
           u^3: &~a_3    &=&~2 \nu,
\\
{\rm 3PN}, u^2: &~q_{42} &=&~  8 \nu - 6 \nu^2,
\\
           u^3: &~d_3   &=&~ 52 \nu - 6 \nu^2,
\\
           u^4: &~a_4   &=&~ \left(\frac{94}{3} - \frac{41}{32} \pi^2\right) \nu,
\\
{\rm 4PN}, u^2: &~q_{62}  &=&~-\frac{9}{5} \nu - \frac{27}{5} \nu^2  + 6 \nu^3,
\\
           u^3: &~q_{43}  &=&~ 20 \nu -83 \nu^2 +10 \nu^3,
\\
           u^4: &~d_4    &=&~   \left(\frac{1679}{9} - \frac{23761}{1536} \pi^2\right) \nu
                      + \left(- 260 + 123 \pi^2\right) \nu^2,
\\
  u^5: &~a_5 &=&~ \left(-\frac{4237}{60} + \frac{2275}{512} 
\pi^2\right) 
\nu + \left(-\frac{221}{6} + \frac{41}{32} \pi^2\right) \nu^2.
\end{align}

The expansion coefficients $a_6, \bar{d}_5, q_{44}, q_{63}$ and $q_{82}$ emerge at 5PN.
In \cite{Bini:2020wpo} the following values were obtained\footnote{The reader should not be
confused with the values given in \cite{Bini:2019nra}, which are in the f--scheme of the authors.
We compare to the h--scheme, cf.~\cite{Bini:2020wpo}.}
\begin{align}
{\rm 5PN}, u^2~:  &~q_{82} \hspace*{-15mm}    &=&~\frac{6}{7} \nu  + \frac{18}{7} \nu^2 +
                                                   \frac{24}{7} \nu^3 - 6 \nu^4,
\\
           u^3~:  &~q_{63} \hspace*{-15mm}   &=&~\frac{123}{10} \nu - \frac{69}{5} \nu^2
+ 116 \nu^3 - 14 \nu^4,
\\
           u^4~:  &~q_{44}   &=&~\left(\frac{1580641}{3150} -
                           \frac{93031}{1536} \pi^2\right) \nu
	                   + \left(-\frac{9367}{15} + \frac{31633}{512} \pi^2 \right) \nu^2
                           + \left(640 - \frac{615}{32} \pi^2\right) \nu^3,
\\
\label{eq:d50}
           u^5:   &~\bar{d}_5     &=&~\left(\frac{331054}{175} -
                       \frac{63707}{512} \pi^2 \right) \nu + \bar{d}_5^{\nu^2} \nu^2
                       + \left(\frac{1069}{3} - \frac{205}{16} \pi^2\right) \nu^3,
\\
\label{eq:a60}
           u^6: &~a_6  &=&~\left(-\frac{1026301}{1575}
           + \frac{246367}{3072} \pi^2\right) \nu + a_6^{\nu^2} \nu^2 + 4 \nu^3.
\end{align}
In Ref.~\cite{Blumlein:2019zku} we have calculated the $\pi^2$ contributions to $\bar{d}_5^{\nu^2}$ and
$a_6^{\nu^2}$ which read
\begin{eqnarray}
\bar{d}_5^{\pi^2 \nu^2}    &=&
\frac{306545}{512} \pi^2 \nu^2,
\\
a_6^{\pi^2 \nu^2}     &=&  \frac{25911}{256} \pi^2 \nu^2.
\end{eqnarray}

All 5PN coefficients can be obtained from 
{the} periastron advance $K^{\rm loc, h}(\hat{E},j)$,
{ expressed in terms of the energy $\hat{E}$ obtained from the reduced Hamiltonian},
\begin{equation}
  \label{eq:Hhat}
{\hat{H} = \frac{H-Mc^2}{\mu c^2}},
\end{equation}
{and the rescaled angular momentum defined in eq. (\ref{j})}.
{Explicit formulas for the calculation of $K^{\rm loc, h}(\hat{E},j)$ from a Hamiltonian
are given e.g.~in \cite{Blumlein:2020pyo}. Comparing the results obtained from the Hamiltonian
(\ref{eq:H}) in harmonic coordinates and the EOB Hamiltonian (\ref{H_EOB}) one finds}
\begin{eqnarray}
\label{eq:VAL1}
q_{82} &=& \frac{6}{7} \nu + \frac{18}{7} \nu^2 + \frac{24}{7} \nu^3 - 6 \nu^4,
\\
\label{eq:VAL2}
q_{63} &=& \frac{123}{10} \nu - \frac{69}{5} \nu^2 + 116 \nu^3 - 14 \nu^4,
\\
\label{eq:VAL3}
q_{44} &=&
             \left(\frac{1580641}{3150} - \frac{93031 \pi^2}{1536} \right) \nu
           + \left({
-\frac{3670222}{4725}}
+ \frac{31633 \pi^2}{512} \right) \nu^2
           +\left(640 -\frac{615}{32} \pi ^2\right) \nu ^3,
\\
\label{eq:VAL4}
\bar{d}_5    &=&  \left( \frac{331054}{175} - \frac{63707}{512} \pi^2 \right) \nu
          + \left({-\frac{31295104}{4725}} +\frac{306545}{512} \pi^2\right) \nu^2
          + \left(\frac{1069}{3} - \frac{205}{16} \pi^2 \right) \nu^3,
\nonumber\\
\\
\label{eq:VAL5}
a_6    &=& \left(- \frac{1026301}{1575} + \frac{246367}{3072} \pi^2\right)\nu +
 \left(
{-\frac{1749043}{1575}}
+ \frac{25911}{256} \pi^2\right) \nu^2 + 4
\nu^3.
\end{eqnarray}
Here $q_{82}$ {is uniquely determined by comparing the respective coefficients of $\hat{E}^4/j^2$,
whereas $q_{63}$ and $q_{44}$ follow from the $\hat{E}^3/j^4$ term}, etc.
Values for $\bar{d}_5$ and $a_6$, which are consistent with (\ref{eq:VAL4}) and (\ref{eq:VAL5}), are
also obtained from
the binding energy and periastron advance in the circular case, which do not depend on the $Q$--potentials.
\subsection{The scattering angle}
\label{sec:32}

\vspace*{1mm}
\noindent
We will now study the expansion coefficients $\chi_k$ of the scattering angle\footnote{It is most useful to perform the integral (\ref{eq:SCA}) directly using the $p.n$ gauge
\cite{Damour:2019lcq} both in
harmonic and in EOB coordinates. It is advisable to use the necessary integrals from
Ref.~\cite{RYGR}, since not all computer algebra systems do perform them correctly.} starting with its local contributions.
We use the well known relations
\begin{eqnarray}
\label{eq:SCA}
\frac{\chi^{\rm loc} + \pi}{2}  = \sum_{k=1}^\infty \frac{\chi_k^{\rm loc}}{j^k}
= {\sf Reg}_{u_{\rm max}} \int_{\ep}^{u_{\rm max}(\ep)} \frac{j 
dy}{(p.n)(y,j)},~~ p.n~:=~\sqrt{p_\infty^2 - j^2 y^2 +
W(y,p_\infty)},
\end{eqnarray}
with $u_{\rm max}(\ep) = (1+\sqrt{1+j^2 \PI^2})/(j \PI) - \ep$,
cf. also \cite{Damour:2019lcq}, Eq.~(3.39). One also has
\begin{eqnarray}
\label{eq:SCA2}
\frac{\chi^{\rm loc}}{2}  = {\sf Reg}_{\varepsilon} \sum_{n=1}^\infty 
\int_\varepsilon^{1-\varepsilon} dx 
\binom{-\tfrac{1}{2}}{n}(1-x^2)^{-\frac{1}{2} -n} \left[\tilde{w}\left(\frac{x}{j}\right)\right]^n,
\end{eqnarray}
with
\begin{eqnarray}
\label{eq:SCA3}
\tilde{w} = \sum_{k=1}^\infty w_k \frac{x^k}{j^k}.
\end{eqnarray}
Here the operator ${\sf Reg}_\varepsilon$ drops the singular terms 
in $\varepsilon$ in the limit $\ep \rightarrow 0$.

Some of the integrals are listed in Appendix~\ref{sec:E}.
Here the following kinematic variables  are used, cf.~\cite{Blumlein:2020znm,Antonelli:2019ytb},
\begin{eqnarray}
p_\infty~&=&~\sqrt{\gamma^2 -1 },
\\
\gamma~&=&~\frac{1}{m_1 m_2} [E_1 E_2 + p^2],~~~E_i = \sqrt{p^2+m_i^2},
  \\
  \label{j}
j~&=&~\frac{J}{G_N m_1 m_2},
\\
\Gamma~&=&~\sqrt{1 + 2 \nu(\gamma-1)},
\end{eqnarray}
$E_i$ are the corresponding energies, $p$ the cms momentum, and $J$ the angular
momentum.
Note that in the definition of the scattering angle in Ref.~\cite{Bini:2020wpo}, Eq.~(9.1), the regularization
operator has not been written, although being necessary, cf.~e.g. Eq.~(3.49) in \cite{Damour:2019lcq}
(in the energy gauge), since both the expansion in $1/j$ and the post--Newtonian expansion is finally carried out.

The quantity $q_{63}$ can be extracted from the component $\chi_3^{\rm h, loc}$ of the scattering
angle from the contribution $O(\PI^7)$
\begin{eqnarray}
\chi_3^{\rm h, loc} &=& -\frac{1}{3 \PI^3} + \frac{4 \eta^2}{\PI} + \eta^4 (24 - 8 \nu) \PI +
 \eta^6 \left(\frac{64}{3} - 36 \nu + 8 \nu^2\right) \PI^3 +
 \eta^8 \left(-\frac{91}{5} \nu + 34 \nu^2 \right.
\nonumber\\ && \left.
- 8 \nu^3\right)
\PI^5 +
 \eta^{10} \left(\frac{96}{35} \nu + \frac{288}{35} \nu^2 - \frac{108}{7} \nu^3 + 6 \nu^4 - \frac{1}{7} q_{63}
 \right) \PI^7.
\label{eq:chi3}
\end{eqnarray}
From
\begin{eqnarray}
\chi_3^{\rm h, loc} &=& \chi_3^{\rm Schw} - \frac{2\nu \PI}{\Gamma^2}
\left[\frac{2}{3} \gamma(14
\gamma^2 +25)+\frac{4(4 \gamma^4 - 12 \gamma^2 -3)}{\PI}
{\rm arcsinh} \left(\sqrt{\frac{\gamma-1}{2}}\right) \right]
\nonumber\\ &=&
-\frac{1}{3 p_\infty^3} + \frac{4 \eta^2}{p_\infty}
+ (24 - 8\nu) \eta^3 \PI + \left(\frac{64}{3} -36 \nu + 8\nu^2 \right) \eta^6 \PI^3
+ \left(- \frac{91}{5} \nu + 34 \nu^2 - 8\nu^3\right)
\nonumber\\ &&
\times \eta^8 \PI^5
+ \left(\frac{69}{70} \nu + \frac{51}{5} \nu^2 - 32 \nu^3 + 8 \nu^4\right) \eta^{10} \PI^7
\end{eqnarray}
and (\ref{eq:chi31}) one obtains $q_{63}$ as in Eq.~(\ref{eq:VAL2}).

In Eq.~(\ref{eq:SCA}) $(p.n)$ is calculated as in the case for periastron advance \cite{Blumlein:2019zku}
and by setting
\begin{eqnarray}
\hat{E} = \frac{\Gamma - 1}{\nu}
\label{eq:hatE}
\end{eqnarray}
and expanding to the respective post--Newtonian order. The operator ${\sf Reg}_{u_{\rm max}}$ removes
all pole terms in $\ep$, which
are implied in the integral (\ref{eq:SCA}) by
\begin{eqnarray}
y = u_{\rm max} := \frac{1}{j p_\infty} \left[1 + \sqrt{1 + j^2 p_\infty^2}\right] -
\ep,
\end{eqnarray}
where we choose $|\ep| \ll 1$. In using (\ref{eq:SCA}) for the calculation of $\chi_k^{\rm loc}$, singular terms
in $\ep$ in the corresponding integrals are disregarded.
Here the expansion coefficients of the Schwarzschild scattering angle $\chi_k^{\rm Schw}$ are given by
\begin{eqnarray}
\frac{\chi^{\rm Schw}}{2} = \sum_{k = 1}^\infty \frac{1}{j^k} \chi_k^{\rm Schw},
\label{eq:SCHW}
\end{eqnarray}
with
\setlength\arraycolsep{1pt}
\begin{align}
\label{eq:SCHW2}
\chi^{\rm Schw}_k &= {\sf Reg}_{\ep} \binom{-\tfrac{1}{2}}{k} \eta^k \int_0^{\sqrt{\eta^2 p_\infty^2  - \ep}} dy
\frac{[2y(1+y^2)]^k}
{(p_\infty^2 \eta^2 - y^2)^{k + 1/2}},&~{k~~\rm even}
\nonumber\\
&=
\frac{(-1)^{k+1} 2^{k-1}}{p_\infty^k} \binom{-\tfrac{1}{2}}{k}
\frac{\Gamma(\tfrac{k+1}{2}) \Gamma(\tfrac{1}{2}-k)}{\Gamma(1-\tfrac{k}{2})}
\pFq{2}{1}{-k,\tfrac{k+1}{2}}{1-\tfrac{k}{2}}{-\eta^2 p^2_\infty},&~{k~~\rm odd}.
\end{align}
\setlength\arraycolsep{0pt}

\noindent
${\sf Reg}_\ep$ projects onto the constant part in the $\ep$-expansion, which correspondingly
implies a regularization scheme dependence. However, the relation
\begin{eqnarray}
\chi_{2k}^{\rm h, loc}(p_\infty,\nu) = \frac{\pi}{2} K^{\rm loc,h}_{2k}(\hat{E}),
\label{eq:chijeven}
\end{eqnarray}
with
\begin{eqnarray}
\label{eq:K2}
K^{\rm loc,h}(\hat{E},j) = \sum_{k = 1}^\infty \frac{1}{j^{2k}} K^{\rm loc,h}_{2k}(\hat{E}),
\end{eqnarray}
cf.~\cite{Kalin:2019inp}, can be used in the limit $\nu \rightarrow 0$, fixing the regularization scheme. In
Eq.~(\ref{eq:chijeven}) it is understood, that $\hat{E}$ of (\ref{eq:hatE}) is
used and an expansion in $p_\infty$ to the respective PN order is performed. In the present case we use the
choice (\ref{eq:SCHW2}).
We label the different post--Newtonian contributions by the parameter $\eta$.
For the first orders in the expansion of (\ref{eq:SCHW}) one obtains
\begin{eqnarray}
\chi^{\rm Schw}_1 &=& \frac{1}{p_\infty} + 2 p_\infty \eta^2,
\\
\chi^{\rm Schw}_2 &=& \pi \left(\frac{3}{2} \eta^2 + \frac{15}{8} p_\infty^2 \eta^4 \right),
\\
\chi^{\rm Schw}_3 &=& -\frac{1}{3 p_\infty^3} + \frac{4 \eta^2}{p_\infty} + 24 p_\infty \eta^4 + \frac{64}{3}
p_\infty^3 \eta^6,
\\
\chi^{\rm Schw}_4 &=& \pi \left(
\frac{105}{8} \eta^4 + \frac{315}{8} p_\infty^2 \eta^6 + \frac{3465}{128} p_\infty^4 \eta^8 \right).
\end{eqnarray}
Higher--order expansion coefficients are listed in
Appendix~\ref{sec:E}.
The coefficients $q_{82}$ and $q_{44}$ can also be extracted from the component $\chi_2^{\rm h}$ and $\chi_4^{\rm h}$
of the scattering angle, respectively, using relation (\ref{eq:chijeven})
by expanding to $O(\PI^6)$.
One obtains the values given in Eqs.~(\ref{eq:VAL1}, \ref{eq:VAL3}). For $q_{44}$ we observe the
difference
\begin{eqnarray}
q_{44} - q_{44}^{\rm Ref. \text{\cite{Bini:2020wpo}}} =
{- \frac{719617}{4725} \nu^2}.
\end{eqnarray}
This coefficient is related to the $\hat{E}^3/j^4$ term of the periastron advance $K^{\rm loc,h}(\hat{E},j)$.
The potential contributions to this quantity are of $O(G_N^4)$ and have been first obtained in
\cite{Blumlein:2020pyo}. They have recently been confirmed by the $O(G_N^4)$ post--Minkowskian
results of \cite{Bern:2021dqo}, cf. also \cite{Blumlein:2021txj,Dlapa:2021npj}. The $O(\nu^2 \pi^2)$
contribution to this term is implied
by $q_{44}$, since $q_{82}$ and $q_{63}$ are purely rational in $\nu$. We also would like to mention that
the $\pi^2$ terms of $\bar{d}_5$ and $a_6$, as determined using the scattering angle in \cite{Bini:2021gat}, using
the potential terms given in \cite{Blumlein:2020pyo}, are equivalently obtained by using Eqs.~(\ref{eq:chijeven},
\ref{eq:K2}), which has been presented before in Ref.~\cite{Blumlein:2020pyo}.

The above expansion coefficients at 5PN were all derived without referring to a specific $\nu$ dependence of
the combination \cite{Damour:2019lcq}
\begin{eqnarray}
\Gamma^{k-1} \chi_k - \chi_k^{\rm Schw}.
\label{eq:Dconst}
\end{eqnarray}

The equations determining the rational $O(\nu^2)$ terms of $q_{44}, \bar{d}_5$ and $a_6$  contain
also local {far-zone} contributions, which are finite and just of $O(\nu^2)$ together
with the $O(\nu^2)$ from singular {far-zone} terms, which are fully predicted in relation to their $O(\nu)$
contributions, cf.~\cite{Blumlein:2020pyo}. This aspect will be also discussed in Section~\ref{sec:4}.

From the results obtained in the previous sections we determine now the  $O(\nu^2)$ contributions to
$q_{44}, a_6$ and $\bar{d}_5$. We first consider $\chi_4^{\rm total,h}$,
\begin{eqnarray}
\chi_4^{\rm total,h} = \chi_4^{\rm nonloc,h} + \chi_4^{\rm loc,h}.
\end{eqnarray}
The non--local contribution, $\chi_4^{\rm nonloc, h}$, has been calculated in \cite{Bini:2020wpo} and
is given by
{
\begin{eqnarray}
\chi_4^{\rm nonloc, h} &=& \nu \eta^8 \PI^4 \pi \Biggl\{- \frac{63}{4} - \eta^2 \left(\frac{2753}{1120}
- \frac{1071}{40} \nu\right) \PI^2 - \ln\left(\frac{\PI}{2}\right)\Biggl[\frac{37}{5}
+ \eta^2\left(\frac{1357}{280} -
\frac{111}{10} \nu
\right)\PI^2\Biggr]
\Biggr\}
\nonumber\\
\end{eqnarray}}

\noindent
Its transform to the f--scheme
for $\Gamma^3 \chi_4^{\rm f-h}$ is given in Eq.~(8.6) there, to eliminate the term of $O(\pi \nu^2
p_\infty^6)$.\footnote{Eq.~(7.24) of \cite{Bini:2020wpo} contains typos. Both $a_{\rm 5PN}$ and $b_{\rm 5PN}$
need an additional factor $\PI^2$. Otherwise Eq.~(8.2) will not hold.}
 One has
\begin{eqnarray}
\Gamma^3 \chi_4^{\rm nonloc, h} &=& \nu \eta^8 \PI^4 \pi \Biggl\{- \frac{63}{4} - \frac{37}{5}
\ln\left(\frac{\PI}{2}\right)
+ \eta^2 \PI^2 \Biggl[ - \frac{2753}{1120} - \frac{1357}{280} \ln\left(\frac{\PI}{2}\right) + \frac{63}{20} \nu \Biggr]
\Biggr\}.
\end{eqnarray}

The local contribution is obtained from
Eq.~(\ref{eq:chijeven}) including all local contributions to the Hamiltonian, and reads
\begin{eqnarray}
\label{eq:CHI4}
\chi_4^{\rm loc, h} &=&  \pi \Biggl\{
\eta^4\Biggl(\frac{105}{8} - \frac{15}{4} \nu\Biggr)
+
\eta^6 \Biggl[\frac{315}{8} - \Biggl(\frac{109}{2} - \frac{123}{256} \pi^2 \Biggr) \nu  + \frac{45}{8} \nu^2 \Biggr]
p_\infty^2
\nonumber\\ &&
+
   \eta^8 \Biggl[\frac{3465}{128} -  \Biggl(\frac{19597}{192} - \frac{33601}{16384} \pi^2 \Biggr) \nu +
\Biggl(\frac{4827}{64} - \frac{369}{512} \pi^2\Biggr) \nu^2  - \frac{225}{32} \nu^3
\Biggr] p_\infty^4
\nonumber\\ &&
+
   \eta^{10} \Biggl[
      -\Biggl (\frac{1945583}{33600} - \frac{93031}{32768} \pi^2\Biggr) \nu
+ \Biggl({\frac{12605197}{100800}}
- \frac{94899}{32768} \pi^2\Biggr) \nu^2
\nonumber\\ &&
- \Biggl(\frac{2895}{32} - \frac{1845}{2048} \pi^2\Biggr) \nu^3 + \frac{525}{64} \nu^4
\Biggr] p_\infty^6 \Biggr\},
\end{eqnarray}
and one obtains
\begin{eqnarray}
\Gamma^3 \chi_4^{\rm loc,h} - \chi_4^{\rm Schw} &=& \pi \nu \Biggl[
-\frac{15 }{4} \eta^4
+ \eta^6 \PI^2 \left(-\frac{557}{16} + \frac{123}{256} \pi^2\right)
+ \eta^8 \PI^4 \left(-\frac{4601}{96} + \frac{33601}{16384} \pi^2\right)
\nonumber\\ &&
+ \eta^{10} \PI^6 \left(
-\frac{3978707}{134400} + \frac{93031}{32768} \pi^2 
+ {\frac{402097}{100800}} \nu
\right)\Biggr].
\end{eqnarray}
Forming the combination (\ref{eq:Dconst}) and expanding to $O(\eta^{10})$ (5PN) we obtain
\begin{eqnarray}
\left[\Gamma^3 \chi_4^{\rm total,h} - \chi_4^{\rm Schw}\right] -
\left[\Gamma^3 \chi_4^{\rm total,h} - \chi_4^{\rm Schw}\right]^{\text{\rm Ref.~\cite{Bini:2020wpo}}}
= {\frac{719617}{100800}} \eta^{10} \pi \nu^2 \PI^6,
\end{eqnarray}
comparing with the result of Ref.~\cite{Bini:2020wpo}, Eqs.~(8.2, 8.6, 10.1c, 10.2d),
and thus observe a  breaking of the rule
\begin{eqnarray}
\left[\Gamma^{k-1} \chi_k^{\rm total,h} - \chi_k^{\rm Schw}\right] = P_k(\nu)
\label{eq:REL1}
\end{eqnarray}
with degree of $[(k-1)/2]$ of the polynomial $P_k(\nu)$.

On the other hand, one observes for $k \in [1,3]$, that (\ref{eq:REL1}) holds. It is trivially obeyed
for $k = 1,2$ because of the post--Minkowskian structure \cite{Westpfahl:1979gu}, see also \cite{Damour:2019lcq},
and is implied for $k = 3$ by
\begin{eqnarray}
\Gamma^3 {\chi}^{\rm h, loc}_3 - \chi_3^{\rm Schw}~&=&~
\nu \left[
-\frac{\eta^2}{3\PI} -\frac{47}{12} \eta^4 \PI
-\frac{313}{24} \eta^6 \PI^3 - \frac{749}{320} \eta^8 \PI^5 - \frac{7519}{4480} \eta^{10} \PI^7 \right]
+ O(\PI^9),\nonumber\\
\label{eq:chi31}
\end{eqnarray}
referring to the $O(G_N^3)$ post--Minkowskian Hamiltonian \cite{Bern:2019nnu}. Note that for $k \leq 3$ no
{far-zone} terms
contribute and the scattering angle stems from the potential contributions
only. The post--Minkowskian result of $O(G_N^3)$ \cite{Bern:2019nnu} has been checked to 6PN in
\cite{Blumlein:2020znm}.

Finally we would like to discuss the hypothetical possibility to accommodate the result of the present paper
on $q_{44}$ with the one in \cite{Bini:2020wpo}. In a later paper, \cite{Bini:2021gat}, the authors
of \cite{Bini:2020wpo} claim to have still a missing term in Eq.~(10.2) of \cite{Bini:2021gat} on radiation reaction,
which is not yet quantified. It is of the type
\begin{eqnarray}
\delta H_{\text{{rad}}} = a \eta^{10} \nu^2 \frac{(p.n)^4}{r^4},
\label{eq:delHt}
\end{eqnarray}
with $a \in \mathbb{R}$.
{We stress that to the best of our knowledge no such contributions are missing in the present EFT approach.}
Terms of the kind (\ref{eq:delHt}) can imply the mapping
{
\begin{eqnarray}
q_{44}^{\rm h} \rightarrow  q_{44}^{\text{\rm Ref.~\cite{Bini:2020wpo}}}
\label{eq:q443}
\end{eqnarray}}

\noindent
here with the value
{\begin{eqnarray}
a =  {\frac{
402097}{9450}},
\label{eq:aa}
\end{eqnarray}}

\noindent
to reach full agreement. It transforms $K(\hat{E},j)^{\rm 5PN}_{\rm loc h}$ into \cite{Bini:2020hmy}, Eq.~(F.5).
This also applies e.g. to $\chi_k^{\rm loc},~k=4,5,6$, while $E^{\rm circ}$ and $K^{\rm circ}$ remain unaffected,
see Section~\ref{sec:4}. We finally note that in the post--Minkowskian apprach $\chi_k(j)$ is derived from the terms
of $O(G_N^k)$ and are presently known for all contributions to $k = 4$. 
\section{Phenomenological results}
\label{sec:4}

\vspace*{1mm}
\noindent
In the following we summarize the results of the present calculation for observables and present numerical
results.
We illustrate the result for the local 5PN contribution to the local contributions to general periastron advance $K^{\rm
loc,h}(\hat{E},j)$ and the
circular limits of the binding energy $E^{\rm circ}(j)$ and periastron advance $K^{\rm circ}(j)$ 
which are given by\footnote{The respective non--local contributions to 5PN and the local contributions
to 4PN are given in Ref.~\cite{Blumlein:2020pyo}.}
\begin{eqnarray}
K(\hat{E},j)_{\rm loc,h}^{\rm 5PN} &=&
\Biggl\{
\Biggl[
\frac{15 \nu^2}{16}
 - \frac{15}{4} \nu^3
 + 3 \nu^4
\Biggr]\frac{ \hat{E}^4}{j^2}
+\Biggl[
\frac{3465}{16}
+ \left(-\frac{12160657}{8400}
+\frac{15829 \pi ^2}{256} \right) \nu
+\Biggl(
{\frac{10575547}{6300}}
\nonumber\\ &&
-\frac{35569 \pi ^2}{1024} \Biggr)\nu^2
+\Biggl(\frac{1107 \pi ^2}{128}
-\frac{7113}{8}\Biggr) \nu^3
+75\nu ^4
\Biggr]
\frac{ \hat{E}^3}{j^4}
+\Biggl[
 \frac{315315}{32}
+\Biggl(-\frac{33023719}{840}
\nonumber\\ &&
+\frac{4899565 \pi ^2}{4096}\Biggr) \nu
+ \Biggl(
-\frac{3289285 \pi^2}{1024}
+ {
\frac{139190707}{2520}}
\Biggr) \nu^2
+\Biggl(\frac{35055 \pi ^2}{256}
-\frac{240585}{32}\Biggr) \nu^3
\nonumber\\ &&
+\frac{1575}{8} \nu^4
{\Biggr]}
\frac{ \hat{E}^2}{j^6}
+\Biggl[ \frac{765765}{16}
+ \Biggl(-\frac{30690127}{240}
+ \frac{16173395 \pi ^2}{8192}\Biggr) \nu
+ \Biggl({
\frac{67444909}{432}}
\nonumber\\ &&
- \frac{77646205 \pi^2}{8192} \Biggr)\nu^2
+ \Biggl(
\frac{121975 \pi ^2}{512}
-\frac{271705}{24}\Biggr) \nu^3
+\frac{2205}{16} \nu^4
\Biggr]
\frac{ \hat{E}}{j^8}
+\Biggl[
 \frac{2909907}{64}
\nonumber\\ &&
+\Biggl(-\frac{61358067}{640}
+
\frac{1096263 \pi^2}{1024} \Biggr)\nu
+\Biggl(
{\frac{169533949}{1920}}
-\frac{87068961 \pi^2}{16384}\Biggr) \nu^2
+\Biggl(\frac{90405 \pi^2}{1024}
\nonumber\\ &&
-\frac{127995}{32}\Biggr) \nu^3
+\frac{3465}{128} \nu^4
\Biggr]
\frac{1}{j^{10}}
\Biggr\} \eta^{10}
+ O(\eta^{12})
\label{eq:PERI2}
\end{eqnarray}
\begin{figure}[H]
  \centering
  \hskip-0.8cm
  \includegraphics[width=.8\linewidth]{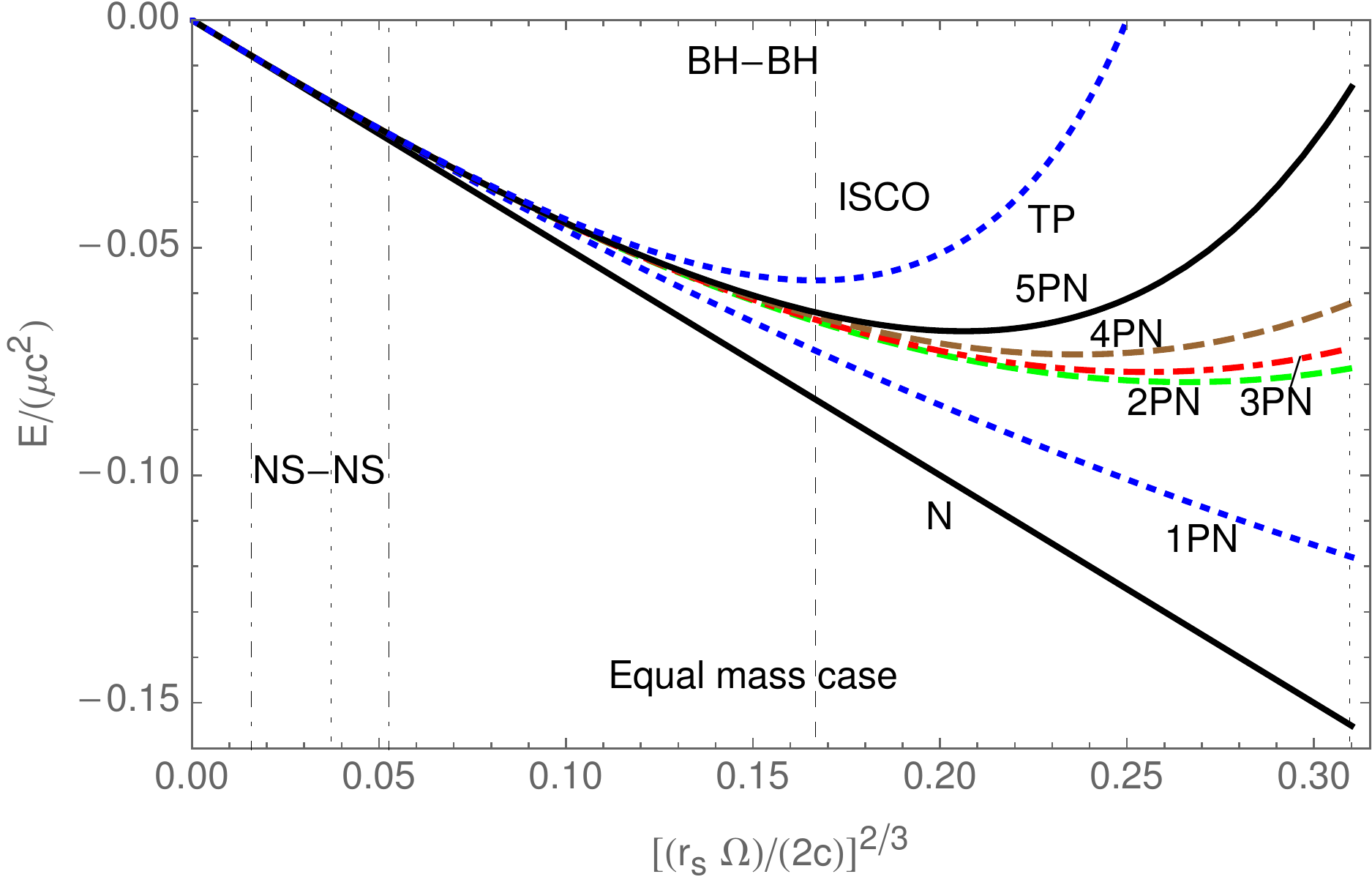}
  \caption[]{\sf The binding energy in the quasi-circular case for equal masses.
  Lower full  line: Newtonian case (N);
  Dotted line: 1PN;
  Dashed line: 2PN;
  Dash-dotted line: 3PN;
  Upper dashed line: 4PN;
  Upper full line: 5PN;
  Upper dotted line: test particle solution (TP).
  Dashed vertical line: the innermost stable circular orbit (ISCO) at 5PN.
  The other vertical lines mark the frequency spectrum for neutron star (NS) and black hole (BH) merging 
  at LIGO.}
  \label{fig:2}
\end{figure}
and
\begin{eqnarray}
\label{Ecirc}
\frac{E^{\rm circ, h}(j)}{\mu c^2}
&=& -\frac{1}{2j^2}
-\left(
\frac{9}{8}
+\frac{\nu }{8}
\right)
\frac{1}{j^4} \eta^2
+\left(
-\frac{81}{16}
+\frac{7 \nu }{16}
-\frac{\nu ^2}{16}
\right)
\frac{1}{j^6} \eta^4
+\Biggl[
-\frac{3861}{128}
+\left(\frac{8833}{384}
-\frac{41 \pi ^2}{64}\right) \nu
\nonumber\\ &&
+\frac{5 \nu ^2}{64}
-\frac{5 \nu ^3}{128}
\Biggr] \frac{1}{j^8} \eta^6
+
\Biggl[
   -\frac{53703}{256}
+\left(\frac{989911}{3840}
-\frac{6581 \pi ^2}{1024}\right) \nu
+\left(
-\frac{8875}{768}
+\frac{41 \pi^2}{128}
\right) \nu^2
\nonumber\\ &&
+\frac{3 \nu ^3}{128}
-\frac{7 \nu ^4}{256}
\Biggr] \frac{1}{j^{10}} \eta^8
+
\Biggl[
-\frac{1648269}{1024}
+\left(\frac{3747183493}{1612800} - \frac{31547 \pi ^2}{1536}\right) \nu
+\left(
-{\frac{400240439}{403200}}
\right.
\nonumber\\ && \left.
+ \frac{132979 \pi^2}{2048}
\right) \nu ^2
+\left(
-\frac{3769}{3072}
+\frac{41 \pi^2}{512}
\right) \nu^3
+\frac{5 \nu ^4}{1024}
-\frac{21 \nu^5}{1024}
\Biggr] \frac{1}{j^{12}} \eta^{10}
+ \frac{E^{\rm circ}_{\rm nl}}{\mu c^2}
+
O\left( \eta^{12} \right),
\nonumber\\
\end{eqnarray}
\begin{eqnarray}
\label{Kcirc}
K^{\rm circ, h}(j)
&=&
1 + 3 \frac{1}{j^2} \eta^2
+\left(\frac{45}{2}-6\nu\right)\frac{1}{j^4} \eta^4
+\left[\frac{405}{2}+\left(-202+\frac{123}{32}\pi^2\right)\nu+3\nu^2\right]\frac{1}{j^6} \eta^6
+
\left[\frac{15795}{8} \right.
\nonumber\\ && \left.
+\left(
-\frac{105991}{36}
+\frac{185767}{3072}\pi^2
\right)\nu
+\left(
\frac{2479}{6}
-\frac{41}{4}\pi^2
\right)\nu^2\right]\frac{1}{j^8} \eta^8
+
\Biggl[\frac{161109}{8}
+\left(-\frac{18144676}{525} \right.
\nonumber\\ && \left.
+ \frac{488373}{2048}\pi^2\right)\nu
+\left({\frac{105496222}{4725}}
- \frac{1379075}{1024} \pi^2
\right)\nu^2
+\left(-\frac{1627}{6}+\frac{205}{32}\pi^2\right)\nu^3\Biggr]\frac{1}{j^{10}} \eta^{10}
\nonumber\\ &&
+ K^{\rm nl}_{\rm 4+5PN}(j) + O\left(\eta^{12} \right).
\end{eqnarray}
The non--local contributions to Eqs.~(\ref{Ecirc}) and (\ref{Kcirc}) have been given in \cite{Blumlein:2020pyo},
Eqs.~(44, 54, 55).
\begin{figure}[H]
  \centering
  \hskip-0.8cm
  \includegraphics[width=.8\linewidth]{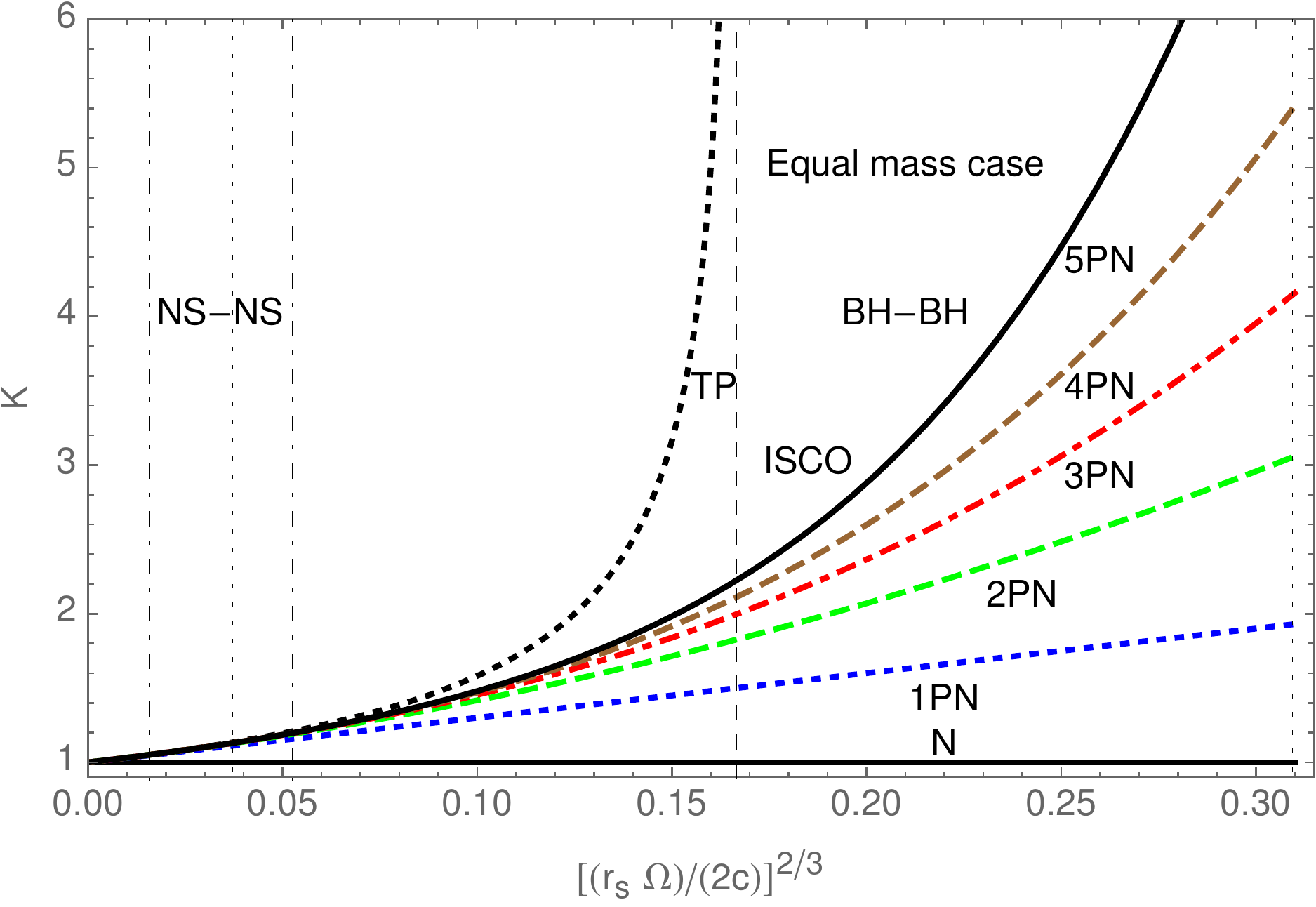}
  \caption[]{\sf The periastron advance in the circular limit.
    The labeling of the curves is the same as in Figure~\ref{fig:2}.
  }
  \label{fig:3}
\end{figure}

We illustrate the different post--Newtonian contributions to the binding energy, $E^{\rm circ}(x)$, and
the periastron advance, $K^{\rm circ}(x)$ for circular orbits, setting $m_1 = m_2,~~(\nu = 1/4)$, with
$x = (G_N M \Omega/c^3)^{2/3}$, $\Omega = \Omega_\phi$ the angular frequency, and $r_s$ the Schwarzschild radius.
The relation between $j$ and $x$ to 5PN is given by
\begin{eqnarray}
j &=& \frac{1}{\sqrt{x}} \Biggl\{
        1
        +\frac{1}{6} \eta ^2 x (9+\nu )
        +\frac{1}{24} \eta ^4  x^2 \big(
                81-57 \nu +\nu ^2\big)
        +\eta ^6 x^3 \Bigg[
                \frac{135}{16}
                +\frac{1}{144}  \big(
                        -6889+246 \pi ^2\big) \nu
\nonumber\\ &&
     +\frac{31 \nu ^2}{24}
                +\frac{7 \nu ^3}{1296}
        \Biggr]
        +\eta ^8 x^4 \Biggl[
                \frac{2835}{128}
                +\nu  \Biggl(
                        \frac{98869}{5760}-\frac{128 \gamma_E }{3}-\frac{6455 \pi ^2}{1536}-\frac{256 \ln(2)}{3}
-\frac{64 \ln(x)}{3}\Biggr)
\nonumber\\ &&
+\frac{5\big(
                        71207-2706 \pi ^2\big)}{3456} \nu^2
                -\frac{215 \nu ^3}{1728}
                -\frac{55 \nu ^4}{31104}
        \Biggr]
        +\eta ^{10} x^5 \Biggl[
                \frac{15309}{256}
                +\Biggl(
                        \frac{59112343}{44800}+\frac{19952 \gamma_E }{105}
\nonumber\\ &&
-\frac{126779 \pi ^2}{768}+\frac{47344 \ln(2)}
{105}-\frac{486 \ln(3)}{7}+\frac{9976 \ln(x)}{105}\Biggr) \nu
                + \Biggl(
{\frac{99980267}{50400}}
+\frac{2624 \gamma_E }{15}
\nonumber\\ &&
-\frac{289595 \pi ^2}{1536}+\frac{6976
\ln(2)}{105}+\frac{1944 \ln(3)}{7}+\frac{1312 \ln(x)}{15}\Biggr) \nu^2
                +\frac{1}{256} \nu ^3 \big(
                        -25189+902 \pi ^2\big)
\nonumber\\ &&
                -\frac{55 \nu ^4}{768}
                -\frac{\nu ^5}{768}
\Biggr]
\Biggr\} + O(\eta^{12}).
\end{eqnarray}
We also add the the test--particle lines (TP), which are given by \cite{Schafer:2018kuf}
\begin{eqnarray}
\frac{E^{\rm circ}_{\rm TP}}{\mu c^2} &=& \frac{1 - 2x \eta^2}{\sqrt{1-3x \eta^2}} -1
\end{eqnarray}
and
\begin{eqnarray}
K_{\rm TP}^{\rm circ} &=& \frac{1}{\sqrt{1-6 \eta^2 x }}
\end{eqnarray}
with
\begin{eqnarray}
j_{\rm TP} = \frac{1}{\sqrt{x(1 - 3 \eta^2 x)}}.
\end{eqnarray}

The different post--Newtonian corrections are all positive correcting lower order results. Even at 5PN the
convergence is not yet perfect both for the binding energy and periastron advance, considering the 
range $x \in [0, 0.30]$, which is calling for effective resummations of these contributions.
We also mention that the post--Minkowskian corrections to $O(G_N^4)$ given in \cite{Bern:2021yeh} recently
represent the bound state dynamics to the 3rd post--Newtonian order.  Starting with 4PN there are 
deviations
both in the potential and tail terms, because of the yet missing higher powers in $1/r$ and also the fact
that these corrections are derived for scattering kinematics, still to be kinematically continued to the
elliptic orbit kinematics. On the other hand, our formalism allowed to predict the post--Newtonian 
expansion terms of the results of \cite{Bern:2021yeh} up to 6PN, 
cf.~\cite{Blumlein:2020znm,Blumlein:2021txj}

Finally, we present the PN--expansion of the local scattering angle, cf.~(\ref{eq:SCA}--\ref{eq:SCA3}).
\begin{eqnarray}
\label{eq:CHIMAIN1}
\chi_1^{\rm h, loc} &=& \frac{w_1}{2 \PI} = 
\frac{1}{\PI }
+2 \eta ^2 \PI,
\\
\chi_2^{\rm h, loc} &=& \pi \frac{w_2}{4} =
\pi  \Biggl[
        \frac{3 \eta ^2}{2}
        +\frac{3}{8} \eta ^4 (5-2 \nu ) \PI ^2
        +\frac{3}{16} \eta ^6 \nu  (-4+3 \nu ) \PI ^4
        +\eta ^8 \Biggl(
                \frac{9 \nu }{64}+\frac{27 \nu ^2}{64}-\frac{15 \nu ^3}{32}\Biggr) \PI ^6
\nonumber\\ &&
        +\eta ^{10} \Biggl(
                -\frac{15 \nu }{256}-\frac{45 \nu ^2}{256}-\frac{15 \nu ^3}{64}+\frac{105 \nu ^4}{256}\Biggr) \PI ^8
\Biggr],
\\
\chi_3^{\rm h, loc} &=&
-\frac{w_1^3}{24 \PI^3} + \frac{w_1 w_2}{2 \PI} + w_3 \PI,
\\
\chi_4^{\rm h, loc} &=& 
\pi \left(\frac{3 w_2^2}{16} + \frac{3 w_1 w_3}{8} + \frac{3 w_4}{8} \PI^2\right),
\\
\chi_5^{\rm h, loc} &=&
 \frac{w_1^5}{160 \PI^5} 
-\frac{w_1^3 w_2}{12 \PI^3}
+\frac{w_1^2 w_3+w_1 w_2^2}{2 \PI}
+ \left(2 w_2 w_3 + 2 w_1 w_4 \right) \PI
+\frac{4 w_5}{3} \PI^3 
\nonumber\\ &=&
\frac{1}{5 \PI ^5}
-\frac{2 \eta ^2}{\PI ^3}
+\frac{8 \eta ^4 (4-\nu )}{\PI }
+\eta ^6 \PI  \left(
        320
        -\left(
                \frac{1168}{3}-\frac{41 \pi ^2}{8}\right) \nu         +24 \nu ^2
\right)
\nonumber\\ &&
+\eta ^8 \PI ^3 \left(
        640
        -\frac{
                3632944-76035 \pi ^2}{2160} \nu
        +\frac{1}{72} \big(
                58736-861 \pi ^2\big) \nu^2
        -40 \nu ^3
\right)
\nonumber\\ &&
+\eta ^{10} \PI ^5 \Biggl(
        \frac{1792}{5}
        -\frac{
                93470656-3886715 \pi ^2}{33600} \nu
       + \left({\frac{64770686}{14175}}
       - \frac{107249}{480}\right) \nu^2
\nonumber\\ &&
-\frac{1}{72} \big(
                88864-1353 \pi ^2\big) \nu^3
        +56 \nu ^4
\Biggr),
\\
\label{eq:CHIMAIN2}
\chi_6^{\rm h, loc} &=&
\pi  \Biggl[
         \frac{5 w_2^3}{32}
        +\frac{15 w_1 w_2 w_3}{16} 
        +\frac{15 w_1^2 w^4}{32} 
        +\left(
                \frac{15 w_3^2}{32}
                +\frac{15 w_2 w_4}{16}
                +\frac{15 w_1 w_5}{16} 
        \right) \PI^2
        +\frac{15 w_6}{32} \PI^4
\Biggr]
\nonumber\\
&=&\pi  \Biggl[
        \eta ^6 \left(
                \frac{1155}{8}
                - \left(
                        \frac{625}{4}-\frac{615 \pi ^2}{256}\right) \nu
                +\frac{105}{16} \nu^2
        \right)
+
        \eta ^8 \PI ^2 \Biggl(
                \frac{45045}{64}
                -\frac{
                        37556864-771585 \pi ^2}{24576} \nu
\nonumber\\ &&
 +\frac{5}{64} \big(
                        7013-123 \pi ^2\big) \nu^2
                -\frac{525}{32} \nu^3
        \Biggr)
        +\eta ^{10} \PI ^4 \Biggl(
                \frac{135135}{128}
-\frac{
                        8099529344-243724425 \pi ^2}{1720320} \nu
\nonumber\\ &&
               + \left({\frac{
288262859}{40320} - \frac{13335615}{32768} \pi^2}\right) \nu^2
-\frac{25 \big(
                        87872-1599 \pi ^2\big)}{2048} \nu^3
                +\frac{3675}{128} \nu^4
        \Biggr)
\Biggr].
\end{eqnarray}
The parameters $w_i$ the expansions need to be expanded up to $O(\PI^8)$ in accordance 
with 5PN accuracy, 
\begin{eqnarray}
w_i &=& \sum_{k=0}^4 v_{i,k}(\nu) \PI^{2k}
\end{eqnarray}
and can be obtained recursively from (\ref{eq:CHIMAIN1}--\ref{eq:CHIMAIN2}).
For $\chi_{3,4}^{\rm h, loc}$ the explicit expressions are given in (\ref{eq:chi31}, 
\ref{eq:CHI4}).
Leaving the 5PN EOB parameters open, one obtains 
\begin{eqnarray}
\chi_2^{{\rm h,loc}, p_\infty^8} &=& - \pi \eta^{10} p_\infty^8 \frac{35}{512} q_{82}
\\
\chi_3^{{\rm h,loc}, p_\infty^7} &=& 
\eta^{10} \PI^7 \left( \frac{18 \nu}{5} + \frac{54 \nu^2}{5} - 12 \nu^3 -  \frac{1}{7} q_{63} - 
q_{82} \right)
\\
\chi_4^{{\rm h,loc}, p_\infty^6} &=& \pi \eta^{10} p_\infty^6 \left(
- \frac{1971}{64} \nu  + \frac{5601}{64} \nu^2 
- \frac{975}{32} \nu^3 - \frac{3}{64} q_{44} 
- \frac{15}{64} q_{63} - \frac{105}{128} q_{82}
\right)
\\
\chi_5^{{\rm h,loc}, p_\infty^5} &=& 
\eta^{10} \PI^5\left(
\frac{1792}{5}
-\frac{415658 \nu }{225}
+\frac{32492 \nu ^2}{15}
-\frac{1896 \nu ^3}{5}
-\frac{1}{120} \nu  (-4085+1722 \nu ) \pi ^2
\right. \nonumber\\ && \left.
-\frac{4}{15} \bar{d}_5
-\frac{4}{5} q_{44}
-2 q_{63}
-\frac{14}{3} q_{82} \right)
\\
\chi_6^{{\rm h,loc}, p_\infty^4} &=& 
\eta^{10} \PI^4 \pi\left[
        \frac{135135}{128}
        +\left(
                -\frac{180377}{48}+\frac{1283765 \pi^2}{16384}\right) \nu
        +\left(
                \frac{190655}{64}-\frac{36285\pi^2}{1024}\right) \nu^2
\right.
\nonumber\\ &&  \left.
        -\frac{9945}{32} \nu ^3 
-\frac{15}{32} a_6
        -\frac{15}{32} \bar{d}_5
        -\frac{45}{64}  q_{44}
        -\frac{75}{64}  q_{63}
        -\frac{525}{256} q_{82}
\right].
\end{eqnarray}
The other parts are determined by the 4PN contributions.

If we consider the phenomenological addition, Eq.~(\ref{eq:delHt}), the following changes are implied.
Let $X$ be one of the following  observables and the change $\delta X$ defined by
\begin{eqnarray}
\delta X = X(H) - X(H + \delta H_{\text{{rad}}}).
\end{eqnarray}
{Then one obtains}
{\begin{eqnarray}
\delta K(\hat{E},j)^{\rm 5PN}_{\rm loc,h}~&=&~  a \eta^{10} \nu^2 \left[
\frac{63}{16 j^{10}} + \frac{105 \hat{E}}{8 j^8} + \frac{45 \hat{E}^2}{4 j^6} + \frac{3 \hat{E}^3}{2 j^4}\right],
\\
\delta E^{\rm circ}(j)   ~&=&~ 0,
\\
\delta K^{\rm circ}(j)                   ~&=&~ 0,
\\
\delta j(x)  ~&=&~ 0,
\\
\label{eq:CHI41}
\delta \chi_4^{\rm h, loc}                    ~&=&~  a \pi \eta^{10} \nu^2 \frac{3}{32}
\PI^6,~~\text{\rm etc.},
\end{eqnarray}}

\noindent
which, again, is a consequence of the transformation of $q_{44}$ {\it only}.
The expressions for $\chi_k^{\rm h},~k = 1,2,3$ remain the same. Yet no consistent solution demanding the
constraint (\ref{eq:REL1}) and referring to the Hamiltonian in the multipole--picture for the {far-zone} 
terms is obtained.
\section{Conclusions}
\label{sec:5}

\vspace*{1mm}
\noindent
We have calculated the 5PN Hamiltonian in the harmonic gauge using an EFT method both for the potential and
the {far-zone} terms. For the memory term {we obtain a  different result comparing to  \cite{Foffa:2019eeb}
and find one more contributing diagram.} From the local contributions to periastron advance
$K^{\rm loc,h}(\hat{E},j)$ it has been possible to derive the five 5PN EOB parameters $q_{82}, q_{63}, q_{44},
\bar{d}_5$ and
$a_6$. The rational contributions of $O(\nu^2)$ to $q_{44}, \bar{d}_5$ and $a_6$ do also depend on the non--singular
multipole moments  $L_k \varepsilon_{ijk} Q_{il} Q_{jl}$ and $Q_{ij} Q_{jk} Q_{ki}$, which enter as $D=4$
dependent quantities and are of $O(\nu^2)$.
We are not aware of other {far-zone} contributions  in the EFT approach.

Our results on the observables $K^{\rm loc,h}(\hat{E},j)$, the total circular peristron advance $K^{\rm tot}(j)$, the
circular binding energy $E^{\rm circ}$ and the contributions to the scattering angle $\chi_k,~~k \in [1,6]$, do agree
with the results of Ref.~\cite{Bini:2020wpo}, in the case of the parameters given there,
except of the rational $O(\nu^2)$ term of $q_{44}$. Furthermore, we have also newly obtained
the rational contributions of $O(\nu^2)$ to $\bar{d}_5$ and $a_6$ for the first time. The $O(\nu^2 \pi^2)$
contributions to
$q_{44}$ were obtained in \cite{Bini:2019nra} and those to $\bar{d}_5$ and $a_6$ in \cite{Blumlein:2020pyo} before.
These terms
stem from the potential contributions. For them the relation (\ref{eq:REL1}) holds. Our calculation of the 5PN Hamiltonian ab
initio leads to a violation of relation (\ref{eq:REL1}) for the $O(\nu^2)$ rational term in $q_{44}$. As has been outlined in
Section~\ref{sec:4}, the results of \cite{Bini:2019nra} could be obtained by invoking the additional term
(\ref{eq:delHt}, \ref{eq:aa}) changing only $q_{44}$, but leaving $\bar{d}_5$ and $a_6$ invariant.
We presented numerical results to 5PN for  $E^{\rm circ}$ and $K^{\rm circ}$.

We are aware of the fact that one may perform resummations of the exact results obtained for the Hamiltonian
dynamics to a certain post--Newtonian order, see e.g. \cite{Damour:2000we,Antonelli:2019ytb}, even with a matching
to results from numerical gravity, to be performed in a gauge invariant way. However, it is well--known that
resummations of this kind, also applied in phenomenological implementations of exact quantum field
theoretic
calculations sometimes, cannot accommodate for exact results, since some of the higher order corrections are
necessarily not included. The latter ones may be of the same order or even larger than the resummed terms.
One example can be found in Ref.~\cite{BK87}.

The level of the 5PN corrections to Hamiltonian dynamics exhibits a complexity which currently could only
be solved using the effective field theory approach. The latter has been originally developed for renormalizable
quantum field theories. There refined algorithms for the automated computer algebraic derivation of all dynamical
contributions and very efficient algorithms for term reduction exist \cite{IBP,CRUSHER}. Likewise, many methods
to compute the contributing integrals analytically have been developed \cite{Blumlein:2018cms}. By these methods
the problem at hand could be solved. Any theory having a path integral representation, including classical
mechanics, can be formulated in this way. Here the path integral \cite{FH} is a consequence of the variational
principles of mechanics \cite{LANCZOS}, the basis of any dynamical physical law. Symbioses of different fields in
science
lead to progress in the present case. Future calculations will use similar technologies at higher post-Newtonian
orders. Given the fast growth of complexity, however, even more refined technologies have to be developed
to solve
these problems.
\appendix

\section{The Feynman rules and integrals}
\label{sec:A}

\vspace*{1mm}
\noindent
In the following we list the Feynman rules, which are necessary to calculate the convergent {far-zone} contributions.
We first present the Feynman rules in the standard case \cite{Kol:2007bc} and turn then to
those in the in--in formalism.
\begin{align}
  \phi:\raisebox{-5.1mm}{\includegraphics{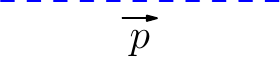}} ={}& - \frac{i}{2c_d} D(p)\,,\displaybreak[0]\\
  A:\raisebox{-5.1mm}{\includegraphics{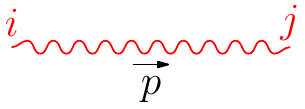}} ={}& \frac{i\delta_{ij}}{2} D(p)\,,\displaybreak[0]\\
  \sigma:\raisebox{-5.1mm}{\includegraphics{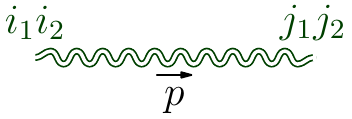}} ={}&
-\frac{i}{2}D(p)[\delta_{i_1j_1}\delta_{i_2j_2} + \delta_{i_1j_2}\delta_{i_2j_1} +
(2-c_d)\delta_{i_1i_2}\delta_{j_1j_2}]\,,\displaybreak[0]\\ %
  \centeredgraphics{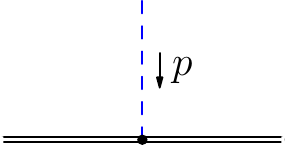} ={}& -\frac{i}{\Lambda} \biggl(
E
{-} \frac{1}{2} Q_{ij}p_i p_j
{-} \frac{i}{6} O_{ijk} p_i p_j p_k
\biggr)\,,\\
  \centeredgraphics{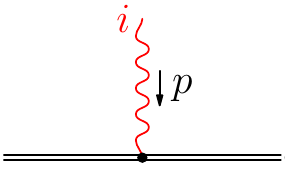} ={}& \frac{i}{\Lambda} \biggl(
- \frac{i}{2} L_{j}\varepsilon_{ijk} p_k
{-} \frac{1}{2} Q_{ij} p_0 p_j
+ \frac{1}{3} J_{jk} \varepsilon_{ijl} p_k p_l
{-} \frac{i}{6} O_{ijk} p_j p_k p_0
\biggr)\,,\\
\label{eq:FR1}
  \centeredgraphics{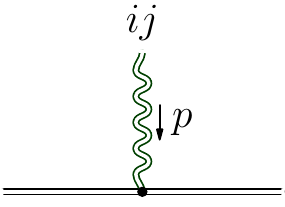} ={}& \frac{i}{\Lambda} \biggl(
- \frac{1}{4} Q_{ij} p_0^2
+ \frac{1}{6}(J_{ik}\varepsilon_{jkl} + J_{jk}\varepsilon_{ikl}) p_l p_0
{-} \frac{i}{12} O_{ijk} p_k p_0^2
\biggr)\,,\\
\label{eq:FR2}
  \centeredgraphics{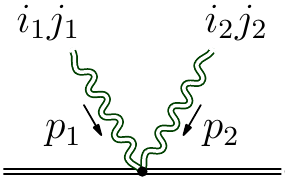} ={}&
\begin{aligned}
\tfrac{i}{16\Lambda^2} p_{1 0} p_{2 0} (&
 \delta_{i_1 i_2} Q_{j_1 j_2}
+ \delta_{i_1 j_2} Q_{i_2 j_1}\\
&+ \delta_{i_2 j_1} Q_{i_1 j_2}
+ \delta_{j_1 j_2} Q_{i_1 i_2}
+ \dots
)\,,
\end{aligned}\\
\label{e114}
  \centeredgraphics{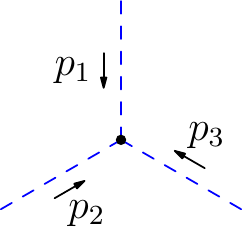} ={}& \frac{2i c_d^2}{\Lambda} (p_{10} p_{20} + p_{10} p_{30} + p_{20} p_{30})\,,\\
\label{e115}
  \centeredgraphics{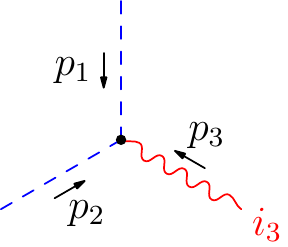} ={}& -\frac{2i c_d}{\Lambda} (p_{10} p_{2i_3} + p_{1i_3} p_{20})\,,\\
\label{e116}
  \centeredgraphics{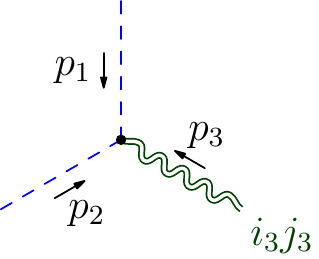} ={}& \frac{i c_d}{\Lambda} (-p_{1_3i} p_{2j_3} - p_{1j_3} p_{2i_3} + \delta_{i_3j_3} p_1\cdot p_2)\,,\\
\label{e117}
  \centeredgraphics{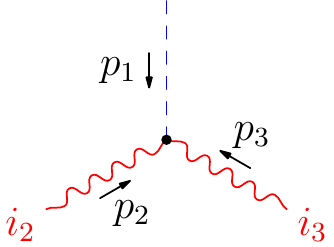} ={}& \frac{2i c_d}{\Lambda}(-p_{2i_2} p_{3i_3} + p_{2i_3} p_{3i_2} - 2 \delta_{i_2
i_3} p_{2i} p_{3i}) \,,\\
\label{e118}
  \centeredgraphics{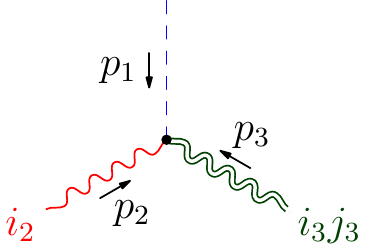} ={}&
\begin{aligned}
\tfrac{i c_d}{\Lambda}(&
- \delta_{i_3 i_2} p_{1 0} p_{2 i_3}
+ \delta_{i_3 i_2} p_{1 j_3} p_{2 0}
+ \delta_{i_3 j_3} p_{1 0} p_{2 i_2}\\
&- \delta_{i_3 j_3} p_{1 i_2} p_{2 0}
- \delta_{i_2 j_3} p_{1 0} p_{2 i_3}
+ \delta_{i_2 j_3} p_{1 i_3} p_{2 0}
)\,,
\end{aligned}
\\
\label{e119}
    \centeredgraphics{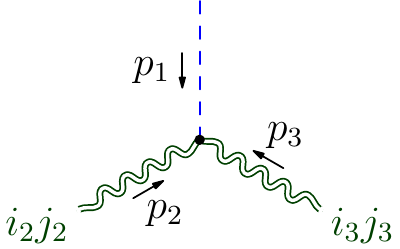} ={}& \frac{i c_d}{2\Lambda} p_{2 0} p_{3 0} (
  \delta_{i_2 i_3} \delta_{j_2 j_3}
- \delta_{i_2 j_2} \delta_{i_3 j_3}
+ \delta_{i_2 j_3} \delta_{i_3 j_2}
)\,,\\
\label{e120}
\centeredgraphics{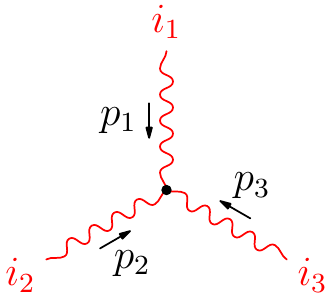} ={}&
\begin{aligned}
\tfrac{2i}{\Lambda} (
&- \delta_{i_1 i_2} p_{1 0} p_{2 i_3}
- \delta_{i_1 i_2} p_{1 0} p_{3 i_3}
+ \delta_{i_1 i_2} p_{1 i_3} p_{2 0}
+ \delta_{i_1 i_2} p_{1 i_3} p_{3 0}\\
&+ \delta_{i_1 i_2} p_{2 0} p_{3 i_3}
+ \delta_{i_1 i_2} p_{2 i_3} p_{3 0}
+ \delta_{i_1 i_3} p_{1 0} p_{2 i_2}
+ \delta_{i_1 i_3} p_{1 0} p_{3 i_2}\\
&- \delta_{i_1 i_3} p_{1 i_2} p_{2 0}
+ \delta_{i_1 i_3} p_{1 i_2} p_{3 0}
- \delta_{i_1 i_3} p_{2 0} p_{3 i_2}
+ \delta_{i_1 i_3} p_{2 i_2} p_{3 0}\\
&+ \delta_{i_2 i_3} p_{1 0} p_{2 i_1}
+ \delta_{i_2 i_3} p_{1 0} p_{3 i_1}
+ \delta_{i_2 i_3} p_{1 i_1} p_{2 0}
- \delta_{i_2 i_3} p_{1 i_1} p_{3 0}\\
&+ \delta_{i_2 i_3} p_{2 0} p_{3 i_1}
- \delta_{i_2 i_3} p_{2 i_1} p_{3 0}
)\,,
\end{aligned}
\\
\label{e121}
  \centeredgraphics{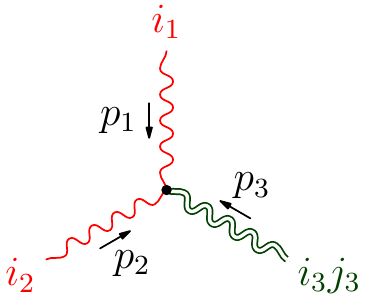} ={}&
\begin{aligned}
\tfrac{i}{\Lambda} (
&p_1\cdot p_2 (
- \delta_{i_1 i_2} \delta_{i_3 j_3}
+ \delta_{i_1 i_3} \delta_{i_2 j_3}
+ \delta_{i_2 i_3} \delta_{i_1 j_3}
)\\
&+ \delta_{i_1 i_2} (p_{1 i_3} p_{2 j_3} + p_{1 j_3} p_{2 i_3})
+ \delta_{i_1 i_3} (p_{1 i_2} p_{2 j_3} - p_{1 j_3} p_{2 i_2})\\
&- \delta_{i_2 i_3} (p_{1 i_1} p_{2 j_3} - p_{1 j_3} p_{2 i_1})
+ \delta_{i_1 j_3} (p_{1 i_2} p_{2 i_3} - p_{1 i_3} p_{2 i_2})\\
&- \delta_{i_2 j_3} (p_{1 i_1} p_{2 i_3} - p_{1 i_3} p_{2 i_1})
+ \delta_{i_3 j_3} (p_{1 i_1} p_{2 i_2} - p_{1 i_2} p_{2 i_1})
)\,,
\end{aligned}
\\
\label{e122}
\centeredgraphics{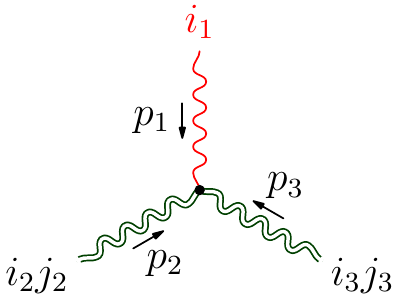} ={}&
\frac{i}{8\Lambda}(V^{A\sigma\sigma}_{i_1,i_2j_2,i_3j_3}
+ V^{A\sigma\sigma}_{i_1,i_2j_2,j_3i_3}
+ V^{A\sigma\sigma}_{i_1,j_2i_2,i_3j_3}
+ V^{A\sigma\sigma}_{i_1,j_2i_2,j_3i_3})\\
V^{A\sigma\sigma}_{i_1,i_2j_2,i_3j_3}={}&\delta_{i_{2} i_{3}} \delta_{j_{2} j_{3}} [4 p_{20} p_{2i_{1}}+2 (p_{1i_{1}} p_{20}+p_{10} p_{2i_{1}})]+\delta_{i_{3} j_{3}} [-2 p_{20} (\delta_{i_{1} i_{2}} p_{1j_{2}}\notag\\
&      +\delta_{i_{2} j_{2}} p_{2i_{1}})-\delta_{i_{2} j_{2}} (p_{1i_{1}} p_{20}+p_{10} p_{2i_{1}})]+\delta_{i_{1} i_{2}} [2 \delta_{i_{3} j_{3}} p_{10} p_{2j_{2}}\notag\\
&   +\delta_{i_{3} j_{2}} (4 p_{1j_{3}} p_{20}-4 p_{10} p_{2j_{3}})]+\delta_{i_{1} i_{3}} [\delta_{i_{2} j_{3}} (-4 p_{1j_{2}} p_{20}\notag\\
&      +4 p_{10} p_{2j_{2}})+\delta_{i_{2} j_{2}} (2 p_{1j_{3}} p_{20}-2 p_{10} p_{2j_{3}})] \\
\label{e124}
  \centeredgraphics{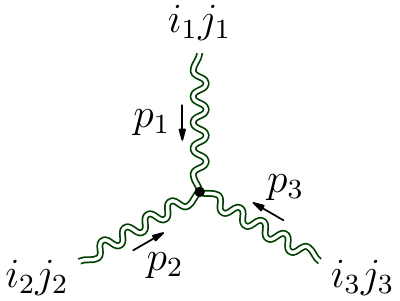} ={}& \frac{i}{32\Lambda}(\tilde{V}^{\sigma\sigma\sigma}_{i_1j_1,i_2j_2,i_3j_3} +
       \tilde{V}^{\sigma\sigma\sigma}_{j_1i_1,i_2j_2,i_3j_3})\\
  \tilde{V}^{\sigma\sigma\sigma}_{i_1j_1,i_2j_2,i_3j_3} ={}&
       V^{\sigma\sigma\sigma}_{i_1j_1,i_2j_2,i_3j_3} +
       V^{\sigma\sigma\sigma}_{i_1j_1,j_2i_2,i_3j_3} +
       V^{\sigma\sigma\sigma}_{i_1j_1,i_2j_2,j_3i_3} +
  V^{\sigma\sigma\sigma}_{i_1j_1,j_2i_2,j_3i_3}
\displaybreak[0]\\
V^{\sigma\sigma\sigma}_{i_1j_1,i_2j_2,i_3j_3}
={}&({p}_1^2+{p}_1\cdot{p}_2+{p}_2^2)\*\Bigl(-\delta_{i_2j_2}\*\bigl(2\*\delta_{i_1i_3}\*
\delta_{j_1j_3}-\delta_{i_1j_1}\*\delta_{i_3j_3}\bigr)
\notag\\
&\quad+2\*\bigl[\delta_{i_1i_2}\*\bigl(4\*\delta_{j_1i_3}\*\delta_{j_2j_3}-\delta_{j_1j_2}\*\delta_{i_3j_3}\bigr)
-\delta_{i_1j_1}\*\delta_{i_2i_3}\*\delta_{j_2j_3}\bigr] \Bigr)
\notag\\
&+2\*\Bigl\{4\*\bigl(p_{1 j_3}\*p_{2 j_1}-p_{1 j_1}\*p_{2 j_3}\bigr)\*\delta_{i_1i_2}\*\delta_{j_2i_3}
\notag\\
&\quad+2\*\bigl[\bigl(p_{1 i_1}+p_{2 i_1}\bigr)\*p_{2 j_1}\*\delta_{i_2i_3}\*\delta_{j_2j_3}-p_{1 i_3}\*p_{2 j_3}\*
\delta_{i_1i_2}\*\delta_{j_1j_2}\bigr]
\notag\\
&\quad+\delta_{i_2j_2}\*\bigl[p_{1 i_3}\*p_{2 j_3}\*\delta_{i_1j_1}+2\*\bigl(p_{1 j_3}\*p_{2 j_1}-p_{1 j_1}\*
p_{2 j_3}\bigr)\*\delta_{i_1i_3}\notag\\
&\quad-\bigl(p_{1 i_1}+p_{2 i_1}\bigr)\*p_{2 j_1}\*\delta_{i_3j_3}\bigr]
\notag\\
&\quad+p_{2 j_2}\*\Bigl(4\*p_{1 j_1}\*\delta_{i_1i_3}\*\delta_{i_2j_3}+p_{1 i_2}\*\bigl(2\*\delta_{i_1i_3}\*
\delta_{j_1j_3}-\delta_{i_1j_1}\*\delta_{i_3j_3}\bigr)
\notag\\
&\qquad+2\*\bigl[\delta_{i_1i_2}\*\bigl(p_{1 j_1}\*\delta_{i_3j_3}-2\*p_{1 j_3}\*\delta_{j_1i_3}\bigr)-p_{1 j_3}\*
\delta_{i_1j_1}\*\delta_{i_2i_3}\bigr] \Bigr)
\notag\\
&\quad+p_{1 j_2}\*\Bigl(p_{1 i_2}\*\bigl(2\*\delta_{i_1i_3}\*\delta_{j_1j_3}-\delta_{i_1j_1}\*\delta_{i_3j_3}\bigr)
-4\*p_{2 j_1}\*\delta_{i_1i_3}\*\delta_{i_2j_3}
\notag\\
&\qquad+2\*\bigl[p_{2 j_3}\*\delta_{i_1j_1}\*\delta_{i_2i_3}+\delta_{i_1i_2}\*\bigl(2\*p_{2 j_3}\*\delta_{j_1i_3}
-p_{2 j_1}\*\delta_{i_3j_3}\bigr)\bigr]
\Bigr)\Bigr\},
\end{align}
with
\begin{eqnarray}
c_d = \frac{2(d-1)}{d-2},
\end{eqnarray}
$\Lambda^{-1} = \sqrt{32 \pi G_N}$ and
$d = 3 -2 \ep$. The scalar propagators $D(p)$ are left generic and will be specified either as
causal, retarded, or advanced propagators, cf.~Section~\ref{sec:B}. Furthermore, in the specific in--in
calculations below one has to replace in (\ref{eq:FR1}, \ref{eq:FR2}) for the electric quadrupole moment
the vertices as given in (\ref{eq:V1}--\ref{eq:V3}).
Furthermore, we list the contributing field combinations in Table~\ref{tab:1}.
\begin{table}[H]\centering
\begin{tabular}{|r|l|}
\hline
  Eq.          & contributions \\
  \hline
  (\ref{e114}), (\ref{e120}), (\ref{e124}) & $++-,---$ \\
  (\ref{e115}), (\ref{e116}), (\ref{e121}) & $++-,+-+,---$ \\
  (\ref{e117}), (\ref{e119}), (\ref{e122}) & $++-,-++,---$ \\
  (\ref{e118}) & $++-,+-+,-++,---$ \\
\hline
\end{tabular}
\caption{
\label{tab:1} \sf The contributing field combinations in the in--in formalism.}
\end{table}

\noindent
Here the bulk vertices have to be rescaled by $1/\sqrt{2}$.
The free--field combinations are defined in Eq.~(\ref{eq:PHI1}).

The $D$--dimensional integral over an Euclidean momentum $k_E$ is given by, cf.~e.g.~\cite{YNDURAIN},
\begin{eqnarray}
\int \frac{d^D \vec{k}_E}{(2\pi)^D} \frac{(\vec{k}^2_E)^r}{(\vec{k}^2_E + R^2)^m} = \frac{(-1)^{r-m}}{(4\pi)^{D/2}}
\frac{\Gamma(r+D/2)}{\Gamma(D/2)} \Gamma(m-r-D/2) (R^2)^{D/2+r-m}.
\end{eqnarray}

\section{Invariant functions}
\label{sec:B}

\begin{figure}[H]
  \centering
  \hskip-0.8cm
  \includegraphics[width=.32\linewidth]{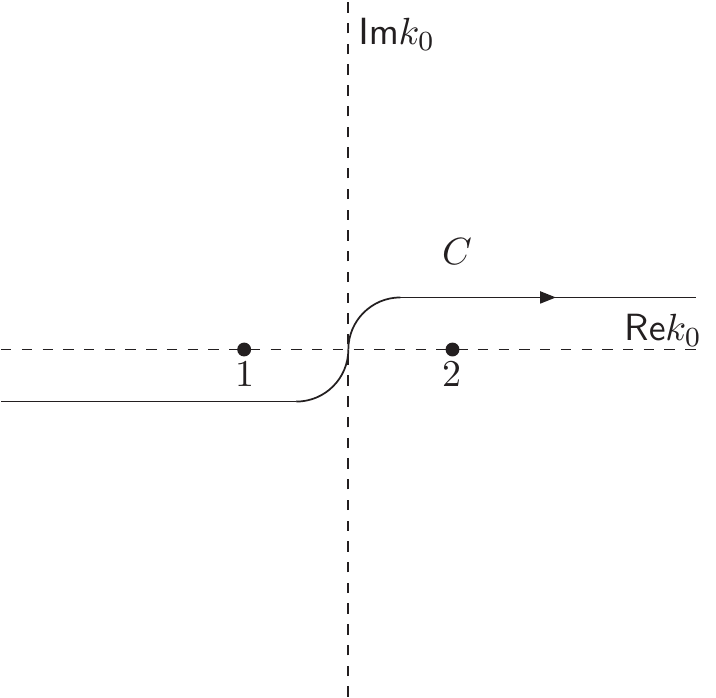}
  \includegraphics[width=.32\linewidth]{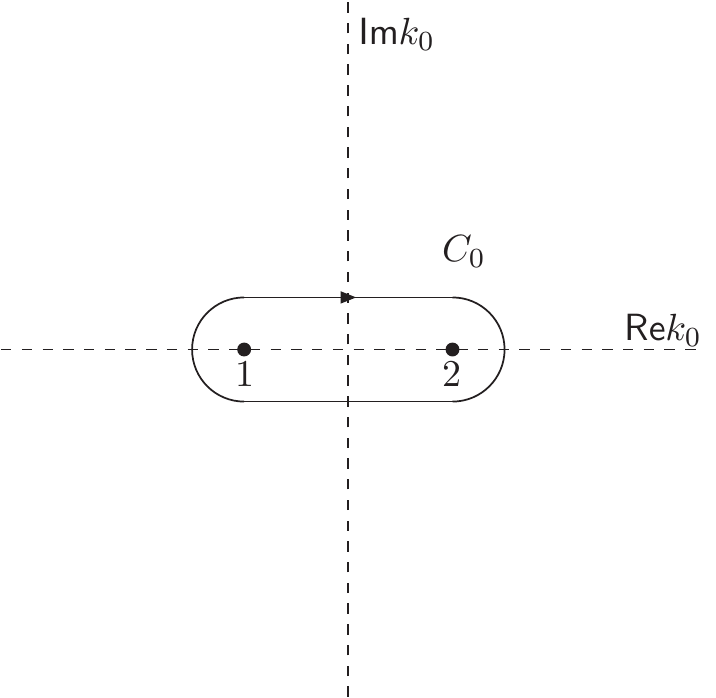}
  \includegraphics[width=.32\linewidth]{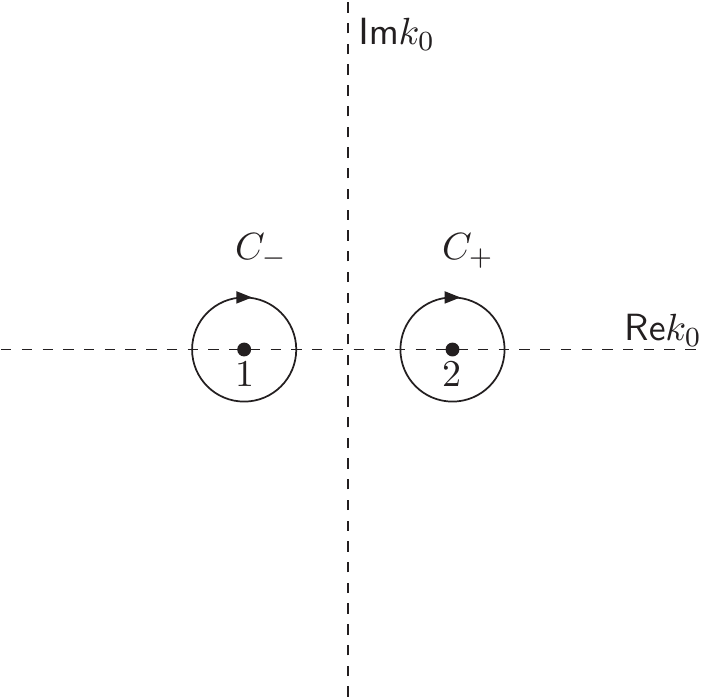} \\
  \includegraphics[width=.32\linewidth]{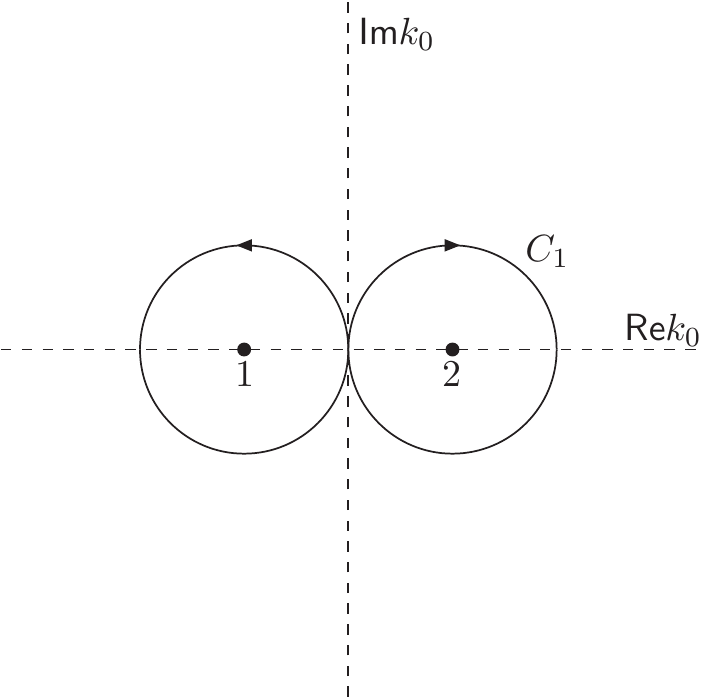}
  \includegraphics[width=.32\linewidth]{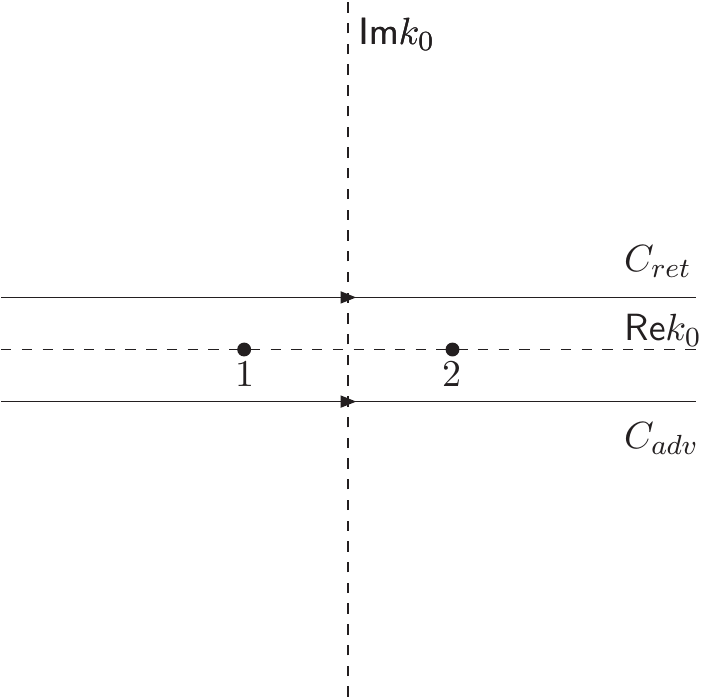}
  \caption[]{\sf The integration paths in the complex $k_0$ plane for the different types of
  propagator functions. The paths correspond to the integrals given in the text and correspond
  to the causal Green's function $(C)$, the Jordan--Pauli function $(C_0)$, the two Wightman
functions $(C_\pm)$, the anticommutator
$(C_1)$, and the retarded and advanced Green's
functions $(C_{\rm ret (adv)})$, where  $C_1$
  denotes a single path.}
  \label{fig:4}
\end{figure}

\vspace*{1mm}
\noindent
In the following we summarize different functions, related to the scalar field operators
$\Phi(x_1)$
and $\Phi(x_2)$, and $x = x_1 - x_2$, leading to the different kind of propagators, \cite{AB},
which are distribution--valued in part \cite{DISTR,GELF}.
The corresponding contours for the defining integrals are shown in Figure~\ref{fig:4}.

We start with the commutator
\begin{eqnarray}
i \Delta(x) = [\Phi(x_1),\Phi(x_2)]_- = - i\int \frac{d^{D-1}k}{(2\pi)^{D-1}} \frac{1}{\omega_k}
\frac{e^{-ikx}}{k^2 - m^2},
\end{eqnarray}
with $\omega_k = \sqrt{\vec{k}^2 + m^2}$, denoting the Jordan--Pauli function \cite{JP}. 
It has the contour integral representation
\begin{eqnarray}
\Delta(x)     &=& \int_{C_0} \frac{d^D k}{(2\pi)^D} \frac{e^{ik.x}}{k^2 - m^2}
= \Delta_+(x) + \Delta_-(x) = \Delta_{\rm ret}(x) -  \Delta_{\rm adv}(x),
\end{eqnarray}
showing also the relation to other quantities
\begin{eqnarray}
\\
\Delta_1(x)     &=& \int_{C_1} \frac{d^D k}{(2\pi)^D} \frac{e^{ik.x}}{k^2 - m^2} = \Delta_+(x)
- \Delta_-(x)
\\
\Delta_\pm(x) &=& \int_{C_{\pm}} \frac{d^D k}{(2\pi)^D} \frac{e^{ik.x}}{k^2 - m^2}
= \frac{1}{2}\left[\Delta(x) \pm  \Delta_1(x) \right],
\\
\Delta_c(x)   &=& -i \int_{C} \frac{d^D k}{(2\pi)^D} \frac{e^{ik.x}}{k^2 - m^2}
=
-i \left[\Theta(t)
\Delta_+(x) + \Theta(-t) \Delta_-(x)\right],
\nonumber\\
\\
\Delta_D(x)   &=& -i\left[\Theta(t) \Delta_-(x) + \Theta(-t) \Delta_+(x)\right],
\nonumber\\
\\
\label{eq:ret}
\Delta_{\rm ret}(x) &=& \int_{C_{\rm ret}} \frac{d^D k}{(2\pi)^D} \frac{e^{ik.x}}{k^2 - m^2}
= \Theta(t) \Delta(x),
\\
\label{eq:adv}
\Delta_{\rm adv}(x) &=& \int_{C_{\rm adv}} \frac{d^D k}{(2\pi)^D} \frac{e^{ik.x}}{k^2 - m^2}
=  - \Theta(-t) \Delta(x),
\end{eqnarray}
with $\Theta(t)$ the Heaviside function.

The causal Green's function $\Delta_c(x)$ \cite{CAUSAL} is also called Feynman 
function $\Delta_F(x)$ and
the Dyson function $\Delta_D(x)$ is also called anti--causal Green's function.

In momentum space the causal, retarded and advanced propagators read
\begin{eqnarray}
\Delta_c(k)         ~&=&~ \frac{1}{k^2 - m^2 + i0},
\\
\Delta_{\rm ret}(k) ~&=&~ \frac{1}{k^2 - m^2 + (2 k_0) i0},
\\
\Delta_{\rm adv}(k) ~&=&~ \frac{1}{k^2 - m^2 - (2 k_0) i0},
\end{eqnarray}
where the distribution relations \cite{SOCH,DISTR,GELF}
\begin{eqnarray}
\frac{1}{x \pm i0} = {\cal P} \frac{1}{x} \mp i \pi \delta(x)
\end{eqnarray}
hold, with ${\cal P}$ Cauchy's principal value.

\section{The in-in formalism}
\label{sec:C}

\vspace*{1mm}
\noindent
The in--in formalism for binary systems in classical gravity refers to a well--defined initial state
at $t = -\infty$, which is also defined to be the final state, and one integrates over the time paths  $t_F$ and
$t_B$
\begin{eqnarray}
t_F~=~(-\infty, +\infty),~~~~~~~t_B~=~(+\infty, -\infty),
\end{eqnarray}
which are linked together, with $t_F = t_1$ and $t_B = t_2$ labeling the time path. One should note that different
authors use a quite different notation. For definiteness we refer to the one by Keldysh \cite{Keldysh:1964ud} also
given in \cite{Chou:1984es}.\footnote{For an application to gravity see also \cite{Foffa:2011np}.}
In the following 2--dimensional representations the Pauli matrices
\begin{eqnarray}
\sigma_1 ~=~ \left(
\begin{array}{rr}
0 ~~&~~ 1 \\ 1 ~~&~~ 0
\end{array}
\right),
~~~~~
\sigma_2 ~=~ \left(
\begin{array}{rr}
0 ~&~ -i \\ i ~&~ 0
\end{array}
\right),
~~~~~
\sigma_3 ~=~ \left(
\begin{array}{rr}
1 ~&~ 0 \\ 0 ~&~ -1
\end{array}
\right)
\end{eqnarray}
are of use. One has
\begin{eqnarray}
\label{eq:KELDpm}
\left(\begin{array}{c} \Psi_1 \\ \Psi_2 \end{array} \right) = \frac{1}{\sqrt{2}}(1 + i \sigma_2)
\left(\begin{array}{c} \Psi_- \\ \Psi_+ \end{array} \right)
= \frac{1}{\sqrt{2}} \left(\begin{array}{c} \Psi_+ + \Psi_- \\ \Psi_+  - \Psi_-\end{array} \right)
\end{eqnarray}
for the fields, coordinates, and Schwinger--parameters, as well as later also the multipole moments, leaving
the action invariant. The free fields are doubled to allow for a single time, i.e. we set
\begin{eqnarray}
\label{eq:PHI1}
\Phi(\vec{x},t_i) = \Phi_i(x,t)
\end{eqnarray}
and use (\ref{eq:KELDpm}) to form the $\pm$ combinations.

We first consider the path integral for the general motion of the gravitating two--body system, not specifying yet
either to the near or far zones. In the in--in formalism it is given by \cite{Galley:2009px}
\begin{eqnarray}
\label{eq:PATH1}
\exp\left[i W(\{\vec{j}_{k,1}\},\{\vec{j}_{k,2}\},J_1,J_2)\right] ~&=&~
\int \prod_{k=1}^2 {\cal D} \vec{x}_{k,1}  {\cal D} \vec{x}_{k,1}
\exp\Biggl\{i \sum_{k=1}^2 \left(
S_{pp}(\vec{x}_{k,1}) -
S_{pp}(\vec{x}_{k,2}) \right)
\nonumber\\ &&
+
i \sum_{k=1}^2 \int dt \left(
\vec{j}_{k,1}.\vec{x}_{k,1} -
\vec{j}_{k,2}.\vec{x}_{k,2} \right)
\nonumber\\ &&
+ i \int d^Dx {\cal L}_{\rm int}\left[
\{\vec{x}_{k,1}\},
\{\vec{x}_{k,2}\}, -i \delta_{J_1}, -i \delta_{J_2}\right]\Biggr\} Z_0\left[J_1,J_2\right].
\nonumber\\
\end{eqnarray}
Here $\vec{x}_k$ denote the positions of the gravitating masses, $\vec{j}_k$ the associated Schwinger
parameters, $S_{pp}$ the point particle action, Eq.~(10) \cite{Blumlein:2019zku}, ${\cal L}_{\rm int}$
the interaction Lagrangian density, $J_i$ are the Schwinger parameters to $\Phi_i$,
with
\begin{eqnarray}
\delta_{J_k} \equiv \frac{\delta}{\delta J_k},
\end{eqnarray}
and $Z_0$ denotes the free gravitational field part of the path integral.

As has been shown in \cite{Galley:2005tj}, Eq.~(2.22), the functional derivative of the coarse grained
effective
action $S_{\rm CGEA}$
\begin{eqnarray}
\left. \frac{\delta S_{\rm CGEA}}{\delta x_{k-}}\right|_{x_1 = x_2 = \bar{x}} = 0,
\end{eqnarray}
which implies that in the {far-zone} terms being dealt with below the contributing functions are at most
$\propto x_{k-}$.
This is implied by the ${\mathbb Z}_-$ operator applied to the path integrals for the {far-zone} term below.

We are now specifying to the calculation of contributions in the far zone. For this purpose the
multipole expansion has to be carried out and the multipole moments appear in the effective interaction
Lagrangian,
cf.~(\ref{eq:Smp}), as new entities.

Let us consider the vertex functions, with $Q_{1(2)}$ the electric quadrupole moments with
indices suppressed, which contribute to the {far-zone} terms shown in Figure~\ref{fig:1} and the graviton
self--interaction vertex. The symbols $h_{1(2)}$
denote the respective field couplings given by $O_k(\Phi_{1(2)})$ as different linear differential
operators
acting on the decomposition of the gravitational field into 10 fields $\Phi$, as described in \cite{Kol:2007bc}.
One obtains
\begin{eqnarray}
\label{eq:V1}
V_{Qh}~&=&~ Q_1 h_1 - Q_2 h_2 ~=~ Q_- h_+ + Q_+ h_-,
\\
\label{eq:V2}
V_{Qh^2}~&=&~Q_1 h_1^2 - Q_2 h_2^2 ~=~ \frac{1}{\sqrt{2}} \left[2 h_+ h_- Q_+ + (h_+^2 + h_-^2) Q_-\right],
\\
\label{eq:V3}
V_{h^3}~&=&~\left[h_1^3 - h_2^3 \right] ~=~ \frac{1}{\sqrt{2}}\left[ 3 h_+^2 h_- +  h_-^3\right].
\end{eqnarray}
This notation is symbolic, but sufficient to derive the corresponding Feynman diagrams by functional
differentiation for the Schwinger parameters. It is understood that the respective Feynman rules given in
Appendix~\ref{sec:A} are used in the final result.
From Eq.~(\ref{eq:V1}) one obtains for the electric quadrupole moments at leading order (0PN)
\begin{eqnarray}
\label{eq:QUADx}
Q_{ij,1(2)} = \sum_{a=1}^2 m_a \left(x_{a,1(2),i} {x}_{a,1(2),j} - \delta_{ij}
\frac{{x}_{a,1(2),k} {x}_{a,1(2),k}}{d}
\right),
\end{eqnarray}
the $\pm$ projections
\begin{eqnarray}
\label{eq:QM}
Q_{ij,-} ~&=&~
\sum_{a=1}^2 \frac{m_a}{\sqrt{2}} \left(
x_{a,-,i} x_{a,+,j} + x_{a,+,i} x_{a,-,j} - 2 \frac{\delta_{ij}}{d} {x}_{a,-,k} {x}_{a,+,k}
\right)
\\
Q_{ij,+} ~&=&~
\sum_{a=1}^2 \frac{m_a}{\sqrt{2}} \left[
x_{a,+,i} x_{a,+,j} +
x_{a,-,j} x_{a,-,j}
- \frac{\delta_{ij}}{d} ({x}_{a,+,k} {x}_{a,+,k} + {x}_{a,-,k} {x}_{a,-,k})
\right],
\end{eqnarray}
with $x_{l} \equiv x_l(t)$.

The contribution of the interaction terms to the path integral read
\begin{eqnarray}
\exp\left[-i S_{\rm int}\right] =
\exp\left[-i \int d^Dz \left(V_{Qh}(z) + V_{Qh^2}(z) + V_{h^3}(z)\right)\right],
\end{eqnarray}
where the fields $\Phi_\pm$ are replaced  by the folloing functional derivative
\begin{eqnarray}
\Phi_{\pm} \rightarrow i \frac{\delta}{\delta J_{\pm}}.
\end{eqnarray}
Equivalently on may use
\begin{eqnarray}
h_{\pm} \rightarrow i \frac{\delta}{\delta J_{\pm}}
\frac{\partial O_k(\Phi_{\pm})}{\partial \Phi_\pm}
\equiv i \delta_{J_{\pm}} \frac{\partial O_k(\Phi_{\pm})}{\partial \Phi_\pm}.
\end{eqnarray}
For the free--field part the path integral can be integrated to \cite{FRADKIN,Edwards:1954cac}
\begin{eqnarray}
\label{eq:free1}
\exp[ -\frac{i}{2} {\bf J}^T \hat{G} {\bf J}]~~~~\text{with}~~~~{\bf J} =
\left( \begin{array}{c} J_1 \\ J_2 \end{array}\right)
\end{eqnarray}
and
\begin{eqnarray}
\label{eq:Ghat}
\hat{G} = \left(\begin{array}{cc} \Delta_{11} ~&~ \Delta_{12} \\ \Delta_{21} ~&~ \Delta_{22}
\end{array}\right),
\end{eqnarray}
where
\begin{eqnarray}
\Delta_{11} ~&=&~ -i\Theta(t) \Delta_- - i\Theta(-t) \Delta_+,
\\
\Delta_{12} ~&=&~ -i \Delta_+
\\
\Delta_{21} ~&=&~ -i \Delta_-
\\
\Delta_{22} ~&=&~ -i\Theta(-t) \Delta_- - i\Theta(t) \Delta_+.
\end{eqnarray}
The following relation holds
\begin{eqnarray}
\Delta_{11} + \Delta_{22} ~=~ \Delta_{12} + \Delta_{21}.
\end{eqnarray}
Eq.~(\ref{eq:Ghat}) can be rewritten by
\begin{eqnarray}
\label{eq:Ghat1}
\hat{G} = \frac{1}{2}\left(\begin{array}{cc}
\Delta_{\rm ret} + \Delta_{\rm adv} + \Delta_{\rm C} ~~&~~
- \Delta_{\rm ret} + \Delta_{\rm adv} + \Delta_{\rm C} \\
\\
\Delta_{\rm ret} - \Delta_{\rm adv} + \Delta_{\rm C}
~~&~~
-\Delta_{\rm ret} - \Delta_{\rm adv} + \Delta_{\rm C}
\end{array}\right),
\end{eqnarray}
where
\begin{eqnarray}
\Delta_{\rm adv} ~&=&~ \frac{1}{2}\left[\Delta_{11} - \Delta_{21} + \Delta_{12} - \Delta_{22}\right],
\\
\Delta_{\rm ret} ~&=&~ \frac{1}{2}\left[\Delta_{11} - \Delta_{12} + \Delta_{21} - \Delta_{22}\right],
\\
\Delta_{\rm C}  ~&=&~ \frac{1}{2}\left[\Delta_{11} + \Delta_{22} + \Delta_{12} + \Delta_{21}\right].
\end{eqnarray}

After functional differentiation retarded and advanced propagators appear in different directions, which one
may synchronize using
\begin{eqnarray}
\Delta_{\rm adv, ret}(x_j - x_i)~=~ \Delta_{\rm ret, adv}(x_i - x_j).
\end{eqnarray}
One further considers the transformed matrix $\tilde{G}$~\cite{Keldysh:1964ud}
and the transformed vectors
\begin{eqnarray}
\label{eq:Gtilde}
\tilde{G} ~&=&~ \frac{1}{\sqrt{2}} (1 - i \sigma_2) \hat{G} \frac{1}{\sqrt{2}}(1 + i \sigma_2)
~=~
\left(\begin{array}{cc} 0 ~&~ \Delta_{\rm adv} \\ \Delta_{\rm ret} ~&~ \Delta_{\rm C}
\end{array}\right),
\end{eqnarray}
appearing in the combination
\begin{eqnarray}
\tilde{\bf J}^T \tilde G \tilde{\bf J} = {\bf J}^T \hat{G} {\bf J},
\end{eqnarray}
with
\begin{eqnarray}
\tilde{\bf J} = \frac{1}{\sqrt{2}} (1 - i \sigma_2) {\bf J} = \left(\begin{array}{c} J_- \\ J_+ \end{array}
\right)~~~\text{and}~~~J_\pm = \frac{1}{\sqrt{2}}
(J_1 \pm J_2).
\end{eqnarray}
This leads to the free propagator contribution (\ref{eq:free1})
\begin{eqnarray}
\exp\left[-i S_{\rm free}\right] &=&
\exp\Biggl[-\frac{i}{2} \int dx \int dy \left(J_-(x) \Delta_{\rm adv}(x-y) J_+(y)
+ J_+(x) \Delta_{\rm ret}(x-y) J_-(y) \right.
\nonumber\\ && \left.
+ J_+(x) \Delta_{\rm C}(x-y) J_+(y)
\right) \Biggr].
\end{eqnarray}

In deriving the Feynman diagrams of Figure~\ref{fig:1} we consider the connected Green's function from the
beginning \cite{STERMAN}, since the disconnected diagrams are canceled by the denominator function and
apply the ${\mathbb Z}_-$ operator,
\begin{eqnarray}
{\mathbb Z}_-\left[\exp\left[-i S_{\rm int}(\delta J_{\pm})\right]
\exp\left[-i S_{\rm free}(J_\pm)\right]\right]_{\rm conn.}
\end{eqnarray}

Within the in--in formalism one ends up with representations in which the multipole moments, $M_I$,
appear in terms of their projections $M_{I,\pm}$. At 5PN the {far-zone} contributions calculated in
Appendix~\ref{sec:C} depend on the electric quadrupole moment only. The ${\mathbb Z}_-$ operator projects
onto contributions containing one multipole moment $Q^{ij}_-$ only. For the translation from $Q^{ij}_\pm$
to $Q^{ij}$ see Ref.~\cite{Galley:2005tj}.
\section{Calculation of {far-zone} diagrams}
\label{sec:D}

\vspace*{1mm}
\noindent
All contributing Feynman diagrams are of two--loop order with maximally three propagators.
We employ integration-by-parts \cite{IBP,CRUSHER}.
With the definition of the integrals
\begin{equation}
I_{i_1,i_2,i_3} = \int d^d \vec{k} \int d^d \vec{q} \frac{1}{(\vec{k}^2
- k_0^2 )^{i_1}(\vec{q}^2 - q_0^2 )^{i_2}( (\vec{k}+\vec{q})^2 -
(k_0+q_0)^2 )^{i_3}}
\end{equation}
we obtain
\begin{equation}
I_{1,1,1}
=   \left ( -\frac{1}{4 \varepsilon}  + \frac{1}{2} \right ) \left(
\frac{I_{0,1,1}}{q_0 \left(k_0+q_0\right)}+\frac{I_{1,0,1}}{k_0 \
\left(k_0+q_0\right)}-\frac{I_{1,1,0}}{k_0 q_0}\right ) \,.
\end{equation}
There are also other possibilities to calculate the three--propagator integrals, like
hypergeometric methods \cite{HYP,Blumlein:2018cms} and/or the use of one Mellin--Barnes integral \cite{MB}.
In all these representations one has to maintain the distribution character of these integrals
in all intermediary steps, which can be technically demanding. The IBP method, on the other hand,
maintains the propagator structure in all these respects and is therefore the method of choice
in the following.

In the following we present the explicit calculation of the diagrams shown in Figure~1.
The 5PN diagram on the r.h.s. is obtained as the closed Green's function
\begin{eqnarray}
I_1 &=& {\mathbb Z}_-\Biggl[\tfrac{1}{3!} \prod_{l=1}^3 \left[-i \int d^Dw_l(V_{Qh}(w_l) +
V_{Qh^2}(w_l)]\right]
 \tfrac{1}{2!} \prod_{k=1}^2 \Biggl[-\frac{i}{2} \left(\int dy_k^D dz_k^D
  J_-(y_k) \Delta_{\rm adv}(y_k - z_k)  \right.
  \nonumber\\ && \hspace*{.7cm} \hspace*{-5mm} \left.
\times J_+(z_k)
+ J_+(y_k) \Delta_{\rm ret}(y_k - z_k) J_-(z_k)
+ J_+(y_k) \Delta_{\rm C}(y_k - z_k) J_+(z_k)\right)\Biggr]\Biggr],
\end{eqnarray}
i.e. integrating over the coordinates $x_1$ to $x_3$ and applying the  ${\mathbb Z}_-$ operator, leading to
\begin{eqnarray}
I_1 &=& \Biggl[
\frac{(-i)^3i^4}{3!} \int dx_1^D
\int dx_2^D
\int dx_3^D
\Bigl[
\delta_{J_-(x_1)} Q_+(x_1) V_+(x_1)
\delta_{J_+(x_2)}^2 Q_-(x_2) V_-(x_2)
\delta_{J_-(x_3)} Q_+(x_3) V_+(x_3)
\nonumber\\ && \hspace*{4.4cm}
+\delta_{J_+(x_1)}^2 Q_-(x_1) V_-(x_1)
\delta_{J_-(x_2)} Q_+(x_2) V_+(x_2)
\delta_{J_-(x_3)} Q_+(x_3) V_+(x_3)
\nonumber\\ && \hspace*{4.4cm}
+\delta_{J_-(x_1)} Q_+(x_1) V_+(x_1)
\delta_{J_-(x_2)} Q_+(x_2) V_+(x_2)
\delta_{J_+(x_3)}^2 Q_-(x_3) V_-(x_3)
\nonumber\\ && \hspace*{2.1cm}
+ 2 \big[
\delta_{J_-(x_1)} Q_+(x_1) V_+(x_1)
\delta_{J_-(x_2)} \delta_{J_+(x_2)} Q_+(x_2) V_+(x_2)
\delta_{J_+(x_3)} Q_-(x_3) V_-(x_3)
\nonumber\\ && \hspace*{2.5cm}
+
\delta_{J_-(x_1)} \delta_{J_+(x_1)} Q_+(x_1) V_+(x_1)
\delta_{J_-(x_2)} Q_+(x_2) V_+(x_2)
\delta_{J_+(x_3)} Q_-(x_3) V_-(x_3)
\nonumber\\ && \hspace*{2.5cm}
+
\delta_{J_+(x_1)} Q_-(x_1) V_-(x_1)
\delta_{J_-(x_2)} \delta_{J_+(x_2)} Q_+(x_2) V_+(x_2)
\delta_{J_-(x_3)} Q_+(x_3) V_+(x_3)
\nonumber\\ && \hspace*{2.5cm}
+
\delta_{J_-(x_1)} \delta_{J_+(x_1)} Q_+(x_1) V_+(x_1)
\delta_{J_+(x_2)} Q_-(x_2) V_-(x_2)
\delta_{J_-(x_3)} Q_+(x_3) V_+(x_3)
\nonumber\\ && \hspace*{2.5cm}
+ \delta_{J_-(x_1)} Q_+(x_1) V_+(x_1)
\delta_{J_+(x_2)} Q_-(x_2) V_-(x_2) \delta_{J_-(x_3)}
\delta_{J_+(x_3)} Q_+(x_3) V_+(x_3)
\nonumber\\ && \hspace*{2.5cm}
+
\delta_{J_+(x_1)} Q_-(x_1) V_-(x_1)
\delta_{J_-(x_2)} Q_+(x_2) V_+(x_2)
\delta_{J_-(x_3)} \delta_{J_+(x_3)} Q_+(x_3) V_+(x_3)
\big] \Bigr] \nonumber\\ &&
~~~~\times \tfrac{1}{2!} \prod_{k=4}^5 \left[-\frac{i}{2} \left(\int dx_k^D dy_k^D
  J_-(x_k) \Delta_{\rm adv}(x_k - y_k) J_+(y_k) \right. \right.
\nonumber\\ && \left. \left. ~~~
+ J_+(x_k) \Delta_{\rm ret}(x_k - y_k) J_-(y_k)
+ J_+(x_k) \Delta_{\rm C}(x_k - y_k) J_+(y_k)\right)\right]\Biggr]_{\rm conn.},
\\
&=& - 2 \int dx_1^D
\int dx_2^D
\int dx_3^D
\Delta_{\rm adv}(x_1 - x_2) Q_{+}(x_1) V_+(x_1)
\nonumber\\ &&
\times \left[
  \Delta_{\rm adv}(x_2 - x_3) Q_{+}(x_2) V_+(x_2) Q_{-}(x_3) V_-(x_3)
+ \frac{\Delta_{\rm ret}(x_2 - x_3)}{2} Q_{-}(x_2) V_-(x_2) Q_{+}(x_3) V_+(x_3)\right].
\nonumber\\
\end{eqnarray}
Here $V_{\pm}(x_i)$ denote the numerator functions at the respective vertices, while $G_{r(a)}(x_i - x_j)$
are the retarded and advanced propagators in configuration space. In the diagrams of Figure~\ref{fig:1} the propagator
$\Delta_{\rm C}(x-y)$ does not contribute.

The Fourier transform from momentum to configuration space and its inverse are defined by \cite{DISTR,GELF}
\begin{eqnarray}
{\bf F}[f](r) &=& \int \frac{d^Dk}{(2\pi)^D} \exp[ik.r] f(k)
\\
{\bf F}^{-1}[g](k) &=& \int d^Dr \exp[-ik.r] g(r).
\end{eqnarray}
One obtains the following contribution to the action in momentum space\footnote{The spatial
coordinates of the multipole moments are kept fixed.}
\begin{eqnarray}
\label{eq:S1a}
{\sf S}_1 &=& {\frac{8}{3} \sqrt{2} \pi^2 G_N^2}
\int_{-\infty}^{+\infty} \frac{dl_1^0}{2\pi}
\int_{-\infty}^{+\infty} \frac{dl_2^0}{2\pi}
\int \frac{d^d {\bf l}_1}{(2 \pi)^d}
\int \frac{d^d {\bf l}_2}{(2 \pi)^d}
   \frac{\tilde{Q}^{ij}_+(l_1^0)}{|{\bf l}_1|^2 - (l_1^0 + i \delta)^2}
\nonumber\\ &&
\times
\Biggl[
   \frac{P_{11}[l_1^0,l_2^0] \tilde{Q}^{jk}_+(-l_1^0-l_2^0) \tilde{Q}^{ki}_-(l_2^0)}{|{\bf l_2}|^2 - (l_2^0 + i
\delta)^2}
+  \frac{1}{2} \frac{P_{12}[l_1^0,l_2^0] \tilde{Q}^{jk}_-(-l_1^0-l_2^0)
\tilde{Q}^{ki}_+(l_2^0)}{|{\bf l_2}|^2 - (l_2^0 - i \delta)^2}
\Biggr],
\end{eqnarray}
with
\begin{eqnarray}
- \int \frac{d^{d} {\mathbf l}_1}{(2\pi)^{d}} \frac{1}{|{\mathbf l}_1|^2 -
(l_1^0 - i \delta)^2}
&=& \frac{i }{4 \pi} l_1^0,
\\
\label{eq:I1a}
- \int \frac{d^{d} {\mathbf l}_1}{(2\pi)^{d}}
\frac{1}{|{\mathbf l}_1|^2 -
(l_1^0 + i \delta)^2}
&=& - \frac{i }{4 \pi} l_1^0~
\end{eqnarray}
for $d = 3$ represented by an analytic continuation in case \cite{WW}. Eq.~(\ref{eq:S1a}) has
been obtained after tensor decomposition and the use of master integrals, such that the three momenta appear
only in the propagators.

We use
\begin{eqnarray}
\tilde{Q}_\pm(k_0) = \int_{-\infty}^{+\infty} dt \exp[-i k_0 t] Q_\pm(t)
\end{eqnarray}
leading to
\begin{eqnarray}
{\sf S}_1 &=&
\frac{\sqrt{2}}{6} G_N^2
\int_{-\infty}^{+\infty} \frac{dl_1^0}{2\pi}
\int_{-\infty}^{+\infty} \frac{dl_2^0}{2\pi}
l_1^0 l_2^0 \tilde{Q}_+(l_1^0)
\nonumber\\ &&
\times
\Biggl[
   P_{11}[l_1^0,l_2^0] \tilde{Q}_+(-l_1^0-l_2^0) \tilde{Q}_-(l_2^0)
-  \frac{1}{2}P_{12}[l_1^0,l_2^0] \tilde{Q}_-(-l_1^0-l_2^0) \tilde{Q}_+(l_2^0)
\Biggr],
\end{eqnarray}
with
{
\begin{eqnarray}
P_{11}(x,y) =  P_{12}(x,y) = x^3 y^3,
\end{eqnarray}}

\noindent
resulting in\footnote{This term would vanish for $P_{11}(x,y) =
P_{12}(x,y)/2.$}
{\begin{eqnarray}
{\sf S}_1 = \frac{\sqrt{2}}{12} G_N^2 \int_{-\infty}^{+\infty} dt~{\rm tr}\left\{
2 Q_+(t) Q_+^{(4)}(t)Q_-^{(4)}(t)
- Q_-(t) (Q_+^{(4)}(t))^2\right\}.
\end{eqnarray}}.

Let us also remark on the mathematical structure in the causal case for completeness, which does not
contribute in
the present case. Here the derivation of the final result requires a different technique. One has
\begin{eqnarray}
\label{eq:ME1}
{\sf \tilde{S}}_1^{\rm c} &=&  8 \pi^2 G_N^2
\int_{-\infty}^{+\infty} \frac{dl_1^0}{2\pi}
\int_{-\infty}^{+\infty} \frac{dl_2^0}{2\pi}
\int \frac{d^{D-1} {\mathbf l}_1}{(2\pi)^{D-1}}
\int \frac{d^{D-1} {\mathbf l}_2}{(2\pi)^{D-1}}
P_2[l_1^0,l_2^0]
\frac{\tilde{Q}(l_1^0) \tilde{Q}(l_2^0) \tilde{Q}(-l_1^0-l_2^0)}{
[(l_1^0)^2 - {\mathbf l}_1^2 + i \delta]
[(l_2^0)^2 - {\mathbf l}_2^2 + i \delta]}.
\nonumber\\
\end{eqnarray}
This integral leads to terms $ \left((l_k^0)^2 + i \delta\right)^{1/2} $ which are Schwartz ${\mathcal S}'$ distributions
\cite{DISTR}.  One has, cf.~\cite{GELF},
\begin{eqnarray}
\label{eq:zp}
(z + i \delta)^\lambda &=& z_+^\lambda + \exp(+i\pi\lambda) z_-^\lambda
\\
\label{eq:zm}
(z - i \delta)^\lambda &=& z_+^\lambda + \exp(-i\pi\lambda) z_-^\lambda,
\end{eqnarray}
with $z, \lambda \in \mathbb{R}$ and $\lambda \neq 0, -1, -2, -3, ...$ and
\begin{eqnarray}
z_+^\lambda &=& \theta(+z) z^\lambda
\\
z_-^\lambda &=& \theta(-z) (-z)^\lambda.
\end{eqnarray}
Due to the polynomial $P_2$ in (\ref{eq:ME1}) we have to consider Fourier-transforms of
$f \in {\cal S}'$ distributions of the kind
\begin{eqnarray}
\label{eq:DERIV}
{\bf F}^{-1}[x^m f(x)](\sigma) = (-i)^m \frac{\partial^m}{\partial \sigma^m} {\bf F}^{-1}[f(x)](\sigma),~~~m \in
\mathbb{N}.
\end{eqnarray}
Since $(l_1^0)^2 > 0$ one has
\begin{eqnarray}
  \left((l_1^0)^2 + i \delta\right)^{1/2} = \left[\sqrt{(l_1^0)^2}\right]_+ = |l_1^0|.
\end{eqnarray}
We consider now the one-dimensional Fourier transform of the distribution $z_\pm^\lambda$, cf.~\cite{GELF}.
\begin{eqnarray}
{\bf F}^{-1}[z_+^\lambda](t) &=& i \frac{\Gamma(\lambda + 1)}{2\pi} \frac{\exp[i \lambda \pi/2]}{(-t + i
\delta)^{\lambda +1}},
\label{eq:Fp}
\\
{\bf F}^{-1}[z_-^\lambda](t) &=& -i \frac{\Gamma(\lambda + 1)}{2\pi} \frac{\exp[-i \lambda \pi/2]}{(-t - i
\delta)^{\lambda +1}}
\label{eq:Fm}
\end{eqnarray}
and one has
\begin{eqnarray}
{\bf F}^{-1}[|z|](t) &=&
{\bf F}^{-1}[z_+](t) +  {\bf F}^{-1}[z_-](t) = -\frac{1}{2 \pi} \Biggl[ \frac{1}{(t + i\delta)^2}
+ \frac{1}{(t - i\delta)^2} \Biggr].
\label{eq:Fm11}
\end{eqnarray}
Only the first term in the r.h.s. contributes to the contour integral using the residue theorem.

As an example we consider
\begin{eqnarray}
P_2(l_1^0, l_2^0) = (l_1^0)^2 (l_2^0)^2 [l_1^0 + l_2^0]^2.
\end{eqnarray}
According to (\ref{eq:DERIV}) the polynomial implies the term
\begin{eqnarray}
-\frac{1}{4 \pi^2} \Biggl[ \frac{1440}{(t_1-t+i\delta)^6 (t_2-t+i\delta)^4} + \frac{1152}{(t_1-t+i\delta)^5
(t_2-t+i\delta)^5} \Biggr].
\end{eqnarray}

Therefore we have
\begin{eqnarray}
{\sf \tilde{S}}_1^{\rm c} &=& 8 \pi^2 G_N^2
\int_{-\infty}^{+\infty} dt Q(t)
\int_{-\infty}^{+\infty} dt_1
\int_{-\infty}^{+\infty} dt_2 \frac{(-1)}{16 \pi^2}  \frac{(-1)}{4 \pi^2}
\nonumber\\ &&
\times  \Biggl\{\frac{1440}{(t_1-t+i\delta)^6 (t_2-t+i\delta)^4}
+ \frac{1152}{(t_1-t+i\delta)^5 (t_2-t+i\delta)^5} \Biggr\} Q(t_1) Q(t_2)
\label{eqQ1}
\\
 &=&  {-}16 \pi^2 G_N^2 \frac{(-1)^2}{64 \pi^4} (2 \pi i)^2
\int_{-\infty}^{+\infty} dt Q(t)
\Biggl\{\left[\frac{d^5}{dt^5} Q(t)\right] \left[\frac{d^3}{dt^3} Q(t)\right] +\left[\frac{d^4}{dt^4} Q(t)\right]^2 \Biggr\}
\nonumber\\
 &=&  G_N^2
\int_{-\infty}^{+\infty} dt Q(t)
\Biggl\{\left[\frac{d^5}{dt^5} Q(t)\right] \left[\frac{d^3}{dt^3} Q(t)\right] +\left[\frac{d^4}{dt^4} Q(t)\right]^2 \Biggr\}
\label{eqQ2}
\\
\label{eq:QQQc}
 &=&  \frac{G_N^2}{2} \int_{-\infty}^{+\infty} dt
\left[\frac{d^2}{dt^2} Q(t)\right] \left[\frac{d^3}{dt^3} Q(t)\right]^2,
\label{eqQ3}
\end{eqnarray}
using the residue theorem, \cite{PRIW2}, since $Q(t)$ is bounded and obeys a Taylor expansion.
The result is given in (\ref{eqQ2}).
In (\ref{eqQ3})
it has been further assumed, that $Q(t)$ and the derivatives of $Q(t)$ vanish in the limit $t \rightarrow \pm
\infty$. (\ref{eq:QQQc}) is again a Riemann integral.

We finally turn to the memory term, Figure~\ref{fig:1} r.h.s. The corresponding Green's function is
obtained by carrying out the functional derivations in $I_2$
\begin{eqnarray}
  \label{eq:I2}
    I_2 ~&=&~ \biggl[\frac{i^{10}}{3!} \int dx^D_1\int dx^D_2\int dx^D_3 \int dx^D_4
  \ \Bigl[\delta_{J_-(x_1)} Q_+(x_1) V_+(x_1) \delta_{J_-(x_2)} Q_+(x_2) V_+(x_2)
\nonumber\\ &&
\times
\delta_{J_+(x_3)} Q_-(x_3) V_-(x_3)\nonumber\\ &&
  + \delta_{J_-(x_1)} Q_+(x_1) V_+(x_1) \delta_{J_+(x_2)} Q_-(x_2) V_-(x_2) \delta_{J_-(x_3)}
Q_+(x_3) V_+(x_3) \nonumber\\ &&
  + \delta_{J_+(x_1)} Q_-(x_1) V_-(x_1) \delta_{J_-(x_2)} Q_+(x_2) V_+(x_2) \delta_{J_-(x_3)} Q_+(x_3)
V_+(x_3)\Bigr]\delta_{J_+(x_4)}^2\delta_{J_-(x_4)}v(x_4)\nonumber\\
    && \times\frac{1}{3!} \prod_{k=5}^8\biggl[-\frac{i}{2} \int dx_k^Ddy_k^D
    \ (J_-(x_k)\Delta_{\text{adv}}(x_k-y_k)J_+(y_k)\nonumber\\
    &&+ J_+(x_k)\Delta_{\text{ret}}(x_k-y_k)J_-(y_k)
    + J_+(x_k)\Delta_{\text{C}}(x_k-y_k)J_+(y_k))\biggr]\biggr]_{\text{conn.}}
\end{eqnarray}
and setting the Schwinger parameters to zero.
Here the function $v$ refers to the contributing triple bulk vertices, including their combinatorics,
and also account for the propagator numerators as also $V_\pm$ at the multipole vertices. Unlike the
case for $I_1$, here various fields are contributing. By analogous operations as in the case of ${\sf
S}_1$ one obtains
\begin{eqnarray}
\label{eq:S2a}
{\sf {S}}_2 &=&  {-}G_N^2 \sqrt{2} \frac{1}{10} \int_{-\infty}^{+\infty} dt~{\rm tr} \Bigl\{
\frac{4}{7}\left[3
Q_+^{(2)}(t)
Q_+^{(3)}(t)
Q_-^{(3)}(t)
- 2 Q_-^{(2)}(t)
Q_+^{(3)}(t)
Q_+^{(3)}(t)\right]
\nonumber\\ &&
+
\left[2
Q_+(t)
Q_+^{(4)}(t)
Q_-^{(4)}(t)
- Q_-(t)
Q_+^{(4)}(t)
Q_+^{(4)}(t)\right]\Bigr\}.
\end{eqnarray}
Finally one obtains from these terms the contributions
(\ref{eq:Mmem1}--\ref{eq:Mmem3}).

To extract the conservative part of the action, we express
$\textsf{S}_1$ and $\textsf{S}_2$ in terms of $Q_1, Q_2$:
\begin{align}
  \label{Si_dec}
  \textsf{S}_i ={}& \int dt\ \bigl(L_i[Q_1] - L_i[Q_2] + R_i[Q_1, Q_2]\bigr)\qquad i=1,2,\\
  \label{L1}
  L_1[Q_j] ={}& \frac{G_N^2}{24} \tr\left[Q_j Q_j^{(4)}Q_j^{(4)}\right],\\
  \label{L2}
  L_2[Q_j] ={}& {-}G_N^2\tr\left[\frac{1}{20} Q_j Q_j^{(4)}Q_j^{(4)} + \frac{1}{35} Q_j^{(2)}
Q_j^{(3)}Q_j^{(3)}\right],\\
  R_1[Q_1, Q_2] ={}& G_N^2\tr\left[{-}\frac{1}{12} Q_1 Q_1^{(4)}Q_2^{(4)} {+} \frac{1}{8} Q_2
Q_1^{(4)}Q_1^{(4)}\right] - (1 \leftrightarrow 2),\\
  R_2[Q_1, Q_2] ={}& G_N^2\tr\left[\frac{1}{10} Q_1 Q_1^{(4)}Q_2^{(4)} {-} \frac{3}{20} Q_2
Q_1^{(4)}Q_1^{(4)} {+} \frac{4}{35} Q_1^{(2)} Q_1^{(3)}Q_2^{(3)} {-} \frac{1}{7} Q_2^{(2)}
Q_1^{(3)}Q_1^{(3)}\right] \nonumber\\
& - (1 \leftrightarrow 2),
\end{align}
cf.~(\ref{eq:QUADx}).
Following~\cite{Galley:2015kus}, we identify $L_1$ and $L_2$ with the
conservative contribution to the Lagrangian. We therefore add $L_1[Q]
+ L_2[Q]$ to the Lagrangian derived in~\cite{Blumlein:2020pyo} by
integrating out the potential modes. After eliminating accelerations
and higher time derivatives as outlined in~\cite{Blumlein:2020pyo}, we
perform a Legendre transformation to obtain a Hamiltonian. The contribution resulting from
$\textsf{S}_1$ is given by equation~\eqref{eq:Mmem1}. For
$\textsf{S}_2$, the first term in equation~(\ref{L2}) gives rise to
equation~\eqref{eq:Mmem2} and the second term to
equation~\eqref{eq:Mmem3}.
\section{Scattering angle integrals}
\label{sec:E}

\vspace*{1mm}
\noindent
In the following we list integrals, which appear in the calculation of the scattering angle, using relations
given in \cite{RYGR}.

At the Newtonian level one obtains the well--known integral
\begin{eqnarray}
I_0 &=& \int_0^{x_{\rm max}} {dx} \frac{1}{\displaystyle \sqrt{1 - x^2 + \frac{2x}{j\PI}}} =
-\frac{\pi}{2} + {\rm arcsin}\left(\frac{1}{\sqrt{1 + j^2 \PI^2}}\right).
\end{eqnarray}
It implies the value $\pi/2$ in (\ref{eq:SCA}). Its $\eta$--expansion delivers the $\nu$ and $\eta$--independent terms
of $\chi_{2k+1}$ with
\begin{eqnarray}
I_0^{\rm (reg)} &=& -\frac{\pi}{2} + \frac{1}{\PI j} - \frac{1}{3 \PI^3 j^3} + \frac{1}{5 \PI^5 j^5}
+ O\left(\frac{1}{j^7}\right).
\end{eqnarray}

The $O(\eta^2)$ term reads
\begin{eqnarray}
I_2 &=&
\eta^2 \Biggl[
\frac{2 \PI}{j}
+ \frac{3 \pi}{2 j^2}
+\frac{4}{j^3 \PI}
-\frac{2}{j^5 \PI^3}
-\frac{10}{7 j^9 \PI^7}
+\frac{8}{5 j^7 \PI^5}
+\frac{4}{3 j^{11} \PI^9}
\Biggr] + O\left(\frac{1}{\PI^{11}}\right) +
O\left(\frac{1}{\sqrt{\ep}}\right).
\nonumber\\
\end{eqnarray}

The  $O(\eta^4)$ is given by
\begin{eqnarray}
I_4 &=& \eta^4 \Biggl[\pi \left(\frac{15}{8} - \frac{3}{4} \nu\right)
\PI^2 \frac{1}{j^2}
+(24-8 \nu) \PI \frac{1}{j^3}
+\pi \left(\frac{105}{8} - \frac{15}{4} \nu \right) \frac{1}{j^4}
+(32-8 \nu) \frac{1}{j^5 \PI}
\nonumber\\ &&
-\left(16-\frac{16}{5}\nu\right) \frac{1}{j^7 \PI^3}
+\left(\frac{96}{7}-\frac{16}{7}\nu\right) \frac{1}{j^9 \PI^5}
-\left(\frac{40}{3}-\frac{40}{21} \nu\right) \frac{1}{j^{11} \PI^7}
\Biggr] +  O\left(\frac{1}{\PI^9}\right) +
O\left(\frac{1}{\ep^{3/2}}\right).
\nonumber\\
\end{eqnarray}
Similar structures are obtained for the higher order terms in $\eta$. Here the singularity in $\ep$ becomes
stronger by one unit going from $I_{2k}$ to $I_{2k+2}$.
One easily sees that these integrals
form the first terms of the coefficients $\chi_k$ given in Eq.~(\ref{eq:CHIMAIN1}--\ref{eq:CHIMAIN2}).

We also list a series of higher expansion coefficients for $\chi^{\rm Schw}_k$ for convenience, which result from
(\ref{eq:SCHW2}),
\begin{eqnarray}
\chi^{\rm Schw}_5 &=&
\frac{1}{5 p_\infty} - \frac{2 \eta^2}{p_\infty^3} + \frac{32 \eta^4}{p_\infty} + 320 p_\infty \eta^6
+ 640 p_\infty^3 \eta^8 +
\frac{1792}{5} p_\infty^5 \eta^{10},
\\
\chi^{\rm Schw}_6 &=& \pi \left(
\frac{1155}{8} \eta^6 + \frac{45045}{64} p_\infty^2 \eta^8 + \frac{135135}{128} p_\infty^4 \eta^{10}
+ \frac{255255}{512} p_\infty^6 \eta^{12}
\right),
\\
\chi^{\rm Schw}_7 &=&
-\frac{1}{7 p_\infty^7} + \frac{8 \eta^2}{5 p_\infty^5} - \frac{16 \eta^4}{p_\infty^3} + \frac{320
\eta^6}{p_\infty} + 4480
p_\infty \eta^8 +
 14336 p_\infty^3 \eta^{10} + \frac{86016}{5} p_\infty^5 \eta^{12} + \frac{49152}{7} p_\infty^7 \eta^{14},
\nonumber\\
\\
\chi^{\rm Schw}_8 &=& \pi \left(
\frac{225225}{128} \eta^8 + \frac{765765}{64} p_\infty^2 \eta^{10} + \frac{14549535}{512} p_\infty^4 \eta^{12} +
\frac{14549535}{512} p_\infty^6 \eta^{14} + \frac{334639305}{32768} p_\infty^8 \eta^{16}
\right),
\nonumber\\ &&
\\
\chi^{\rm Schw}_9 &=&
\frac{1}{9 p_\infty^9} - \frac{10 \eta^2}{7 p_\infty^7} + \frac{96 \eta^4}{7 p_\infty^5} - \frac{448 \eta^6}{3
p_\infty^3}
+ \frac{3584 \eta^8}{p_\infty} +
 64512 p_\infty \eta^{10} + 286720 p_\infty^3 \eta^{12} + 540672 p_\infty^5  \eta^{14}
\nonumber\\ &&
+ \frac{3244032}{7} p_\infty^7 \eta^{16}
+ \frac{9371648}{63} p_\infty^9 \eta^{18},
\\
\chi^{\rm Schw}_{10} &=& \pi \left(
\frac{2909907}{128} \eta^{10} + \frac{101846745}{512} p_\infty^2 \eta^{12}  + \frac{334639305}{512} p_\infty^4
\eta^{14}
+ \frac{8365982625}{8192} p_\infty^6 \eta^{16}
\right. \nonumber\\ && \left.
+ \frac{25097947875}{32768} p_\infty^8 \eta^{18}
+ \frac{29113619535}{131072} p_\infty^{10} \eta^{20}
\right),
\\
\chi^{\rm Schw}_{11} &=&
- \frac{1}{11 p_\infty^{11}}
+ \frac{4 \eta^2}{3 p_\infty^9}
- \frac{40 \eta^4}{3 p_\infty^7}
+ \frac{128 \eta^6}{p_\infty^5}
- \frac{1536\eta^8}{p_\infty^3}
+ \frac{43008 \eta^{10}}{p_\infty}
+ 946176 p_\infty \eta^{12}
+ 5406720 p_\infty^3 \eta^{14}
\nonumber\\ &&
+ 14057472 p_\infty^5 \eta^{16}
+ 18743296 p_\infty^7 \eta^{18}
+ \frac{37486592 p_\infty^9}{3} \eta^{20}
+ \frac{109051904 p_\infty^{11}}{33} \eta^{22},
\\
\chi^{\rm Schw}_{12} &=& \pi \left(
\frac{156165009}{512} \eta^{12} + \frac{1673196525}{512} p_\infty^2 \eta^{14} + \frac{225881530875}{16384}
p_\infty^4 \eta^{16}
+ \frac{242613496125}{8192} p_\infty^6 \eta^{18} \right.
\nonumber\\ && \left.
+ \frac{4512611027925}{131072} p_\infty^8 \eta^{20}
+ \frac{2707566616755}{131072} p_\infty^{10} \eta^{22}
+ \frac{10529425731825}{2097152} p_\infty^{12} \eta^{24}
\right),
\\
\chi^{\rm Schw}_{13} &=&
  \frac{1}{13 p_\infty^{13}}
- \frac{14 \eta^2}{11 p_\infty^{11}}
+ \frac{448 \eta^4}{33 p_\infty^9}
- \frac{128 \eta^6}{p_\infty^7}
+ \frac{1280 \eta^8}{p_\infty^5}
- \frac{16896 \eta^{10}}{p_\infty^3}
+ \frac{540672 \eta^{12}}{p_\infty}
+ 14057472 p_\infty \eta^{14}
\nonumber\\ &&
+ 98402304 p_\infty^3 \eta^{16}
+ 328007680 p_\infty^5 \eta^{18}
+ 599785472 p_\infty^7 \eta^{20}
+ \frac{1853882368 p_\infty^9}{3} \eta^{22}
\nonumber\\ &&
+ \frac{3707764736 p_\infty^{11}}{11} \eta^{24}
+ \frac{10838081536 p_\infty^{13}}{143} \eta^{26},
\\
\chi^{\rm Schw}_{14} &=& \pi \left(
   \frac{2151252675}{512} \eta^{14}
 + \frac{436704293025 p_\infty^2}{8192} \eta^{16}
 + \frac{4512611027925 p_\infty^4}{16384} \eta^{18}
\right.
\nonumber\\ && \left.
 + \frac{49638721307175 p_\infty^6}{65536} \eta^{20}
 + \frac{157941385977375 p_\infty^8}{131072} \eta^{22}
 + \frac{1168766256232575 p_\infty^{10}}{1048576} \eta^{24}
\right. \nonumber\\ && \left.
 + \frac{1168766256232575 p_\infty^{12}}{2097152} \eta^{26}
 + \frac{977947275623175 p_\infty^{14}}{8388608}  \eta^{28} \right).
\end{eqnarray}

\vspace*{5mm}
\noindent
{\bf Acknowledgment.}
We thank Th.~Damour, S.~Foffa, R.~Sturani, Z.~Bern, M.~Ruf, G.~Kaelin,
L.~Blanchet, C.~Kavanagh, and B.~Wardell for
discussions. This work has been funded in part by EU TMR network SAGEX agreement No. 764850
(Marie Sk\l{}odowska--Curie). G.~Sch\"afer has been supported in part by Kolleg Mathematik Physik
Berlin (KMPB) and DESY. The Feynman-type diagrams shown have been drawn
using {\tt axodraw}, \cite{Vermaseren:1994je} and  {\tt Asymptote} \cite{ASYMPTOTE}.

\vspace*{3mm}
\noindent
{\bf Note added.}~After completion of this paper two preprints appeared \cite{BLANC}, in which the
representation of the electric quadrupole moment have been given in $d$ dimensions, without the previously
used additional Hadamard regularization \cite{HADAMARD}. This new representation has no impact on the results of 
the
present calculation.

S.~Foffa communicated to us a note in preparation \cite{Almeida:2021xwn}, in which it is shown that the previously
necessary finite renormalization
of the magnetic quadrupole term {\sf JEJ}, using a representation containing Levi-Civita symbols, can be avoided
by utilizing a dual representation, free of them, giving the same result.

After submission of the present paper, Ref.~\cite{Bern:2021yeh} appeared, confirming Eq.~(6.17) of 
\cite{Bini:2021gat} in the post-Minkowskian approach to $O(G_N^4/r^4)$. The tail terms
resulting from the multipole--expansion alone, within the Hamiltonian approach to the bound state 
problem, however, do not lead to this result, as has been shown by the explicit calculation
in the present paper in detail. Ref.~\cite{Bern:2021yeh} compares to Ref.~\cite{Bini:2021gat} not in 
the scattering angle defined in \cite{Bini:2020wpo} but in the quantity ${\cal M}_4^{\rm radgrav, f}$.
We mention that, on the other hand, the potential terms do fully agree between
Refs.~\cite{Bern:2021dqo} and \cite{Blumlein:2021txj}.

For future work it would be highly desirable to have a consistent field theoretic description of the 
tail terms ab initio with Feynman rules to all orders in the effective field theory approach. First 
steps in this direction have been undertaken in Ref.~\cite{Blumlein:2020pyo} already, by using 
expansion by regions,~cf.~\cite{Jantzen:2011nz} for a survey.

Very recently a phenomenological analysis on the impact of the numerical difference described in 
Section~\ref{sec:32} on the scattering angle
has been made in Ref.~\cite{Khalil:2022ylj}, coming to the result of an effect of $10^{-3}$ in $120$ and larger.
Here velocities of $v \leq 1/2$ have been assumed. Despite of this, however, we will aim at the determination of 
the exact analytic result in the future.


\begin{thebibliography}{99}
%
\bibitem{LIGO}
B.P.~Abbott et al. (Virgo, LIGO Scientific),
Phys. Rev. Lett. {\bf 116} (2016) 061102 [arXiv:1602.03837 [gr-qc]];
Phys. Rev. X {\bf 6} (2016) 041015 [arXiv:1606.04856 [gr-qc]];
Phys. Rev. Lett. {\bf 119} (2017) 161101 (2017), [arXiv:1710.05832 [gr-qc]];
Phys. Rev. X {\bf 9} (2019) 031040 [arXiv:1811.12907 [astro-ph.HE]].
%
\bibitem{PROJECT}
Y.~Aso, Y.~Michimura, K.~Somiya, M.~Ando, O.~Miyakawa, T.~Sekiguchi, D.~Tatsumi,
and H.~Yamamoto (KAGRA), Phys. Rev. D {\bf 88} (2013) 043007 [arXiv:1306.6747 [gr-qc]];\\
F.~Acernese et al. (VIRGO), Class. Quant. Grav. {\bf 32} (2015) 024001 [arXiv:1408.3978 [gr-qc]];\\
J.~Aasi et al. (LIGO Scientific), Class. Quant. Grav. {\bf 32} (2015)  074001 [arXiv:1411.4547 [gr-qc]];\\
B.~Iyer et al. (LIGO Collaboration), {\it LIGO-India,
Proposal of the Consortium for Indian Initiative in
Gravitational-wave Observations} (2011), LIGO Document M1100296-v2.
%
\bibitem{1PN}
A.~Einstein,
Sitzungsber. Preuss. Akad. Wiss. (1915) 831--839;\\
J.~Droste,
 Proc.\ Acad.\ Sci.\ Amst.\ {\bf 19} (1916) 447--455; \\
H.~Lorentz and J.~Droste, in: {\sf The motion of a system of bodies under the influence of their
mutual attraction, according to Einstein's theory}, (Nijhoff, The Hague, 1937) pp.~330--355;\\
  A.~Einstein, L.~Infeld and B.~Hoffmann,
  Annals Math.\  {\bf 39} (1938) 65--100;\\
H.P. Robertson, Annals Math. {\bf 39} (1938) 101--104;\\
A. Eddington and G.L. Clark, Proc. R. Soc. London A {\bf 166} (1938) 465--475.
%
\bibitem{Kol:2007bc}
B.~Kol and M.~Smolkin,
Class. Quant. Grav. \textbf{25} (2008) 145011
[arXiv:0712.4116 [hep-th]].
%
\bibitem{2PN}
  S.~Chandrasekhar,
  Astrophys.\ J.\  {\bf 142} (1965) 1488--1512; \\
Astrophys. J. {\bf 158} (1969) 45--54;\\
S.~Chandrasekhar and Y.~Nutku, Astrophys. J. {\bf 158} (1969) 55--79;\\
S.~Chandrasekhar and F.~Esposito, Astrophys. J. {\bf 160} (1970) 153--179;\\
C. Hoenselaers, Prog. Theor. Phys. {\bf 56} (1976) 324--326;\\
T. Damour and G. Sch\"afer, C.R. Acad. Sci. Paris {\bf 305}, s\'erie II, (1987) 839--842;
Nuovo Cim. B \textbf{101} (1988) 127--176;\\
T.~Ohta and T.~Kimura,
Prog. Theor. Phys. \textbf{81} (1989) 679--689;\\
G. Sch\"afer and N. Wex, Phys. Lett. A {\bf 174} (1993) 196--205;\\
T. Ohta, H. Okamura, K. Hiida and T. Kimura, Prog. Theor. Phys. {\bf 50} (1973)
492--514; {\bf 51} (1974) 1598--1612; {\bf 51} (1974) 1220--1238;\\
T. Damour and N. Deruelle, C. R. Acad. Sci. s\'erie II {\bf 293} (1981) 537--540;\\
T. Damour, C.R. Acad. Sci. s\'erie II {\bf 29}4 (1982) 1355--1357;
and in: {\sf Gravitational Radiation}, eds. N. Deruelle and T. Piran (NATO ASI, North-Holland, Amsterdam, 1983)
59--144;\\
T.~Damour and G.~Sch\"afer,
Gen. Rel. Grav. \textbf{17} (1985) 879--905;\\
S.M. Kopeikin, Sov. Astron. {\bf 29} (1985) 516--524; \\
L.P. Grishchuk and S.M. Kopeikin, in: {\sf Relativity in Celestial Mechanics and Astrometry: High Precision
Dynamical Theories and Observational Verifications}, eds. J. Kovalevsky and V. A. Brumberg
(D. Reidel Publishing, Dordrecht, 1986) 19--34;\\
M.E. Pati and C.M. Will, Phys. Rev. D {\bf 65} (2002) 104008 [arXiv:gr-qc/0201001 [gr-qc]];\\
J.B.~Gilmore and A.~Ross,
Phys. Rev. D \textbf{78} (2008) 124021
[arXiv:0810.1328 [gr-qc]].
%
\bibitem{3PN}
P.~Jaranowski and G.~Sch\"afer,
Phys. Rev. D \textbf{57} (1998) 7274--7291
[Erratum: Phys. Rev. D \textbf{63} (2001) 029902]
[arXiv:gr-qc/9712075 [gr-qc]];
Phys. Rev. D \textbf{60} (1999) 124003
[arXiv:gr-qc/9906092 [gr-qc]];\\
T.~Damour, P.~Jaranowski and G.~Sch\"afer,
Phys. Rev. D \textbf{62} (2000)  044024
[arXiv:gr-qc/9912092 [gr-qc]];
Phys. Rev. D \textbf{62} (2000) 021501
[Erratum: Phys. Rev. D \textbf{63} (2001), 029903]
[arXiv:gr-qc/0003051 [gr-qc]];
Phys. Lett. B \textbf{513} (2001) 147--155
[arXiv:gr-qc/0105038 [gr-qc]];\\
L.~Blanchet and G.~Faye,
Phys. Lett. A \textbf{271} (2000) 58--64
[arXiv:gr-qc/0004009 [gr-qc]];\\
V.C.~de Andrade, L.~Blanchet and G.~Faye,
Class. Quant. Grav. \textbf{18} (2001) 753--778
[arXiv:gr-qc/0011063 [gr-qc]];\\
L.~Blanchet, T.~Damour and G.~Esposito-Farese,
Phys. Rev. D \textbf{69} (2004) 124007
[arXiv:gr-qc/0311052 [gr-qc]];\\
Y.~Itoh and T.~Futamase,
Phys. Rev. D \textbf{68} (2003) 121501
[arXiv:gr-qc/0310028 [gr-qc]];\\
Y.~Itoh,
Phys. Rev. D \textbf{69} (2004) 064018
[arXiv:gr-qc/0310029 [gr-qc]];\\
R.M.~Memmesheimer, A.~Gopakumar and G.~Sch\"afer,
Phys. Rev. D \textbf{70} (2004) 104011
[arXiv:gr-qc/0407049 [gr-qc]];\\
S.~Foffa and R.~Sturani,
Phys. Rev. D \textbf{84} (2011) 044031
[arXiv:1104.1122 [gr-qc]].
%
\bibitem{Damour:2014jta}
  T.~Damour, P.~Jaranowski and G.~Sch\"afer,
  Phys.\ Rev.\ D {\bf 89} (2014) no.6,  064058
  [arXiv:1401.4548 [gr-qc]].
%
\bibitem{4PN}
P.~Jaranowski and G.~Sch\"afer,
Phys. Rev. D \textbf{92} (2015) no.12, 124043
[arXiv:1508.01016 [gr-qc]];\\
L.~Bernard, L.~Blanchet, A.~Boh\'e, G.~Faye and S.~Marsat,
Phys. Rev. D \textbf{93} (2016) no.8, 084037
[arXiv:1512.02876 [gr-qc]];\\
T.~Damour, P.~Jaranowski and G.~Sch\"afer,
Phys. Rev. D \textbf{93} (2016) no.8, 084014
[arXiv:1601.01283 [gr-qc]];\\
  T.~Damour and P.~Jaranowski,
  Phys.\ Rev.\ D {\bf 95} (2017) no.8,  084005
  [arXiv:1701.02645 [gr-qc]];\\
S.~Foffa, P.~Mastrolia, R.~Sturani and C.~Sturm,
Phys. Rev. D \textbf{95} (2017) no.10, 104009
[arXiv:1612.00482 [gr-qc]];\\
L.~Bernard, L.~Blanchet, A.~Boh\'e, G.~Faye and S.~Marsat,
Phys. Rev. D \textbf{95} (2017) no.4, 044026
[arXiv:1610.07934 [gr-qc]];\\
  T.~Marchand, L.~Bernard, L.~Blanchet and G.~Faye,
  Phys.\ Rev.\ D {\bf 97} (2018) no.4,  044023
  [arXiv:1707.09289 [gr-qc]];\\
  L.~Bernard, L.~Blanchet, G.~Faye and T.~Marchand,
  Phys.\ Rev.\ D {\bf 97} (2018) no.4,  044037
  [arXiv:1711.00283 [gr-qc]].
%
\bibitem{Foffa:2019rdf}
S.~Foffa and R.~Sturani,
Phys. Rev. D \textbf{100} (2019) no.2, 024047
[arXiv:1903.05113 [gr-qc]].
%
\bibitem{Foffa:2019yfl}
S.~Foffa, R.A.~Porto, I.~Rothstein and R.~Sturani,
Phys. Rev. D \textbf{100} (2019) no.2, 024048
[arXiv:1903.05118 [gr-qc]].
%
\bibitem{Blumlein:2020pog}
J.~Bl\"umlein, A.~Maier, P.~Marquard and G.~Sch\"afer,
Nucl. Phys. B \textbf{955} (2020) 115041
[arXiv:2003.01692 [gr-qc]].
%
\bibitem{Foffa:2019hrb}
  S.~Foffa, P.~Mastrolia, R.~Sturani, C.~Sturm and W.J.~Torres Bobadilla,
  Phys.\ Rev.\ Lett.\  {\bf 122} (2019) no.24,  241605
  [arXiv:1902.10571 [gr-qc]].
%
\bibitem{Blumlein:2019zku}
  J.~Bl\"umlein, A.~Maier and P.~Marquard,
  Phys.\ Lett.\ B {\bf 800} (2020) 135100
  [arXiv:1902.11180 [gr-qc]].
%
\bibitem{Bini:2019nra}
  D.~Bini, T.~Damour and A.~Geralico,
  Phys.\ Rev.\ Lett.\  {\bf 123} (2019) no.23,  231104
  [arXiv:1909.02375 [gr-qc]].
%
\bibitem{Bini:2020wpo}
D.~Bini, T.~Damour and A.~Geralico,
Phys. Rev. D \textbf{102} (2020) no.2, 024062
[arXiv:2003.11891 [gr-qc]].
%
\bibitem{Blumlein:2020pyo}
J.~Bl\"umlein, A.~Maier, P.~Marquard and G.~Sch\"afer,
Nucl. Phys. B \textbf{965} (2021) 115352
[arXiv:2010.13672 [gr-qc]].
%
\bibitem{Foffa:2020nqe}
S.~Foffa, R.~Sturani and W.~J.~Torres Bobadilla,
JHEP \textbf{02} (2021) 165
[arXiv:2010.13730 [gr-qc]].
%
\bibitem{Bini:2020nsb}
D.~Bini, T.~Damour and A.~Geralico,
Phys. Rev. D \textbf{102} (2020) no.2, 024061
[arXiv:2004.05407 [gr-qc]].
%
\bibitem{Bini:2020hmy}
D.~Bini, T.~Damour and A.~Geralico,
Phys. Rev. D \textbf{102} (2020) no.8, 084047
[arXiv:2007.11239 [gr-qc]].
%
\bibitem{Blumlein:2020znm}
J.~Bl\"umlein, A.~Maier, P.~Marquard and G.~Sch\"afer,
Phys. Lett. B \textbf{807} (2020), 135496
[arXiv:2003.07145 [gr-qc]].
%
\bibitem{PM}
J.~Bl\"umlein, A.~Maier, P.~Marquard, G.~Sch\"afer and C.~Schneider,
Phys. Lett. B \textbf{801} (2020) 135157
[arXiv:1911.04411 [gr-qc]];\\
M.~Accettulli Huber, A.~Brandhuber, S.~De Angelis and G.~Travaglini,
Phys. Rev. D \textbf{102} (2020) no.4, 046014
[arXiv:2006.02375 [hep-th]];\\
G.~K\"alin, Z.~Liu and R.A.~Porto,
Phys. Rev. Lett. \textbf{125} (2020) no.26, 261103
[arXiv:2007.04977 [hep-th]];\\
G.~K\"alin and R.A.~Porto,
JHEP \textbf{11} (2020), 106
[arXiv:2006.01184 [hep-th]];
G.~K\"alin and R.A.~Porto,
JHEP \textbf{01} (2020) 072
[arXiv:1910.03008 [hep-th]];\\
T.~Damour,
Phys. Rev. D \textbf{102} (2020) no.12, 124008
[arXiv:2010.01641 [gr-qc]].
%
\bibitem{Kalin:2019inp}
G.~K\"alin and R.~A.~Porto,
JHEP \textbf{02} (2020) 120
[arXiv:1911.09130 [hep-th]].
%
\bibitem{Damour:2019lcq}
T.~Damour,
Phys. Rev. D \textbf{102} (2020) no.2, 024060
[arXiv:1912.02139 [gr-qc]].
%
\bibitem{Bern:2019nnu}
  Z.~Bern, C.~Cheung, R.~Roiban, C.H.~Shen, M.P.~Solon and M.~Zeng,
  Phys.\ Rev.\ Lett.\  {\bf 122} (2019) no.20,  201603
  [arXiv:1901.04424 [hep-th]];\\
  Z.~Bern, C.~Cheung, R.~Roiban, C.H.~Shen, M.P.~Solon and M.~Zeng,
  JHEP {\bf 1910} (2019) 206
  [arXiv:1908.01493 [hep-th]];
%
\bibitem{Bini:2014nfa}
D.~Bini and T.~Damour,
Phys. Rev. D \textbf{89} (2014) no.10, 104047
[arXiv:1403.2366 [gr-qc]];
Phys. Rev. D \textbf{91} (2015), 064050
[arXiv:1502.02450 [gr-qc]];\\
C.~Kavanagh, A.C.~Ottewill and B.~Wardell,
Phys. Rev. D \textbf{92} (2015) no.8, 084025
[arXiv:1503.02334 [gr-qc]].
%
\bibitem{WARD1}
Electronic archive of post-newtonian coefficients,\newline
{\tt http://www.barrywardell.net/research/code}
%
\bibitem{Blanchet:2019rjs}
L.~Blanchet, S.~Foffa, F.~Larrouturou and R.~Sturani,
Phys. Rev. D \textbf{101} (2020) no.8, 084045
[arXiv:1912.12359 [gr-qc]].
%
\bibitem{SF}
S.L.~Detweiler,
Phys. Rev. D \textbf{77} (2008) 124026
[arXiv:0804.3529 [gr-qc]];\\
L.~Barack and N.~Sago,
Phys. Rev. Lett. \textbf{102} (2009) 191101
[arXiv:0902.0573 [gr-qc]];\\
T.~Damour,
Phys. Rev. D \textbf{81} (2010) 024017
[arXiv:0910.5533 [gr-qc]];\\
L.~Blanchet, S.L.~Detweiler, A.~Le Tiec and B.F.~Whiting,
Phys. Rev. D \textbf{81} (2010), 084033
[arXiv:1002.0726 [gr-qc]];\\
L.~Barack and A.~Pound,
Rept. Prog. Phys. \textbf{82} (2019) no.1, 016904
[arXiv:1805.10385 [gr-qc]].
%
\bibitem{Goldberger:2004jt}
  W.D.~Goldberger and I.Z.~Rothstein,
  Phys.\ Rev.\ D {\bf 73} (2006) 104029
  [hep-th/0409156].
%
\bibitem{Foffa:2019eeb}
S.~Foffa and R.~Sturani,
Phys. Rev. D \textbf{101} (2020) no.6, 064033
[Erratum: Phys. Rev. D \textbf{103} (2021) no.8, 089901]
[arXiv:1907.02869v5 [gr-qc]].
%
\bibitem{Nogueira:1991ex}
  P.~Nogueira,
  J.\ Comput.\ Phys.\  {\bf 105} (1993) 279--289.
%
\bibitem{FORM}
  J.A.M.~Vermaseren,
  {\it New features of FORM},
  math-ph/0010025;\\
  M.~Tentyukov and J.A.M.~Vermaseren,
  Comput.\ Phys.\ Commun.\  {\bf 181} (2010) 1419--1427
  [hep-ph/0702279].
%
\bibitem{CRUSHER}
P.~Marquard and D.~Seidel, {\it The {\tt Crusher} algorithm}, unpublished.
%
\bibitem{IBP}
J. Lagrange, {\sf Nouvelles recherches sur la nature et la propagation
du son}, Miscellanea Taurinensis, t. II, 1760-61; Oeuvres t. I, p. 263;\\
C.F. Gau\ss{}, {Theoria attractionis corporum sphaeroidicorum ellipticorum
homogeneorum methodo novo tractate}, Commentationes societas scientiarum
Gottingensis recentiores, Vol III, 1813, Werke Bd. {\bf V} pp. 5--7;\\
G. Green, {\sf Essay on the Mathematical Theory of Electricity and
Magnetism}, Nottingham, 1828 [Green Papers, pp. 1--115];\\
M. Ostrogradski, Mem. Ac. Sci. St. Peters., {\bf 6}, (1831) 39--53;\\
  K.G.~Chetyrkin and F.V.~Tkachov,
  Nucl.\ Phys.\ B {\bf 192} (1981) 159--204;\\
  S.~Laporta,
  Int.\ J.\ Mod.\ Phys.\ A {\bf 15} (2000) 5087--5159
  [hep-ph/0102033].
%
\bibitem{Thorne:1980ru}
K.S.~Thorne,
Rev. Mod. Phys. \textbf{52} (1980) 299--339.
%
\bibitem{TAIL}
L.~Blanchet and T.~Damour,
Phil. Trans. Roy. Soc. Lond. A \textbf{320} (1986) 379--430; \\
L.~Blanchet and T.~Damour,
Phys. Rev. D \textbf{37} (1988) 1410--1435; \\
L.~Blanchet and G.~Sch\"afer,
Class. Quant. Grav. \textbf{10} (1993) 2699--2721; \\
O.~Poujade and L.~Blanchet,
Phys. Rev. D \textbf{65} (2002) 124020
[arXiv:gr-qc/0112057 [gr-qc]];\\
W.D.~Goldberger and I.Z.~Rothstein,
Phys. Rev. D \textbf{73} (2006) 104030
[arXiv:hep-th/0511133 [hep-th]]; \\
G.L.~Almeida, S.~Foffa and R.~Sturani,
JHEP \textbf{11} (2020), 165
[arXiv:2008.06195 [gr-qc]].
%
\bibitem{Ross:2012fc}
A.~Ross,
Phys. Rev. D \textbf{85} (2012) 125033
[arXiv:1202.4750 [gr-qc]].
%
\bibitem{Marchand:2020fpt}
T.~Marchand, Q.~Henry, F.~Larrouturou, S.~Marsat, G.~Faye and L.~Blanchet,
Class. Quant. Grav. \textbf{37} (2020) no.21, 215006
[arXiv:2003.13672 [gr-qc]].
%
\bibitem{Lins:2020omt}
A.N.~Lins and R.~Sturani,
Phys. Rev. D \textbf{103} (2021) no.8, 084030
[arXiv:2011.02124 [gr-qc]].
%
\bibitem{Larin:1993tq}
S.A.~Larin,
Phys. Lett. B \textbf{303} (1993) 113--118
[arXiv:hep-ph/9302240 [hep-ph]].
%
\bibitem{Henry:2021cek}
Q.~Henry, G.~Faye and L.~Blanchet,
Class. Quant. Grav. \textbf{38} (2021) no.18, 185004
[arXiv:2105.10876 [gr-qc]].
%
\bibitem{Bini:2021gat}
D.~Bini, T.~Damour and A.~Geralico,
Phys. Rev. D \textbf{104} (2021) no.8, 084031
[arXiv:2107.08896 [gr-qc]].
%
\bibitem{BLANCHET21}
L.~Blanchet, private communication, 22.09.21.
%
\bibitem{Heisenberg:1925zz}
W.~Heisenberg,
Z. Phys. \textbf{33} (1925) 879--893;\\
J.~von~Neumann, {\sf Mathematische Grundlagen der Quantenmechanik}, (Springer, Berlin, 1932).
%
\bibitem{Einstein:1918btx}
A.~Einstein,
Sitzungsber. Preuss. Akad. Wiss. Berlin (Math. Phys. ) \textbf{1918} (1918) 154--167
%
\bibitem{PAPAPETROU}
A.~Papapetrou,
C.R. Acad Sci Paris 255 (1962) 1578--1580;
Annales de l'I. H. P., section A, {\bf 14} (1971) 79--95.
%
\bibitem{Blanchet:1989cu}
L.~Blanchet and G.~Sch\"afer,
Mon. Not. Roy. Astron. Soc. \textbf{239} (1989), 845-867
[Erratum: Mon. Not. Roy. Astron. Soc. \textbf{242} (1990), 7
%
\bibitem{Blanchet:2020ngx}
L.~Blanchet, G.~Comp\`ere, G.~Faye, R.~Oliveri and A.~Seraj,
JHEP \textbf{02} (2021) 029
[arXiv:2011.10000 [gr-qc]].
%
\bibitem{Foffa:2021pkg}
S.~Foffa and R.~Sturani,
Phys. Rev. D \textbf{104} (2021) no.2, 024069
[arXiv:2103.03190 [gr-qc]].
%
\bibitem{Schwinger:1960qe}
J.S.~Schwinger,
J. Math. Phys. \textbf{2} (1961) 407--432;
Proc. Nat. Acad. Sci. USA {\bf 46} (1961) 1401--1415.
%
\bibitem{Bakshi:1962dv}
P.M.~Bakshi and K.T.~Mahanthappa,
J. Math. Phys. \textbf{4} (1963) 1--11;
J. Math. Phys. \textbf{4} (1963) 12--16
%
\bibitem{Keldysh:1964ud}
L.V.~Keldysh,
Zh. Eksp. Teor. Fiz. \textbf{47} (1964) 1515--1527 [JETP {\bf 20} (1965) 1018--1026].
%
\bibitem{Korenman:66}
V.~Korenman, Ann. Phys. (NY) {\bf 39} (1966) 72--126.
%
\bibitem{Buchbinder:1981hu}
I.L.~Buchbinder, D.M.~Gitman and E.S.~Fradkin,
Fortsch. Phys. \textbf{29} (1981) 187--218;\\
E.S.~Fradkin and D.M.~Gitman,
Fortsch. Phys. \textbf{29} (1981) 381--412.
%
\bibitem{Chou:1984es}
K.C.~Chou, Z.B.~Su, B.L.~Hao and L.~Yu,
Phys. Rept. \textbf{118} (1985) 1--131.
%
\bibitem{DEWITT}
B.~DeWitt, in: {\it Quantum Concepts in Space and Time}, eds. R.~Penrose and C.J.~Isham,
Calendron Press, Oxford, Calendron, 1986).
%
\bibitem{Jordan:1986ug}
R.D.~Jordan,
Phys. Rev. D \textbf{33} (1986) 444--454.
%
\bibitem{Hu:2003qn}
B.L.~Hu and E.~Verdaguer,
Living Rev. Rel. \textbf{7} (2004) 3--89.
[arXiv:gr-qc/0307032 [gr-qc]].
%
\bibitem{DEWITT1}
B.~DeWitt, {\sf The Global Approach to Quantum Field Theory}, Vol.~{\bf 1,2}, (Calendron Press,
Oxford, 2003).
%
\bibitem{KLEINERT}
H.~Kleinert,
{\sf Path Integrals in Quantum Mechanics, Statistics, Polymer Physics, and Financial Markets},
(World Scientific, Singapore, 2009), 5th Ed.
%
\bibitem{Galley:2009px}
C.R.~Galley and M.~Tiglio,
Phys. Rev. D \textbf{79} (2009), 124027
[arXiv:0903.1122 [gr-qc]].
%
\bibitem{Galley:2015kus}
C.R.~Galley, A.K.~Leibovich, R.A.~Porto and A.~Ross,
Phys. Rev. D \textbf{93} (2016), 124010
[arXiv:1511.07379 [gr-qc].
%
\bibitem{FH}
R.P.~Feynman and A.R.~Hibbs, {\sf Quantum Mechanics and Path Integrals}, (McGraw--Hill, New York, 1965).
%
\bibitem{STERMAN}
G.~Sterman, {\sf An Introduction to Quantum Field Theory}, (Cambridge University Press, Cambridge,UK, 1993).
%
\bibitem{Lehmann:1954rq}
H.~Lehmann, K.~Symanzik and W.~Zimmermann,
Nuovo Cim. \textbf{1} (1955) 205--225.
%
\bibitem{FC}
A.~J.~Buras,
Rev. Mod. Phys. \textbf{52} (1980) 199--276;\\
E.~Reya,
Phys. Rept. \textbf{69} (1981) 195--333;\\
J.~Bl\"umlein,
Prog. Part. Nucl. Phys. \textbf{69} (2013) 28--84
[arXiv:1208.6087 [hep-ph]].
%
\bibitem{Veltman:1963th}
M.J.G.~Veltman,
Physica \textbf{29} (1963) 186--207.
%
\bibitem{Bern:2021yeh}
Z.~Bern, J.~Parra-Martinez, R.~Roiban, M.S.~Ruf, C.H.~Shen, M.P.~Solon and M.~Zeng,
Phys. Rev. Lett. \textbf{128} (2022) no.16, 161103
[arXiv:2112.10750 [hep-th]].
%
\bibitem{FW}
J.A.~Wheeler and R.P.~Feynman,
Rev. Mod. Phys. \textbf{17} (1945) 157--181;
Rev. Mod. Phys. \textbf{21} (1949) 425--433.
%
\bibitem{Dirac:1938nz}
P.A.M.~Dirac,
Proc. Roy. Soc. Lond. A \textbf{167} (1938) 148--169.
%
\bibitem{Passarino:1978jh}
G.~Passarino and M.J.G.~Veltman,
Nucl. Phys. B \textbf{160} (1979) 151--207.
%
\bibitem{HYP}
 F.~Klein, {\sf Vorlesungen \"uber die hypergeometrische Funktion},
  Wintersemester 1893/94, Die Grundlehren der Mathematischen Wissenschaften
  {\bf 39}, (Springer, Berlin, 1933);\\
W.N.~Bailey, {\sf Generalized Hypergeometric Series}, (Cambridge University Press, Cambridge, 1935).
 L.J.~Slater, {\sf Generalized hypergeometric functions}, (Cambridge University Press, Cambridge, 1966);\\
 P. Appell and J. Kamp\'{e} de F\'{e}riet, {\sf Fonctions
  Hyperg\'{e}om\'{e}triques et Hypersph\'{e}riques, Polynomes D' Hermite},
  (Gauthier-Villars, Paris, 1926);\\
 P. Appell, {\sf Les Fonctions Hyperg\"{e}om\'{e}triques de Plusieur Variables}, (Gauthier-Villars, Paris, 1925);\\
 J. Kamp\'{e} de F\'{e}riet, {\sf La fonction hyperg\"{e}om\'{e}trique},
  (Gauthier-Villars, Paris, 1937);\\
 H. Exton, {\sf Multiple Hypergeometric Functions and Applications}, (Ellis
  Horwood, Chichester, 1976);
{\sf Handbook of Hypergeometric Integrals}, (Ellis Horwood,
  Chichester, 1978).
%
\bibitem{Blumlein:2018cms}
J.~Bl\"umlein and C.~Schneider,
Int. J. Mod. Phys. A \textbf{33} (2018) no.17, 1830015
[arXiv:1809.02889 [hep-ph]].
%
\bibitem{RYGR}
I.M.~Ryshik and I.S.~Gradstein, {\sf Summen-, Produkt- und Integraltafeln}, (DVW, Berlin, 1963), 2nd
Edition.
%
\bibitem{Antonelli:2019ytb}
A.~Antonelli, A.~Buonanno, J.~Steinhoff, M.~van de Meent and J.~Vines,
Phys. Rev. D \textbf{99} (2019) no.10, 104004
[arXiv:1901.07102 [gr-qc]].
%
\bibitem{Bern:2021dqo}
Z.~Bern, J.~Parra-Martinez, R.~Roiban, M.S.~Ruf, C.H.~Shen, M.P.~Solon and M.~Zeng,
Phys. Rev. Lett. \textbf{126} (2021) no.17, 171601
[arXiv:2101.07254 [hep-th]].
%
\bibitem{Blumlein:2021txj}
J.~Bl\"umlein, A.~Maier, P.~Marquard and G.~Sch\"afer,
Phys. Lett. B \textbf{816} (2021) 136260
[arXiv:2101.08630 [gr-qc]].
%
\bibitem{Dlapa:2021npj}
C.~Dlapa, G.~K\"alin, Z.~Liu and R.A.~Porto,
{\it Dynamics of Binary Systems to Fourth Post-Minkowskian Order from the Effective Field Theory
Approach}
[arXiv:2106.08276 [hep-th]].
%
\bibitem{Westpfahl:1979gu} K.~Westpfahl and M.~Goller,
Lett. Nuovo Cim. \textbf{26} (1979) 573--576.
%
\bibitem{Schafer:2018kuf}
G.~Sch\"afer and P.~Jaranowski,
Living Rev. Rel. \textbf{21} (2018) no.1, 7
[arXiv:1805.07240 [gr-qc]].
%
\bibitem{Damour:2000we}
T.~Damour, P.~Jaranowski and G.~Sch\"afer,
Phys. Rev. D \textbf{62} (2000) 084011
[arXiv:gr-qc/0005034 [gr-qc]].
%
\bibitem{BK87}
J.~Bl\"umlein and W.D.~Kraeft, Ann. Phys. (Leipzig) {\bf 44} (1987) 461--467.
%
\bibitem{LANCZOS}
C.~Lanczos, {\sf The Variational Principles of Mechanics}, (Dover, New York, 1970), 4th Ed.
%
\bibitem{YNDURAIN}
F.J.~Yndurain, {\sf The Theory of Quark and Gluon Interactions} (Springer, Berlin, 2006), 4th Ed.
%
\bibitem{AB}
A.I.~Achieser and W.B.~Berestezki, {\sf Quantenelektrodynamik}, (H.~Deutsch, Frankfurt a.M.,
1962);\\
N.N.~Bogoliubov and D.V. Shirkov, {\sf Introduction to the Theory of Quantized Fields}, (Interscience Publ.,
New York, 1959);\\
W.~Pauli, {\sf Lectures on Physics}, Vol.~{\bf 6} {\it Selected topics in field quantization},
Ed.~Ch.P.~Enz, (MIT Press, Cambridge,MA,  1977).
%
\bibitem{DISTR}
L.~Schwartz, {\sf Th\'eorie des distributions}, Vol.~{\bf 1,2}, (Hermann, Paris, 1951);\\
V.S.~Vladimirov, {\sf Gleichungen der mathematischen Physik}, (DVW, Berlin, 1972).
%
\bibitem{GELF}
I.M.~Gelfand, G.E.~Schilow und N.J.~Wilenkin, {\sf Verallgemeinerte Funktionen}, Vol.~{\bf 1--4}, (DVW,
Berlin, 1960--1964).
%
\bibitem{JP}
P.~Jordan and W.~Pauli, 1928 Z. Phys. {\bf 47} (1928) 151--173.
%
\bibitem{CAUSAL}
E.C.G.~St\"uckelberg and D.~Rivier, Helv. Phys. Acta {\bf 23} (1950) 215--222.
%
\bibitem{SOCH}
J.-K.V.~Sokhotski, {\it On definite integrals and functions used in series expansions}, Dissertation,  St.
Petersburg State University, 1873 (in Russian).
%
\bibitem{Foffa:2011np}
S.~Foffa and R.~Sturani,
Phys. Rev. D \textbf{87} (2013) no.4, 044056
[arXiv:1111.5488 [gr-qc]].
%
\bibitem{Galley:2005tj}
C.R.~Galley and B.L.~Hu,
Phys. Rev. D \textbf{72} (2005) 084023
[arXiv:gr-qc/0505085 [gr-qc]].
%
\bibitem{FRADKIN}
E.S.~Fradkin, Dokl. Akad. Nauk. SSSR {\bf 98} (1954) 47--50; {\bf 100} (1955) 897--900.
%
\bibitem{Edwards:1954cac}
S.F.~Edwards and R.E.~Peierls,
Proc. Roy. Soc. Lond. A \textbf{224} (1954) no.1156 24--33.
%
\bibitem{MB}
E.W.~Barnes, Quarterly Journal of Mathematics {\bf 41} (1910) 136--140;\\
H.~Mellin, Math. Ann. {\bf 68}, no. 3 (1910) 305--337.
%
\bibitem{WW}
E.T.~Whittaker and G.N.~Watson, {\sf A course of modern analysis}, (Cambridge University Press, Cambridge, 1963),
pp.~256.
%
\bibitem{PRIW2}
I.I~Priwalow, {\sf Einf\"uhrung in die Funktionentheorie}, Vol.~{\bf II}, (Teubner, Leipzig, 1969).
%
\bibitem{Vermaseren:1994je}
J.A.M.~Vermaseren,
Comput. Phys. Commun. \textbf{83} (1994) 45--58.
%
\bibitem{ASYMPTOTE}
J. C. Bowman and A. Hammerlindl, {\it
Asymptote: A vector graphics language},
TUGBOAT: \newline The Communications of the TeX Users Group, 29:2 (2008)
288--294; \newline {\tt https://asymptote.sourceforge.io/}
%
\bibitem{BLANC}
F.~Larrouturou, Q.~Henry, L.~Blanchet and G.~Faye,
{\it The Quadrupole Moment of Compact Binaries to the Fourth post-Newtonian Order: I. Non-Locality in Time and 
Infra-Red Divergencies},
arXiv:2110.02240 [gr-qc];
{\it The Quadrupole Moment of Compact Binaries to the Fourth post-Newtonian Order: II. Dimensional Regularization 
and Renormalization},
[arXiv:2110.02243 [gr-qc]].
\bibitem{HADAMARD}
J. Hadamard, {\sf Lectures on Cauchy's Problem in Linear Partial Differential Equations}, (Yale Univ. Press,
New Haven, CT, 1923), (Dover, New York, 1952);
\\
J.~L\"utzen, {\sf The prehistory of the theory of distributions}, (Springer, New York, 1982).
%
\bibitem{Almeida:2021xwn}
G.L.~Almeida, S.~Foffa and R.~Sturani,
Phys. Rev. D \textbf{104} (2021) no.12, 124075
[arXiv:2110.14146 [gr-qc]]; private communication Oct. 26, 2021.
%
\bibitem{Jantzen:2011nz}
B.~Jantzen,
JHEP \textbf{12} (2011), 076
[arXiv:1111.2589 [hep-ph]].
%
\bibitem{Khalil:2022ylj}
M.~Khalil, A.~Buonanno, J.~Steinhoff and J.~Vines,
{\it Energetics and scattering of gravitational two-body systems at fourth post-Minkowskian order},
[arXiv:2204.05047 [gr-qc]].
\end{thebibliography}
\end{document}